# Design and implementation of an Out of order execution engine of floating point arithmetic operations

Centro de Investigación en Computación

FACULTAT D'INFORMÀTICA DE BARCELONA

**Ing. Cristóbal Ramírez Lazo**

**Director:**
**PhD. Osman Sabri Unsal**
**Codirector:**
**PhD. Luis Alfonso Villa Vargas**
**Ponent:**
**PhD. Adrián Cristal Kestelman**

Barcelona School of Informatics - Universitat Politècnica de Catalunya
Computing Research Center - Instituto Politécnico Nacional

This thesis is submitted for the degree of
*Master of Science*

# Abstract

A floating point unit (FPU), also known as a math coprocessor, is a part of a processor to perform operations on floating point numbers.
Nowadays, almost all processors include a Floating point unit in the chip, this unit is more complex and consume more area in the chip and for this reason many processors share this unit between a pair of cores.
When a CPU is executing a program that calls for floating point operations and this is not supported by the hardware, the CPU emulates it using a series of simpler fixed point arithmetic operations that run on the integer arithmetic logic unit, causing low performance in this kind of applications.

The *Centro de Investigación en Computación* of the *Instituto Politécnico Nacional* work in a project currently in development called Lagarto to create intellectual property in embedded high performance processor architectures and operating systems to research and teach.
Lagarto II is a superscalar processor which fetches, decodes and dispatches up to two instructions per clock cycle, which will support a complete instruction set of 32-bits that operate in 64-bits data, this architecture is synthesizable in FPGAs devices.

In this thesis, work is undertaken towards the design in hardware description languages and implementation in FPGA of an out of order execution engine of floating point arithmetic operations. A first proposal covers the design of a low power consumption issue queue for out of order processors, register bank, bypass network and the functional units for addition/subtraction, multiplication, division/reciprocal and Fused Multiply Accumulate (FMAC) confirming with the IEEE-754 standard. The design supports double precision format and denormalized numbers; A second proposal is based on a pair of FMAC as functional units which can perform almost all Floating-point operations, this design is more beneficial in area, performance and energy efficiency compared with the first version.

## Acknowledgments

I would like to express my gratitude to my advisors Luis Alfonso Villa, Osman Unsal and professor Marco Antonio Ramírez who shared much of their time and knowledge to I could finish my thesis work. Furthermore, to Professors from the Computer Research Center of IPN and Barcelona School of Informatics of UPC, who walked me through this interesting research line.

In addition, I express my gratitude to CONACYT, to Computer Research Center of IPN and the project: SIP: 20150957 *"Desarrollo de Procesadores de Alto Desempeño para Sistemas en Chips Programables"* who financed part of my master degree.

And of course to my family: my parents Cristóbal Ramírez Salinas and Florina Lazo Osorio and my sister Itzel Ramírez Lazo who have always encouraged me to keep going.

# Table of contents

# List of Figures

# List of Tables

# Glossary

*Dynamic scheduling:* Strategies and techniques applied to the superscalar processors in order to exploit the instructions level parallelism given by a program.

*IEEE 754 Standard:* Technical standard for floating-point computation established in 1985 by the Institute of Electrical and Electronics Engineers (IEEE)

*High performance techniques:* Techniques applied to the superscalar processors in order to exploit the instructions level parallelism given by a program.

*In Order processors:* Processors which processes the instructions in the order that they appear in the binary (according to the sequential semantics of the instructions).

*Issue Queue:* Structure use in superscalar processors to allocate instructions that not comply with all conditions to be executed.

*Low power consumption techniques:* Techniques applied to the superscalar processors in order to save energy. These techniques usually are applied to processors for mobile devices where de autonomy is important.

*Out of Order processors:* Processors which processes the instructions in an order that can be different (and usually is) from the one in the binary. The purpose of executing instructions out of order is to increase the amount of ILP by providing more freedom to the hardware for choosing which instructions to process in each cycle.

*Pipeline:* Technique applied to the processors in order to increase the performance. Basically split the execution of each instruction into multiple phases and allow different instructions to be processed in different phases simultaneously. Pipelining increases instruction level parallelism (ILP), and due to its cost-effectiveness, it practically is used by all processors nowadays.

*Scalar processors:* Processor that cannot execute more than 1 instruction in at least one of its pipeline stages. In other words, a scalar processor cannot achieve a throughput greater than 1 instruction per cycle for any code.

*Superscalar processors:* Superscalar processor can execute more than 1 instruction at the same time in all pipeline stages and therefore can achieve a throughput higher than 1 instruction per cycle for some codes.

# Chapter 1

# 1. Introduction

## 1.1. Motivation

The incessant search of methods and techniques to improve the performance in the processors, which are the basic elements for the functionality of all types of modern devices, from super computers to cellphones, has led to the development of a kind of microarchitecture called superscalar, which has the capacity to perform fetch, decode and dispatch of two or more instructions per clock cycle.

To obtain high performance, modern superscalar processors use many building blocks implemented in hardware as register renaming, dynamic branch predictors and speculative instruction execution. These techniques have the objective to perform dynamic scheduling to expose the maximum amount of instruction level parallelism found in a program, keep busy at maximum the functional units of the processors and for that superscalar processors must be able to perform out of order execution.

The *Centro de Investigación en Computación* of the *Instituto Politécnico Nacional* work in a project currently in development called *Lagarto* to create intellectual property in embedded high performance processors architectures and operating systems to research and teach. The first model (*Lagarto I*) is a scalar pipelined processor, which executes one instruction per clock cycle and is based in MIPS 32-bits architecture developed by PhD. John Hennesy with some modifications. A second version is a superscalar processors called *Lagarto II*, which fetches, decodes and dispatches up to two instructions per clock cycle, which supports a complete instruction set of 32-bits that operate on 64-bits data, both architectures are synthesizable in FPGAs devices. [1]

Superescalar architectures include a large number of components to support out of order execution. Instructions are fetch, decoded, renamed, and if the instruction queue has free locations, are dispatch in order. These instructions waiting for their source operands are ready, and that the corresponding functional unit is found free; should comply with these conditions to be issued to the execution units. *Lagarto II* can execute operations out of order; it will exploit instruction-level parallelism given by the superscalar architectures.

Until today, Lagarto architecture can't execute the instruction set of floating point operations because it lacks a hardware floating point unit, and for this reason it performs floating-point operations by software, causing poor performance in applications that require computing with numbers in floating point format.

## 1.2. Objectives

### General objective

Design and implement an out of order execution engine of floating point arithmetic operations for the super scalar processor *Lagarto II*.

### Specific objectives

Design and implement in hardware description languages of:

- A Floating Point Instruction Queue with out of order Issue.
- A Floating Point Register bank.
- A set of FP Functional units for arithmetic operations:
    - **Add/Subtract**
    - **Multiply**
    - **Divide**

  All functional units should support double precision format (64 bits) and Sub-normal numbers.
- Forwarding unit (Bypass).

Using techniques for high performance and low power consumption.

## 1.3. Justification

The design of the hardware structures that constitute the functional blocks of the processor architecture is not a trivial task. This task becomes even more complicated when the processor architecture is designed to be superscalar and dynamic scheduling with out of order execution. Because, while higher performance is achieved, the complexity of the structures that compose it increases.

Initially the trend of superscalar processors was to obtain the best possible performance, regardless of the energy cost, but now that trend has changed and this because its use in mobile devices is overwhelming nowadays, for that reason it is required to have high performance processors but with low power consumption. Therefore, in this thesis, we propose to design and implement the out of order execution engine of floating point arithmetic operations, using techniques of high performance and low power consumption.

Although there are several companies of processors (AMD, Intel, Nvidia, Siemens, Texas, etc.) the design of their architectures is the intellectual property of companies and techniques implemented to improve the performance are trade secrets.

According as a country is able to develop its own technology, will be able to eliminate not only economic dependence, also knowledge dependence on other countries and generate wealth that is reflected in its population. A clear example is Chinese, who for two decades established a state policy aimed at giving a strong impulse to the development of science and technology in that country. A particular effort made as part of that policy was the research and development of a microprocessor manufactured locally resulting in 2002 of a CPU which became known as Godson1, which is based on the architecture MIPS and is capable of running the Linux operating system. In 2008 it was announced market entry of a low-cost laptop called Yeelong with a Loongson 2F processor. With this project the Chinese government made the proposal that everyone with low purchasing power can have access to a personal computer. Thus China is able to use these processors as an engine of its growing electronics industry, to use them in all sorts of devices ranging from cars to mobile devices, which would allow China to obtain all the benefits that entails eliminating dependence on technology foreign [2].

Other similar project is called RISC-V, this project was originated in 2010 by researchers in the Computer Science Division of the EECS Department at the University of California, Berkley. RISC-V is a new instruction set architecture (ISA) that was originally designed to support computer research and education and this group hope will become a standard open architecture for industry implementations. Also they provide a high performance, energy-efficient processor called Rocket, which is a 6-stage single-issue in-order pipeline that executes the 64-bit scalar RISC-V ISA. Furthermore, implements an MMU that supports page-based virtual memory and is able to boot modern operating systems such as Linux. Rocket also has an optional IEEE 754-2008-compliant FPU, which implements both single- and double-precision floating-point operations, including fused multiply-add. Developing a CPU requires expertise in several specialties: Computer architecture, compiler design and operating system design [3].

México could follow this example and get the same benefits by doing a similar effort. However, research and development in the field of computer architecture design has not been exploited enough, which is why this thesis will be a great contribution to what will become the first superscalar processor designed in Mexico, and further, as any modern processor, must have an execution engine of floating point arithmetic operations.

## 1.4. Organization

The rest of the thesis is composed of the following chapters. Chapter 2 provides a brief background about superscalar architectures and deepens in issue stage, read register stage and execution stage; also we introduce the Floating Point Arithmetic and IEEE 754 standard. Chapter 3, describes the state of the art of the issue queues, register files, Floating-point functional units and a pair of examples of current microprocessors and its FP engine. Chapter 4 describes two designs of the out of order execution engine, second design is an improvement of the first design. In Chapter 5 are shown the implementation results. Chapter 6 presents the testing of the final design and finally Chapter 7 provides the conclusion of this thesis work, future work and research's products.

# Chapter 2

# 2. Background

In this chapter first we introduce the superscalar architecture in order to place this work in context with, making emphasis in the Issue stage, Read Register Stage and Execution Stage which are the goals of this thesis work. After that, we talk about IEEE-754 standard which is a technical standard for floating point computation, define the arithmetic formats, interchange formats, rounding rules, operations and exception handling, and finally, we talk about the instruction set architecture which will be supported by Lagarto II processors, which is for academic and research the MIPS 64 revision 6.

## 2.1. Superscalar Architectures

Nowadays, we can find embedded processors in many components, such as smartphones, game consoles, cars, etc.
Superscalar architectures can be classified in in-order and out-of-order execution fashion.

An in-order processor executes the instructions in the order that they appear in the binary, whereas an out-of-order processor executes the instructions in an order that can be different from the one in the binary. The purpose of executing instructions out of order is to exploit the ILP and increase the performance by the superscalar architectures [4].

Modern superscalar architectures include a large number of elements in order to support the out of order execution. Instructions are fetched in order from the instruction cache, are decoded to understand their semantics. After, most processors apply some type of renaming to the register operands to remove the false dependences introduced by the compiler in order to identify and exploit parallelism in the instruction stream [5]. Then, instructions are dispatched to various buffers, depending of the kind of instruction. Non-memory instructions are dispatched to the integer issue queue or FP issue queue, whereas memory instructions are dispatched to the load/store queue. These instructions waiting for their source operands are ready, and that the corresponding functional unit is found free; should comply with these conditions to be issued to the execution units.

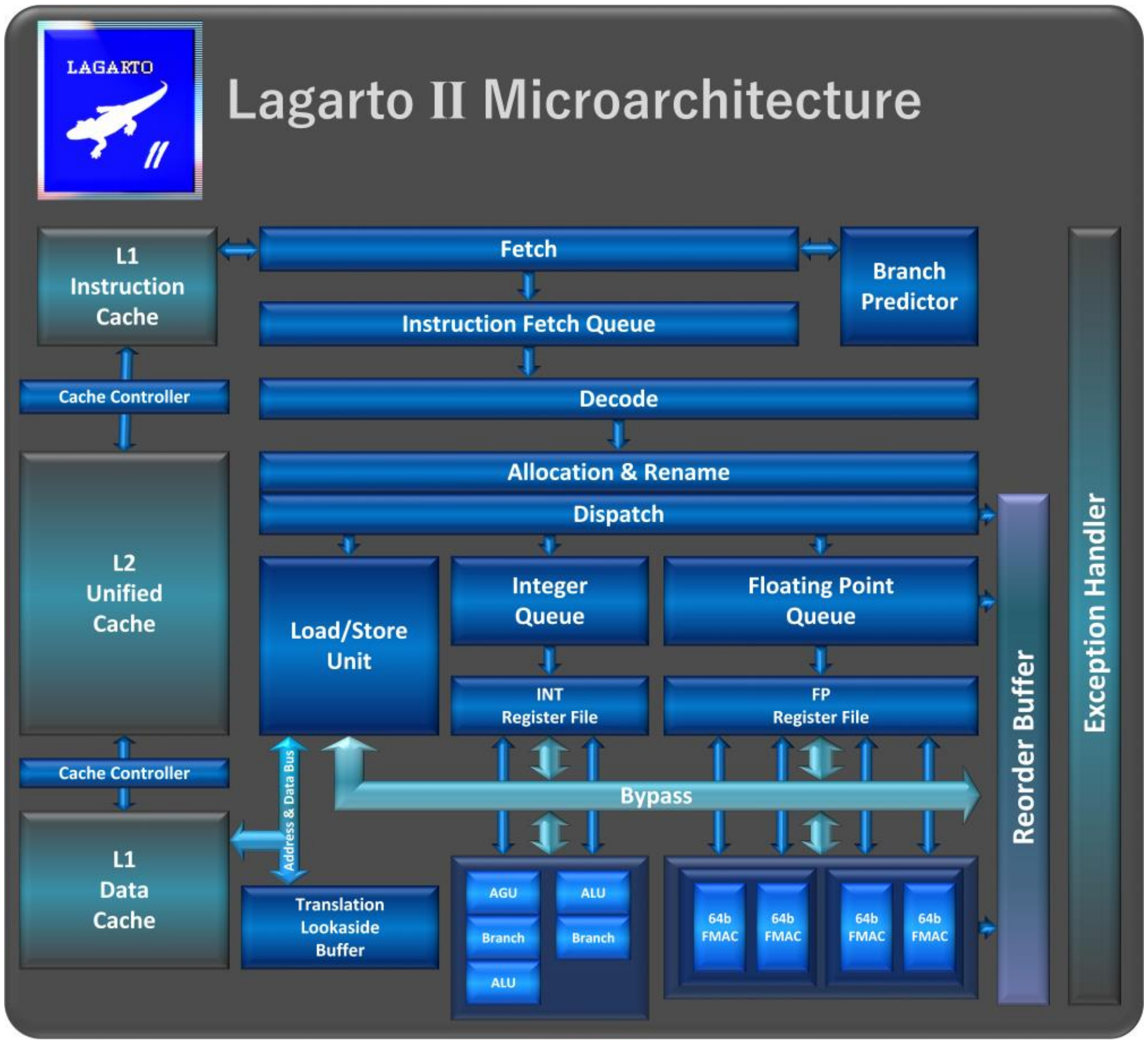

**Fig. 2.1** Lagarto II Microarchitecture

An instruction remains in the reorder buffer until it commits. The goal of the reorder buffer is to preserve the order until the instruction finalizes and to store information about the instruction that is useful for its execution but also to recovery of some error if it is necessary. Finally, non-speculative instructions commit their results in program order. In Figure 2.1 is shown the microarchitecture of Lagarto II which is a superscalar processor.

Basically the superscalar processors can be divided in the Frontend and Backend, where the first one always is in-order and include the Fetch, decode and rename stages, whereas the second, could be in-order or out-of-order and include the issue, execute/writeback and finally the commit which is always in order.  In Figure 2.2 we show this division.

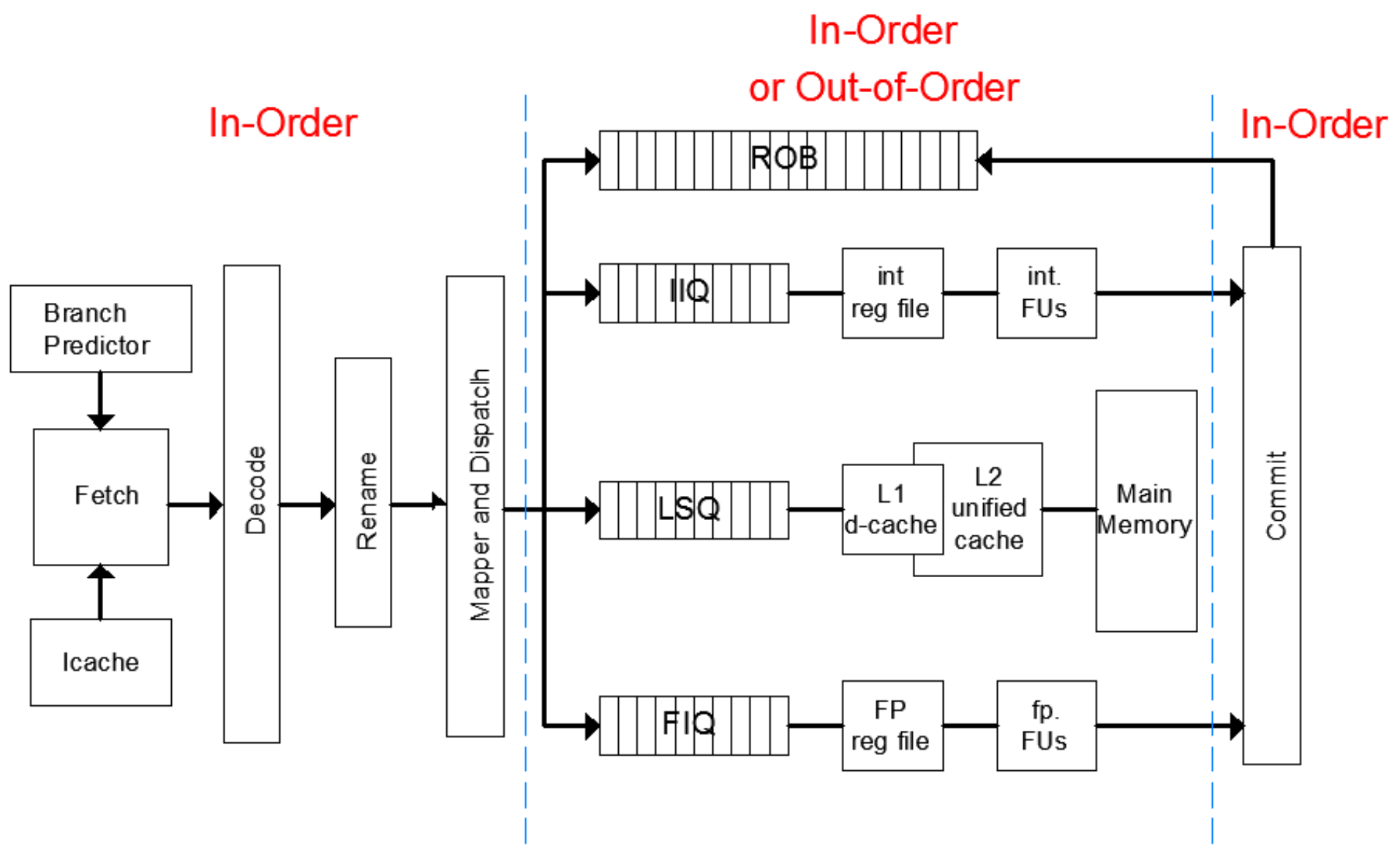

**Fig. 2.2** Backend and Frontend in a superscalar processors

The first part of the pipeline is responsible of the fetching instructions. The main components of this stage are the instruction cache memory, where the instructions are allocated, and a branch predictor, which one determines if the current instruction is a branch but it also takes the decision whether about take or not take this branch while the fetch stage determines the address of the next fetch cycle.

The second stage is the instruction decode. The main components of this part are ROM decoders and ad-hoc circuitry; the main objective is to identify the main attributes of the instruction such as type and resources that it will require for their execution.

The third stage is Register renaming which goal is to change the names of the logical source registers by its corresponding physical register tags mapped in last cycles, also assign new physical register tags to the logical destination with the purpose of removing all false dependences. This is done normally though a set of tables that contain information about the current mapping of logical names to physical ones and what names or tags are not being used, together with some logic to analyze dependences among the multiple instructions.

The fourth stage is the instruction dispatch, is responsible of reserve different resources that the instruction will use, including entries in the reorder buffer, issue queue and load/store buffers. If resources are not available, the processor performs a stall until these resources become free. All the above steps are performed in order [4].

From here on we will emphasize the backend part, deepening in the Issue Stage, read register and execution which are the goals of this work, also to understand the recovery in case of miss speculation we will talk a little about the commit stage.

### 2.1.1. Issue Stage

The Issue stage is in charge of sending instructions to the execution units. There are two types of issue schemes: In order and out of order. The first one scheme sends the instructions in the program order, whereas the second scheme sends the instruction out of the program order as soon as their source operands become available. Most of the latest processors implement out-of-order schemes. There are many different ways of implementing an out-of-order issue.

#### In-order issue logic

In-order issue logic issues the instructions for execution in the same order they were fetched. Therefore, instructions wait until all previous instructions have been issued. Then, the instruction is issued as soon as its source operands are available and the resources it needs for execution are ready. This kind of issue logic is sometimes implemented at the decode stage of the processor due to its simplicity using scoreboarding. A typical scoreboard comprises two tables, a data dependence table and a resource table [4].

#### Out of order issue logic

The issue logic is a key component that determines the amount of instruction level parallelism that processors are able to exploit. It allows out-of-order execution by issuing instructions to the functional units as soon as its source operands become available. However, the hardware components involved in the issue process sit in the critical path of the processor pipeline [4]. Researches have used a variety of schemes to implement the issue queue; also several recent proposals have attempted to reduce the issue logic's complexity and power.

One of the most common ways to implement the issue logic is based on random access memory RAM and content-addressable memory (CAM) array structures called RAM-CAM Schemes as we can see in the Figure 2.3. These structures can store several instructions, but generally fewer than the total number of in-flight instructions. Each entry contains an instruction that has not been issued or has been issued speculatively but not yet validated and thus might need to be rescheduled.

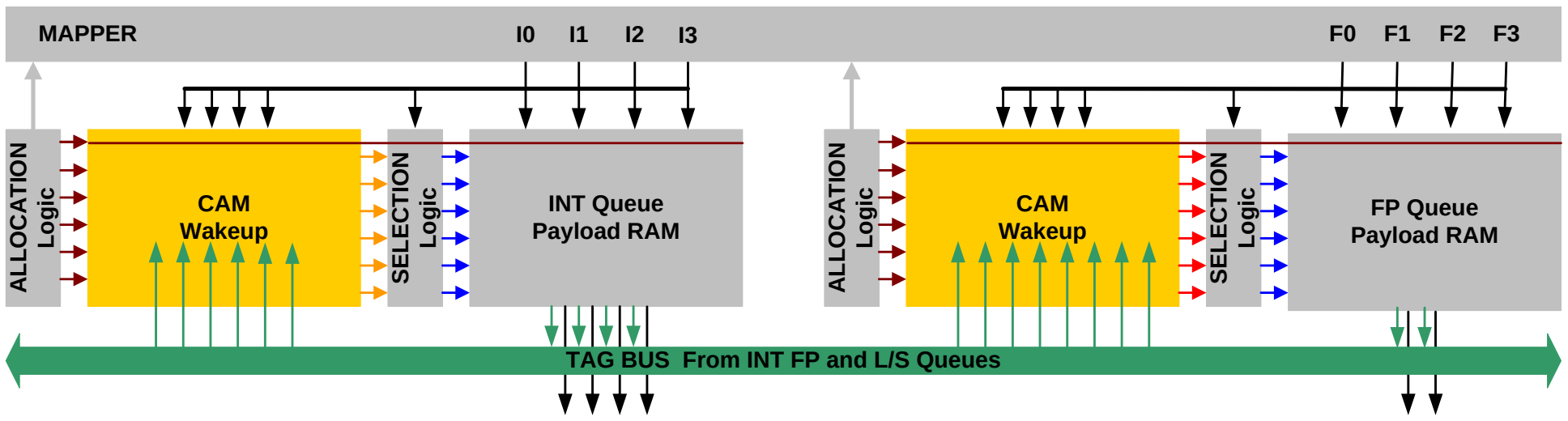

**Fig. 2.3** Issue Queue - RAM/CAM Scheme.

After the issue logic selects an instruction for execution, it broadcast the instruction's destination tag to all the instructions in the issue queue. The wakeup logic compares each source tag in the queue with the broadcast tag and, if there is a match, marks the operand as ready. This process is known as wakeup. A superscalar processor can broadcast and compare multiple tags in parallel. Figure 2.4 shows a block diagram of the issue logic for one entry of the issue queue [6].

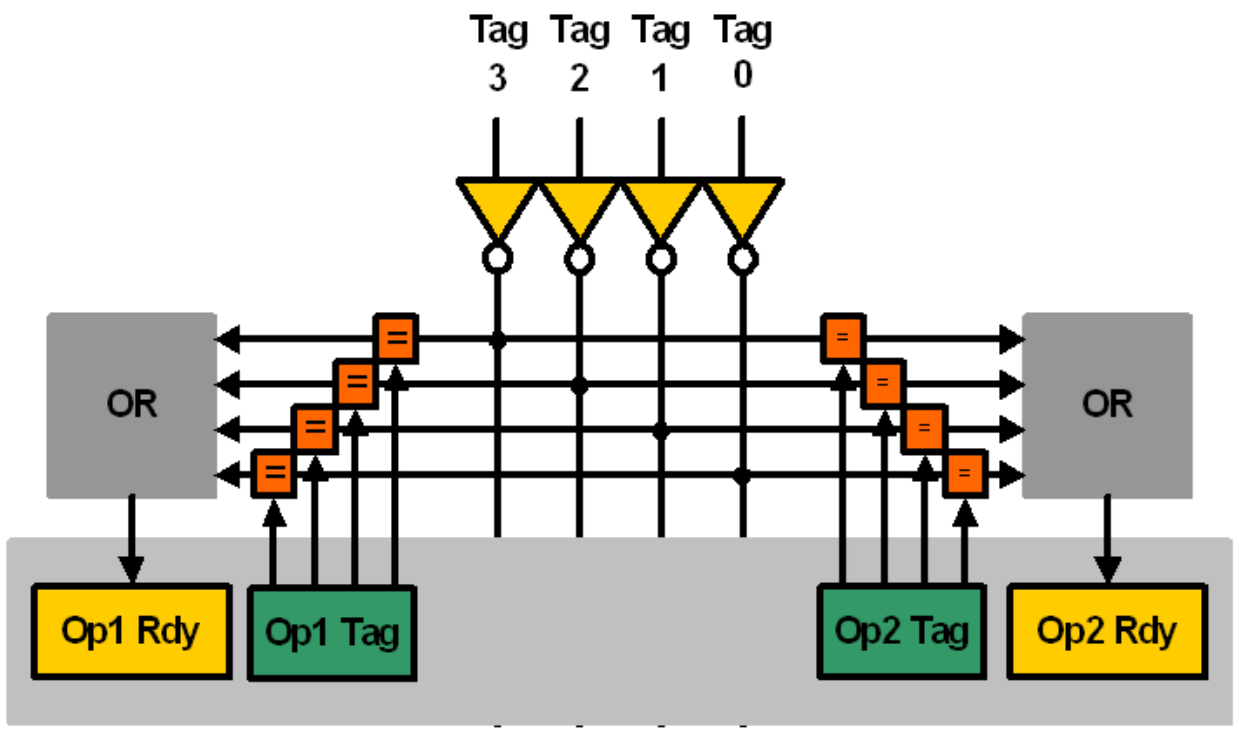

**Fig. 2.4** Issue logic for an entry in a CAM/RAM array.

The selection process identifies instructions whose source operands are ready and whose required resources are available, and then issues them for execution. When more than one instruction is ready and competes for the same resource, the selection logic chooses one of them according to some heuristic like the oldest first or the longest latency first [6].

Overall, the issue logic's main source of complexity and power dissipations the many tag comparisons it must perform every cycle. Researches have proposed several approaches to improve the issue logic's power efficiency, these are described in the following chapter.

### 2.1.2. Read Register stage

Once the instruction is issued, some bits (tag) go to read the source operands to the Floating point register file to send all needed information to the execution stage. The instruction set architecture of a CPU almost always defines a set of registers, which are used to exchange data between memory and the functional units on the chip. In simpler CPUs, these architectural registers correspond one-for-one to the entries in a physical register file within the CPU. Superscalar CPUs use register renaming, so that the mapping of which physical entry stores a particular architectural register changes dynamically during execution, as is explained in section 2.1.

Processors in general have a small number of architectural registers (32 integers and 32 FP) and, as consequence, name dependences through registers are very common, and the benefits of getting rid of them in an out-of-order processor are huge.

High performance processors use an out of order execution scheme in order to exploit the instruction level parallelism (ILP) existent in the program's code. Processor examines a large window of in-flight instructions to find all possible ready instructions capable to execute every cycle. The size of the windows is some of the key determinants of the IPC achieved by the processor. However, if the processors support a large window of in-flight instructions, it requires a large register file and issue queues, which can compromise the cycle time [7].
Another important aspect to take account is that a large issue-width in the processors also requires a large number of read/write ports in the register file. The access time of the register file basically depends on both the number of entries and the number of ports. The register file is a heavily ported RAM structure. A processor capable of issuing eight floating point instructions each cycle may need a floating point register file with sixteen read ports to read two source operands per instruction and eight write ports to write the result of each functional unit, also need other extra ports to interchange information with the integer Register file and one more port to write the new data from the load/store unit.

Register files in dynamic superscalar processors have been a very modestly sized. The Alpha 21264 processor has as many as 80 integer physical registers and 72 floating-point physical registers and use a clustered organization to reduce the number of ports and the access time [8]. Many researches were performed in order to improve the access time and the energy consumption in the register file; some of those are described in the next chapter.

### 2.1.3. Execution Stage

In this stage the instructions results are calculated, the values of the source operands are send to the execution units with extra information like the kind of operation, also, in case of floating point operations send the precision, format and round method. There are several types of operations that the processor can perform in the execution stage. The most common are the arithme-tic operations like addition, multiplication, etc. Memory Instructions operate on data either by loading them from memory to registers or by storing them from registers to memory. Control-flow instructions change the value of the Program Counter (PC) register. More infrequently, specialized instructions can change the machine state by operating on control registers (special registers that define how the processor behaves).

Naturally, the different types of operations have different complexity and, as a consequence, different latency. For this reason, in contemporary microprocessors, the execution stage is not a single pipeline stage. Also, there are usually several different paths in the processor pipeline that an instruction can follow when it reaches the execute stage. The most obvious ones are the integer path, memory path, and the floating-point path.

Another important aspect of the execution stage is the bypassing network. This is the network responsible for forward value results of the computation among the various functional units, the data cache and the register file. In high performance microprocessors, some form of bypass is necessary if we want to provide back-to-back execution of dependent instructions. Because of its importance to performance and its complexity, the bypass network is one of the critical components of the execution stage.

### Floating Point Unit

This unit operates on two floating-point values coming from the floating-point register file or the memory across of bypass network and produces a floating-point result. A floating-point unit (FPU) performs arithmetic operations such as addition, subtraction and multiplication. Depending of the implementation, it can also perform division, square root and other complex operation as trigonometric functions, exponentials, etc. Normally, floating point and integer register file state is kept in separate structures. Depending on the architecture, there may be instructions that convert the floating point values to integers and vice versa. Conversion operations are also implemented in the floating-point unit.

IEEE-754 standard is a technical standard for floating point computation, define the arithmetic formats, interchange formats, rounding rules, operations and exception handling, later we will deepen in this standard. The FPU is a very complex unit, and it is generally several times bigger than the integer units.

### Bypass Logic

When executing instructions in a pipeline, the result of a computation does not update the machine state until the commit stage, which may be many cycles after the result was generated. The result of the computation becomes speculatively available after the write-back stage. The write-back stage is when the result of a functional unit is sent to the architectural register file, to the merged register file, to the reorder buffer, the rename buffer and so on, depending on the machine design (in-order, out-of-order, etc.).

Having bypasses improves the executed instructions per cycle metric (IPC), but it may affect the cycle time and/or power of the microprocessor. Most processors today implement some form of bypass. The notable exception is the IBM POWER5 [38] processor, where the designers opted to not implement a bypass network in order to keep complexity low and in consequence the frequency high.

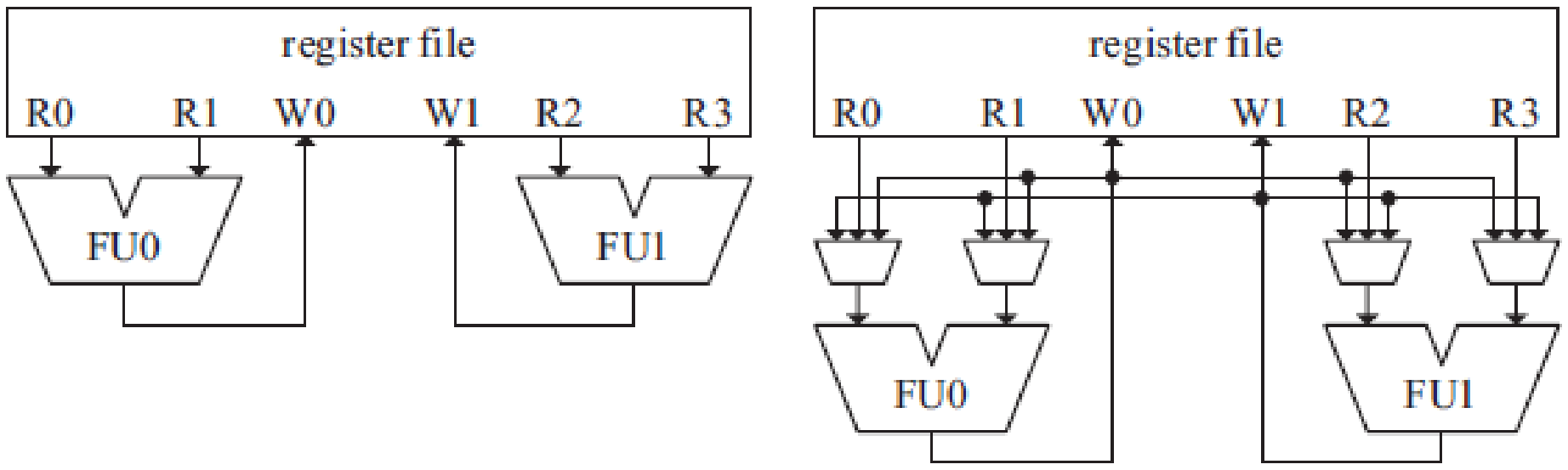

**Fig. 2.5** Simple execution engine with two functional units, without (left) and with (right) value bypassing.

If the processor does not implement a bypass, each input of a functional unit is connected directly to a read port of the register file to read the source value. Similarly, the result of the functional unit is connected directly to a write port of the register file. If we want to implement value bypassing, the source value of a functional unit can come from three different places in the machine in this design: the register file (i.e., no bypass), the functional unit itself and other functional unit. Thus, we need a 3:1 multiplexor at the input of each functional unit. Also, the results of the functional units, instead of connecting directly to the register file, now form a bus that spans the width of the execution engine (called the result bus) and connect to all the functional unit input multiplexors [4].

### 2.1.4. Commit Stage

A processor operates with two separate states: the architectural state and the speculative state. The architectural state is updated at commit as if the processor would execute instructions in sequential order. By contrast, the speculative state implies the architectural state plus the modifications performed by the instructions that are in-flight in the processor. This latter state is called speculative because it is not guaranteed that these modifications will become part of the architectural state. Note that conventional processors rely on speculative techniques like branch prediction or speculative memory disambiguation in order to keep executing instructions. Thus, if some of these speculations fail or an exception occurs, the speculative state becomes invalid, and it never turns into architectural state.

In out of order processors where the execute instructions is accomplished out of the original order, we need to emulate the sequential execution of instructions through the implementation of an additional stage called commit at the end of the pipeline. Instructions flow through this stage in the original program order. Then, any changes that instructions do on previous pipeline stages are considered speculative and do not become part of the architectural state until they reach commit. At this point, we say that the instruction finalizes.

Finally, since the commit is the last stage on the execution path of an instruction, this is the place where hardware resources allocated by the decode stage, like reorder buffer (ROB) entries or physical registers are recycled. Note that an instruction should only reclaim those resources that are not used anymore. Therefore, for those configurations where the instructions write their outcome in a physical register, the reclamation of this physical register should be done by the time we know for sure that the content of the register would not be needed anymore. Thus, before reclaiming a register, we need to be sure that all instructions that may require the value of this register in the future have already read it or they will be able to read it from a different place.

Processors that implement a merged register file like Intel Pentium 4 [9], Alpha 21264 [8] or MIPS R10000 [10] use the same register file for the values belonging to the architectural state and the speculative values. Basically, a physical register is allocated by an instruction to store the result's value, and this register will hold this value until it is not needed anymore, even if the instruction commits. The compiler uses again this register when its value is not necessary for the program flow and it is recycled.

### Recovery in case of Misspeculation

Instructions that are in flight have sometimes to be flushed due to multiple reasons (e.g., branch misprediction, exceptions). If these instructions have gone through the allocate stage, then the resources that they reserved must be released. Besides, the modifications that these instructions did in the register alias tables (RAT) must be undone so that RAT reflect the same state they would have if these instructions never would have been executed.

In the event of a branch mispredition, the speculative state of the machine is incorrect because the processor has been fetching, renaming and executing instructions from wrong path. Therefore, when we identify a branch misprediction, the speculative processor state and the program counter should be restored to the point where start the correct path.

Recovery mechanism after a branch misprediction is typically split into two separate tasks: front-end recovery and back-end recovery. The front-end recovery is usually simpler than the backend recovery. In general, recovering the front-end implies flushing all intermediate buffers where instructions fetched from the wrong path are waiting to be renamed, restoring the history of the branch predictor and updating the program counter to resume fetching instructions from the correct path. By contrast, recovering the back-end implies removing all instructions belonging to the wrong path residing on any buffer like the Issue queue, Reorder buffer, etc. Moreover, RAT's should be restored as well in order to properly rename instructions from the correct path. Finally, back-end resources like physical registers or issue queue entries allocated by wrong-path instructions should also be reclaimed.

Therefore, processors like the MIPS R10000 or Alpha 21264 rely on a checkpoint mechanism in order to reduce the distance of the traversal between the current execution point to mispredicted branch. These processors periodically take a snapshot of the content of the RAT so that the log does not have to be fully traversed, but the traversal begins on an instruction where a checkpoint was taken.

For example, in case of MIPS R10000 processor, the first checkpoint younger than the branch is copied into the RAT, and the renaming log is traversed backwards until the mispredicted branch is found. Every entry on the log includes the previous mapping of the logical register that the renamed mechanism overwrote. Then, the RAT is restored based on this information in order to reflect the precise state of the processor at the moment when the branch was mispredicted.

Besides the Rename scheme, other information like, for instance, the list of available physical register identifiers should be reclaimed to free those registers allocated by the instructions in the wrong path. Some processors like Alpha 21264 implement the list of free physical registers as part of the Rename Structures. Then, this list is restored starting the traversal from the checkpoint in the same form that it is done for the renaming table [4].

## 2.2. Floating Point Numbers

In many scientific and engineering computations, numbers in a wide range, from very small to extremely large, are processed. Fixed-point number representations can represent fractions but the fundamental problem is that this notation is severely limited, not have enough range for this application. For example, a fixed point decimal number system capable of representing both $10^{-20}$ and $10^{20}$ would require at least 40 decimal digits and even then, would not offer much precision with numbers close to $10^{-20}$. Most of the real values will have to be represented in an approximate manner; the scientific notation used to overcome this issue, and is called floating-point representation. In general, floating point numbers are generally of the form

$$(-1)^s x F x 2^E$$

Where F represents the value in the fraction field, E represents the value in the exponent field and S is the sign and its codification is shown in Figure2.6. This chosen size of exponent and fraction give an extraordinary range. Fractions almost as small as $2.0_{ten}x10^{-38}$ and number almost as large as $2.0_{ten}x10^{38}$ can be represented in a computer for a single precision format.

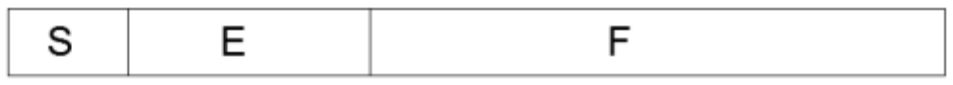

**Fig. 2.6** Codification of Floating Point Numbers

An overflow interrupts can occur in floating point arithmetic as well as in integer arithmetic operations. One way to reduce chances of underflow or overflow is to offer another format that has a larger exponent. In C language this number is called double, and operations on doubles are called double precision floating point arithmetic; single precision floating point arithmetic is the name of the earlier format which in C is called float. They are part of the IEEE 754 floating point standard.

### 2.2.1. IEEE 754 standard

IEEE-754 standard which is a technical standard for floating point computation established in 1985 by the Institute of Electrical and Electronics Engineers (IEEE) [11] in order to improve the portability of floating-point computations. The standard can be implemented in hardware, software or a combination of both.

IEEE 754 standard defines much more than just the representation; the main aspects are list below:

- Basic and extended floating-point formats
- Operations
- Rounding rules
- Exception handling

**Floating Point Formats**

IEEE 754 standard define three floating point formats:

- 32-bit single-precision floating point (Figure 2.7).
- 64-bit double-precision floating point (Figure 2.8).
- 80-bit Extended-precision floating point

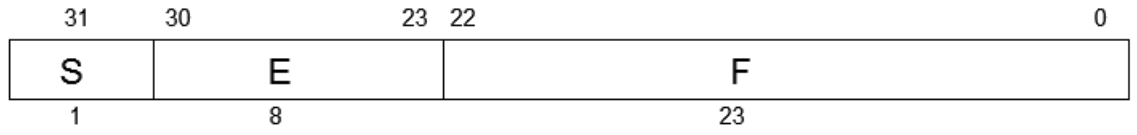

**Fig. 2.7 .** Single-Precision Floating Point Format (S)

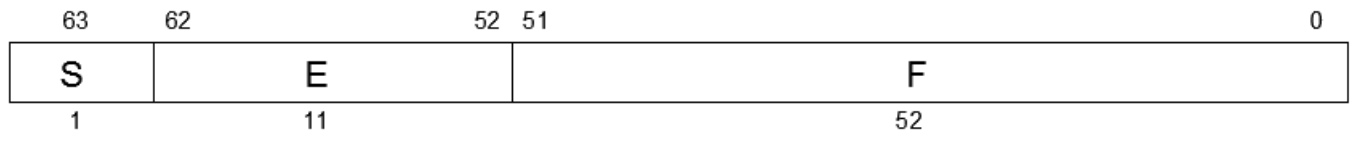

**Fig. 2.8** Double-Precision Floating Point Format (D)

The floating point data types represent numeric values as well as other special entities, such as the following:

- Two zero representations, -0 and +0.
- Two infinities, +∞ and -∞.
- Signaling non-numbers (SNaNs).
- Quiet non-numbers (QNaNs).
- Normal Numbers
- Subnormal Numbers

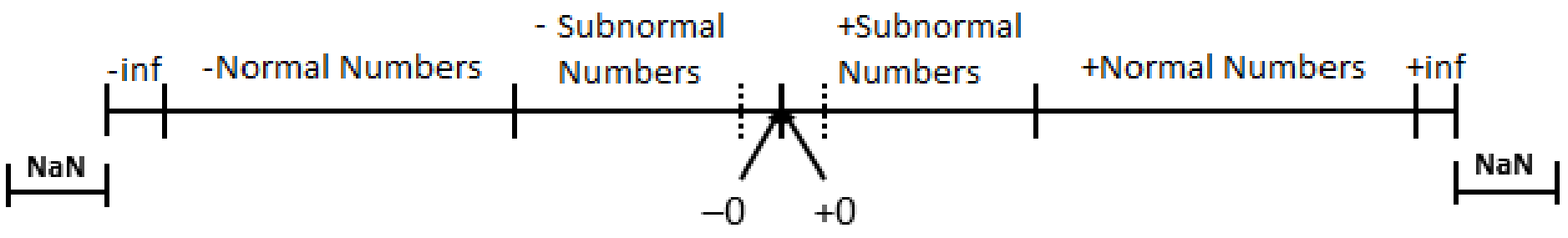

**Fig. 2.9** Floating point representations.

The set of finite floating-point numbers representable within a particular format is determined by the following parameters:

$b$: The radix
$p$: The number of digits in the significant (precision)
$emax$: The maximum exponent
$emin$: The minimum exponent, which is equal to 1- $emax$ for all formats

The smallest positive normal floating-point number is $b^{emin}$ and the largest is $b^{emax} \times (b - b^{1-p})$. The non-zero floating point numbers for a format with magnitude less than $b^{emin}$ are called subnormal because their magnitudes lie between zero and the smallest normal magnitude.

Table 2.1 defines the parameters of basic floating-point formats.

**Table 2.1** Parameters defining basic format floating point numbers

| Parameter | Binary32 | Binary64 |
|---|---|---|
| $p$ | 24 | 53 |
| $emax$ | +127 | +1023 |

Table 2.2 shows an example of representations of floating-point data types.

**Table 2.2** Floating-point representations

| Data Type | Representation for double precision format | | |
|---|---|---|---|
| | **Sign** | **Exponent** | **Fraction** |
| -Zero | 1 | 00000000000 | 0000000000000000000000000000000000000000000000000000 |
| +Zero | 0 | 00000000000 | 0000000000000000000000000000000000000000000000000000 |
| -Infinity | 1 | 11111111111 | 0000000000000000000000000000000000000000000000000000 |
| +Infinity | 0 | 11111111111 | 0000000000000000000000000000000000000000000000000000 |
| QNaN | x | 11111111111 | 1xxxxxxxxxxxxxxxxxxxxxxxxxxxxxxxxxxxxxxxxxxxxxxxxxxx |
| SNaN | x | 11111111111 | 0xxxxxxxxxxxxxxxxxxxxxxxxxxxxxxxxxxxxxxxxxxxxxxxxxxx |
| -Subnormal | 1 | 00000000000 | 0000010000000000000000010000000000000000000010000000 |
| +Subnormal | 0 | 00000000000 | 0000000000000000000000000000000000000000000000000000 |
| -Normal | 1 | 00100000100 | 1000000000100000000000000000001000000000000000000000 |
| +Normal | 0 | 00100000100 | 1000000000100000000000000000001000000000000000000000 |

Table 2.3 exhibits the span of each floating-point format.

**Table 2.3** Span of IEEE 754 Floating Point Formats

| Format | Min Subnormal | Min Normal | Max Normal |
|---|---|---|---|
| Single | 1.4E-45 | 1.2E-38 | 5.96E38 |
| Double | 4.9E-324 | 2.2E-308 | 1.8E308 |

**Floating Point Rounding**

Most arithmetic operations do not result in a number that can be represented exactly. In such cases the result need to be rounded to a number that can be represented in a given format. IEEE-754 standard define four rounding modes listed in Table 2.8.

The most popular mode is round toward nearest, ties to even. This rounding mode generally introduces the smallest error, as the result of round toward nearest is the number closest to the exact value. However, certain applications such as interval arithmetic perform better on simpler rounding mode like round toward zero. For this reason, IEEE-754 includes directed rounding modes as well.

**Table 2.4** Rounding Definitions

| Round | Meaning |
|---|---|
| Round to Nearest | Rounds the result to the nearest representable value. When two representable values are equally near, the result is rounded to the value whose least significant bit is zero (that is, even) |
| Round Toward Zero | Rounds the result to the value closest to but not greater than in magnitude than the result. |
| Round Towards Plus Infinity | Rounds the result to the value closest to but not less than the result. |
| Round Towards Minus Infinity | Rounds the result to the value closest to but not greater than the result. |

**Exceptions**

The following five exception conditions defined by the IEEE standard are described below:

- *Invalid Operation Exception*: Some arithmetic operations are invalid, such a division by zero or square root of a negative number. The result of an invalid operation is a NaN.
- *Division by Zero Exception:* The division of any number by zero gives infinity as a result.
- *Underflow Exception:* Two evens cause this exception, smallness and loss of accuracy.
- *Overflow Exception:* Is signaled whenever the result exceeds the maximum value that can be represented due to the restricted exponent range.
- Inexact Exception: This exception should be signaled whenever the result of an arithmetic operation is not exact due to the precision range.

# Chapter 3

# 3. State of the Art

A typical superscalar processor fetches and decodes more than one instruction at a time. As part of the instruction fetch process, the conditional branch instructions are usually predicted to ensure an uninterrupted stream of the code. The incoming instruction stream is then analyzed for data dependences, and instructions are distributed to functional units, often according to instruction type. Next, instructions are initiated for execution in parallel, based primarily on the availability of operand data, regardless of the order of the original program. This important feature to exploit the instruction level parallelism (ILP), present in the code of applications, is referred in general as *dynamic instruction scheduling* [12].

In hardware terms, the processor needs hardware resources to execute multiple instructions in parallel, more precisely a superscalar processor with dynamic instruction scheduling implements:

- Fetch strategies that simultaneously fetching multiple instructions.
- Branch prediction strategies that predict the execution path of the instructions and fetching speculative code.
- Methods for determining true dependences involving register values, and mechanism for communicating these values to where they are need during execution.
- Methods for issue multiple instructions in parallel (Out of order execution).
- Resources for parallel execution of many instructions.
- Methods for committing the process state in correct order; these mechanisms maintain an outward appearance of sequential execution.

Modern superscalar architectures include a large number of elements in order to support the out of order execution.

Superscalar processors like Pentium 4 were processors which was designed to operate with very high clock frequency up to 10 GHZ [13], (with very good cooling system), the performance was the main point back then. However, with the entry of mobile devices in the market like smartphones, tablets, Laptops, etc., the trend of the design of superscalar processors has changed, the new trend was the low power consumption in order to have grater energy independence in these devices. Nowadays, this trend continues, new Laptops as ASUS Ultrabook series, put a low power consumption processors AMD like A6, A8 or A10, which work in normal mode at 1.6 GHZ. If the operating system detects that the load work is little, the frequency is scaled down to 1GHZ or less and have an autonomy proven of HDVT (720p) video playing for 7 hours [14]. Many researches have been performed in order to obtain a good performance but with low power consumption in superscalar processors.

Following we shown the state of the art of the issue queue which in processors like Pentium 4 is one of the main consumers of energy responsible for approximately 25% of the total energy consumption [15], also we discus about the state of the art of the register banks and the floating point functional units.

## 3.1. Issue Queue

Researches have proposed several approaches to improve the issue logic's power efficiency. This approach can be classified in two groups: Static approaches, which use fixed structures, and dynamic approaches, which dynamically adapt some structures according to the properties of the executed code.

Buyoktusunoglu, Shuster, Brooks, Albonesi and Cook [16] proposed a circuit design to adaptively resize an instruction queue partitioned into fixed size blocks (32 entries and 4 blocks were studied). The resizing was based on IPC monitoring. The use of self-timed circuits allowed delay reduction for smaller queue size.

Moshnyaga Vasily [17] improved this design using voltage scaling. The supply voltage was scaled down when only a single queue block was enabled.

Folegnani and Gonzáles [15] proposed a design, which divided the IQ into blocks (16 blocks of 8 entries). Blocks which did not contribute to the IPC were dynamically disabled using a monitoring mechanism based on the IPC contribution of the last active bank in the queue. In addition, their design dynamically disabled the wake up function for empty entries and ready operands.

The energy consumption of a dynamically scheduled superscalar processor like Pentium 4 is between 50 and 100 Watts. At the micro-architecture level, the issue logic is one of the main consumers of energy responsible for approximately 25% of the total energy consumption of the overall processor [15].

Ramírez, Cristal, Valero, Villa and Veidenbaum [18] proposed a design, which reduce the energy consumption in the wakeup logic by eliminating unnecessary comparisons. They proposed a new element called block mapping table mechanism, design uses a multi-block instruction queue. The blocks are inactive until the mechanism determines which blocks to access on wakeup using a simple successor tracking mechanism.
Results presented are shown in Figure 3.1 for comparisons per committed instructions for floating-point benchmarks for a 32- and 64-queue size. The averages are 12 and 17 comparisons per committed instruction, there are unnecessary comparisons that can be avoided, and using a proposed design only require 1.5 comparisons per committed instruction achieve a reduction near of 73 % for SPEC2000 benchmarks.

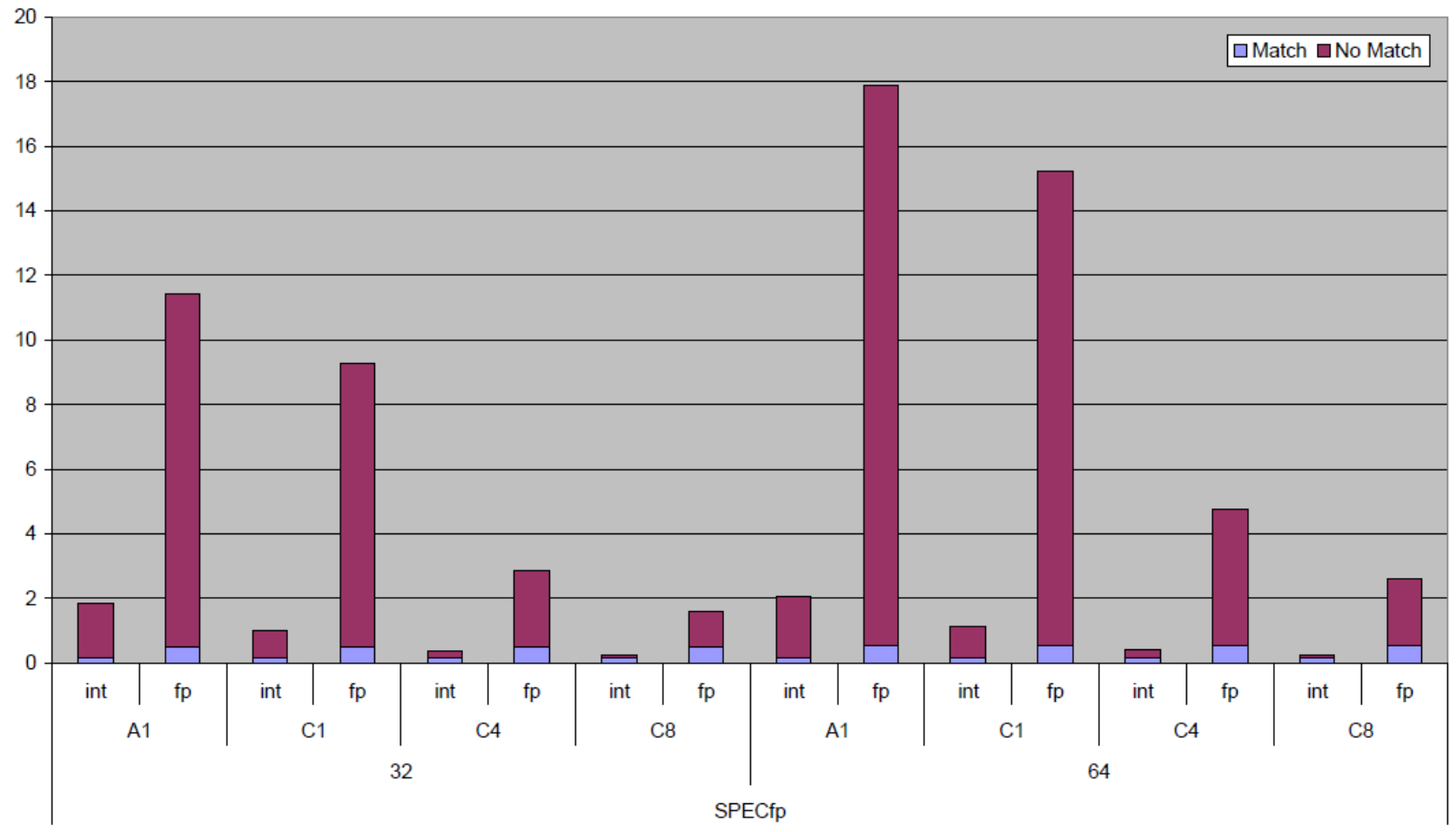

**Fig. 3.1** Average number of comparisons per instruction for FP Benchmarks.

New processors for mobiles devices such as Nvidia Tegra 4, based on ARM Cortex-A15 micro-architecture (superscalar architecture), delivers high performance for mobile applications and improved battery life. Processors such as Tegra 4 uses techniques of high performance and low power consumption, Tegra 4 has an energy consumption close to 10 Watts. Specifically, the issue queue consumes approximately 18 % of the total energy consumption of the overall processor as is shown in Figure 3.2 [19].

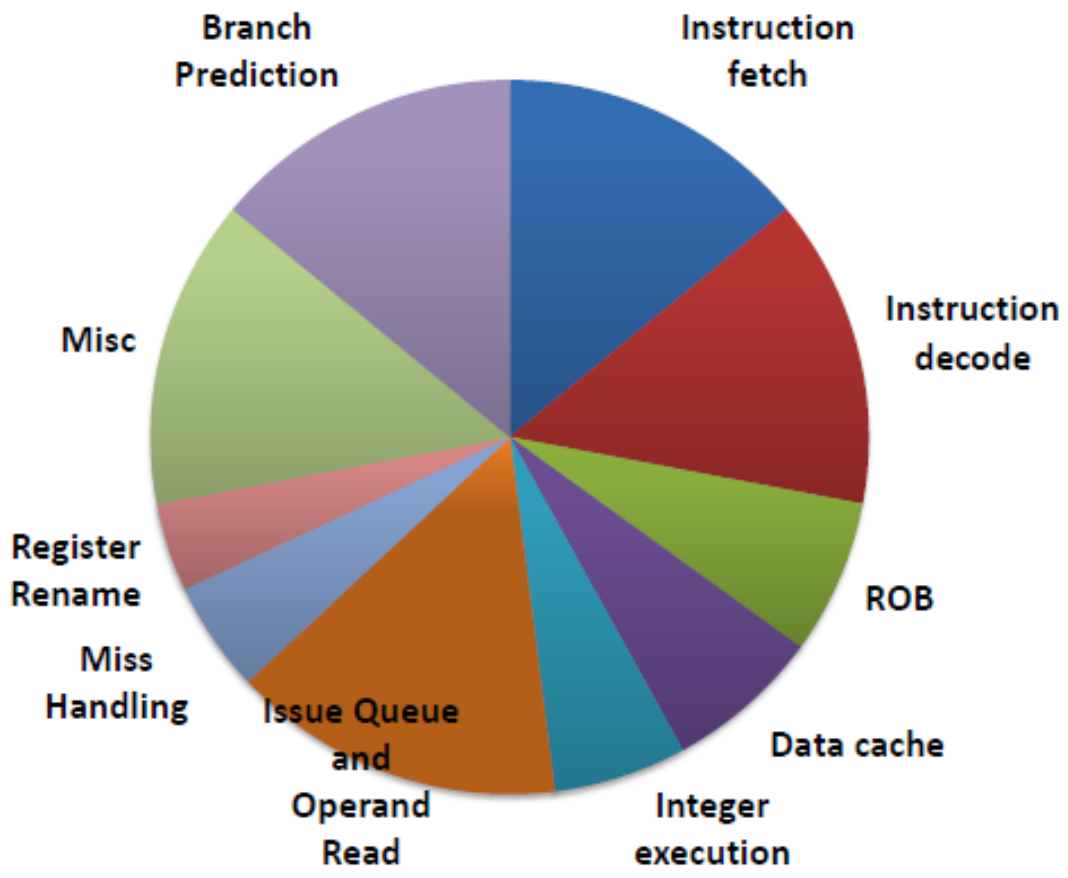

**Fig. 3.2** Power consumption in Nvidia Tegra 4 processors.

## 3.2. Register File

Register files in modern dynamic superscalar processors have been a very modestly sized. The Alpha 21264 processor has as many as 80 integer physical registers and 72 floating-point physical registers and use a clustered organization to reduce the number of ports and the access time [8].

Many researches were performs in order to improve the access time and the energy consumption in the register file.

Lorenzo, Gonzales and Valero [20] proposed a multiple banked register file that can achieve an IPC rate much higher than a multi-cycle file and close to a single-cycle file, but at the same time it requires a single level of bypass. Multiple-banked register file architecture consists of several banks of physical registers with a heterogeneous organization: each bank may have a different number of registers, a different number of ports and therefore, a different access time. Basically authors propose a run-time mechanism, which allocate the values in the registers, where the most critical values are in the fast bank, whereas the remaining values are in the slower bank.

Balasubramonian, Dwarkadas and Albonesi [7] proposed a two level register file to reduce the register file size and a banked organization to reduce the port requirements. If the processor is capable of issue eight integer instructions and simultaneously write back eight integer instructions theoretically could use a register file with 24 ports, but the number of ports required on average is less for several reasons:

- Many operands are read in the bypass network, not from the register file.
- Many instructions have a single register operand.
- Not all instructions write the result in the register file, instructions like branch that send the result to the branch predictor module do not need save the result in the register file, stores instructions save the result in memory.

Then they reduce from 24- ported structure to an 8-ported structure. Their two level register file uses an allocation policy that leaves values that have potential readers in the level one. When using the *instructions per cycle* metric, the two-level organization performs 17% better than the best single-level organization. Using a banked single-porter- bank register file organization reduces access times by a factor of more than two and energy consumption by a factor of more than 18 when compared to a conventional organization, also these improvements are obtained without a significant degradation in IPC.

As we can see these previous proposals are interesting, but to implement a Multiport Register file in the configurable devices is not a trivial task. Altera Co. FPGA devices provide an Em-bedded Memory with Single and Dual port configurations (single port one read, single port one write, dual ports one read and one write, dual ports two reads or dual ports two writes) [21].

There are two ways to implement multiport memories in FPGA, using logic elements, or using memory embedded in the device.

**Implement Multiport Memories Using logic elements**

Implement a Multiport Memory using the logic elements of the FPGA have some inconvenient, the number of logic elements increases according to the number of read and write ports and the size of the memory. Figure 3.3 shows a 6R/6W 64-bits x 128-entries Memory block using logic elements of the FPGA cyclone IV; the advantage is the easy implementation.

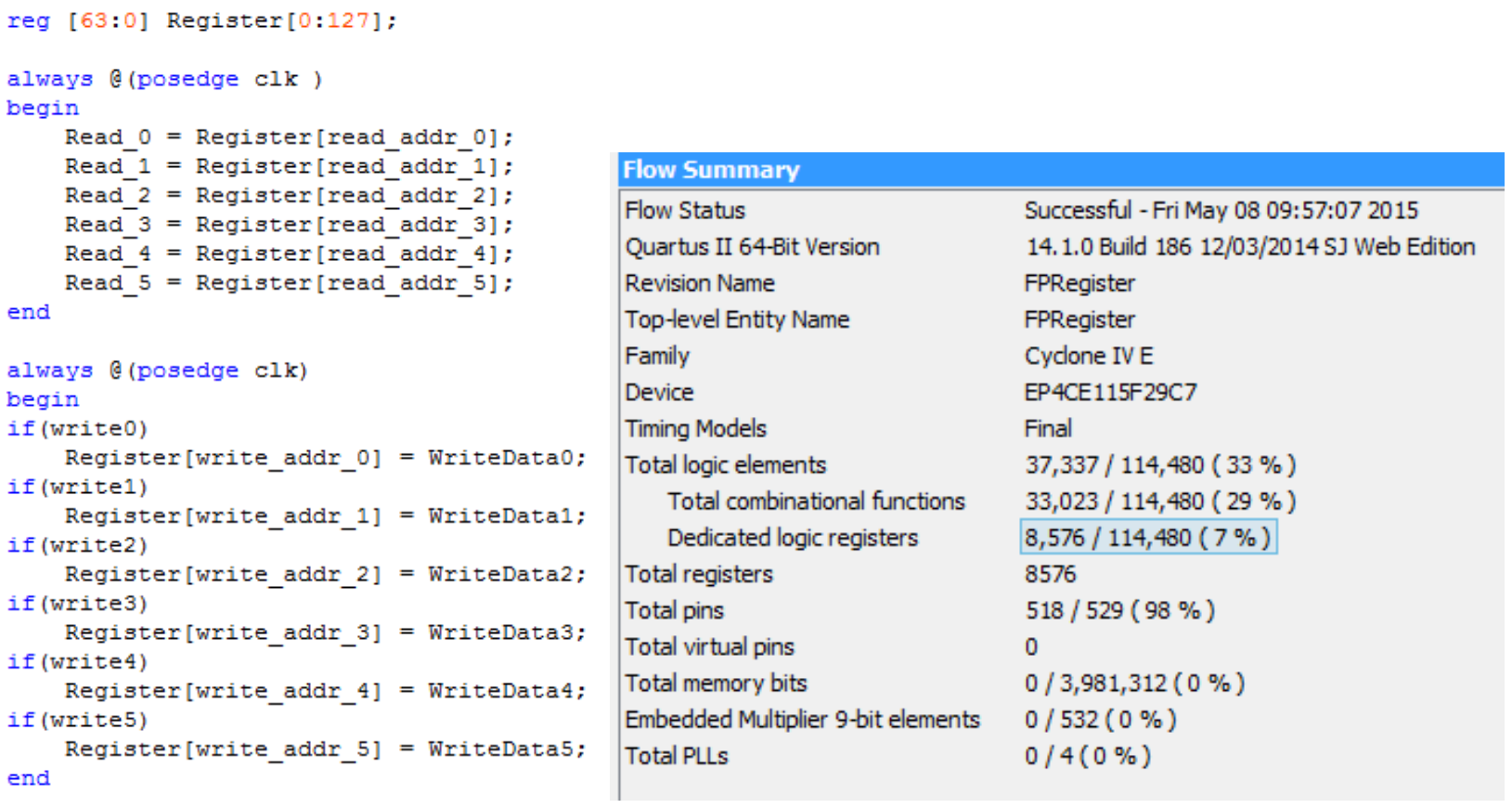

```
reg [63:0] Register[0:127];

always @(posedge clk )
begin
    Read_0 = Register[read_addr_0];
    Read_1 = Register[read_addr_1];
    Read_2 = Register[read_addr_2];
    Read_3 = Register[read_addr_3];
    Read_4 = Register[read_addr_4];
    Read_5 = Register[read_addr_5];
end

always @(posedge clk)
begin
if(write0)
    Register[write_addr_0] = WriteData0;
if(write1)
    Register[write_addr_1] = WriteData1;
if(write2)
    Register[write_addr_2] = WriteData2;
if(write3)
    Register[write_addr_3] = WriteData3;
if(write4)
    Register[write_addr_4] = WriteData4;
if(write5)
    Register[write_addr_5] = WriteData5;
end
```

| Flow Summary | |
|---|---|
| Flow Status | Successful - Fri May 08 09:57:07 2015 |
| Quartus II 64-Bit Version | 14.1.0 Build 186 12/03/2014 SJ Web Edition |
| Revision Name | FPRegister |
| Top-level Entity Name | FPRegister |
| Family | Cyclone IV E |
| Device | EP4CE115F29C7 |
| Timing Models | Final |
| Total logic elements | 37,337 / 114,480 ( 33 % ) |
| Total combinational functions | 33,023 / 114,480 ( 29 % ) |
| Dedicated logic registers | 8,576 / 114,480 ( 7 % ) |
| Total registers | 8576 |
| Total pins | 518 / 529 ( 98 % ) |
| Total virtual pins | 0 |
| Total memory bits | 0 / 3,981,312 ( 0 % ) |
| Embedded Multiplier 9-bit elements | 0 / 532 ( 0 % ) |
| Total PLLs | 0 / 4 ( 0 % ) |

**Fig. 3.3** Multiport Memory using Logic elements

**Implement Multiport Memories Using Embedded Memory**

Latest Altera FPGA devices as cyclone and Stratix series provide an Embedded Memory which are blocks of dedicated memory resources. Following table list and describes the memory operation modes that are supported for embedded memory blocks [22].

**Table 3.1** Supported Memory Operations Modes

| Memory Operation Mode | Description |
|---|---|
| **Single-port RAM** | Single-port mode supports non-simultaneous read and write operations from a single address. |
| **Simple dual-port RAM** | Simultaneously perform one read and one write operations to different locations where the write operation happens on port A and the read operation happens on port B. |
| **True dual-port RAM** | Perform any combination of two port operations:<br>▪ Two reads, two writes, or,<br>▪ One read and one write at two different clock frequencies |
| **Single-port ROM** | Only one address port is available for read operation. the memory blocks are used as a ROM.<br>▪ Initialize the ROM contents of the memory blocks using a .mif or .hex file. |
| **Dual-port ROM** | The dual-port ROM has almost similar functional ports as single-port ROM. The difference is dual-port ROM has an additional address port for read operation. |

Because the above specifications, many techniques have been proposed to implement a multiport memory in the FPGA in [23]. Figure 3.4 show the conventional techniques for provide more ports; the first is replication, which can increase the number of read ports by maintaining a replica of the memory for each additional read port. However, this technique alone cannot support more than only one write port.

The second approach splits the deep memory bank among multiple RAM blocks (sub-banks), allowing each sub-bank to support an additional read and an additional write port. However, with this approach each read-port or write-port can only access locations of its corresponding memory sub-bank.

The third is called multipumping, where the core is an asynchronous memory block with single read port and single write port, to increase the ports number, a Mux (1:N) and Demux (N:1) are used to read and write all ports in one cycle (external frequency lower than internal frequency) providing the illusion of a multiple number of ports, also include a register per port to temporarily hold the addresses and data of pending reads and writes. This approach reduces the operative frequency of the multiport memory.

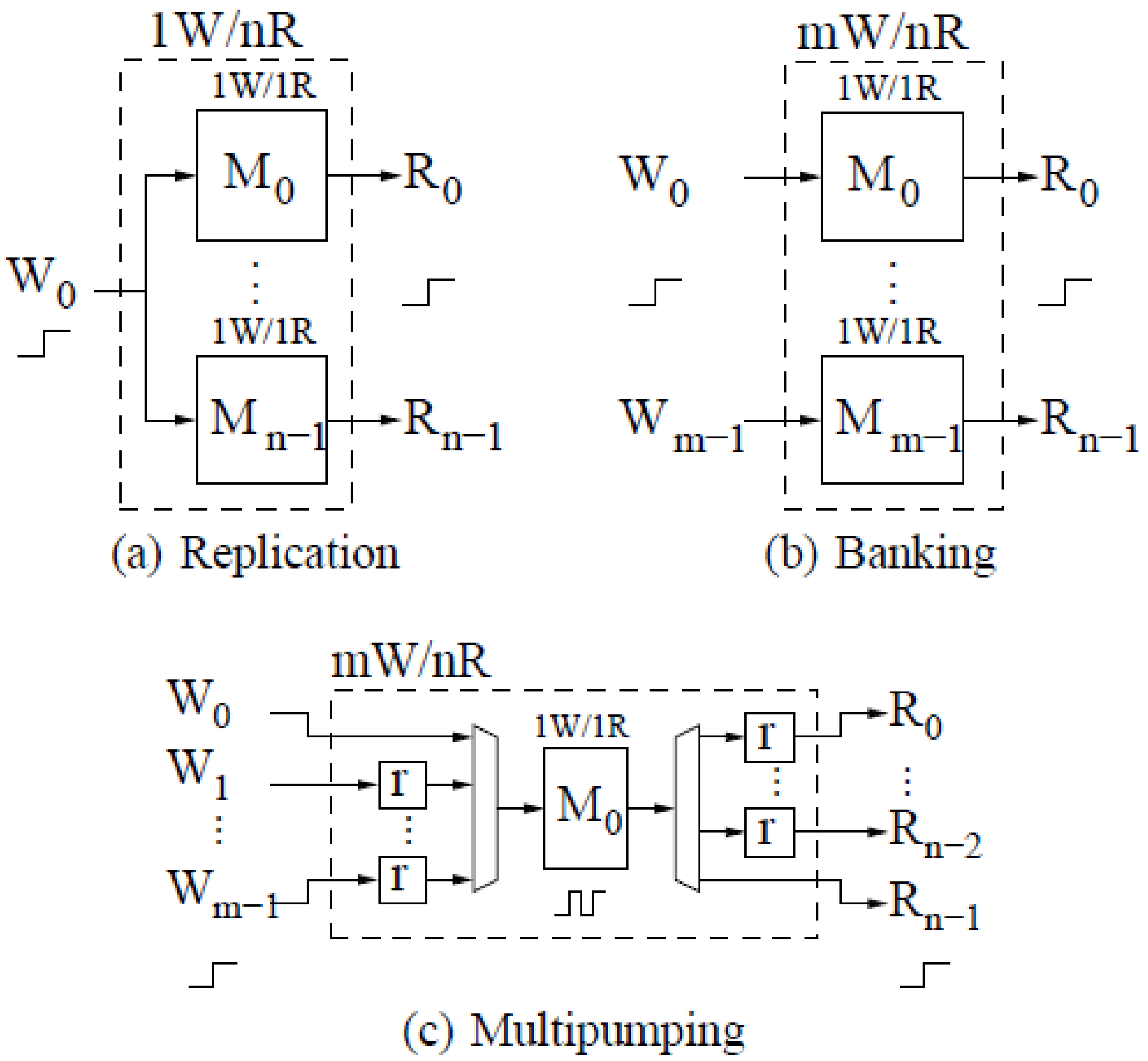

**Fig. 3.4** Conventional Techniques for providing more ports given a 1W/1R memory.

LaForest and Steffan [23] propose a design for true multi-ported memories that uses the FPGA block RAMs. They propose a structure called *Live Value Table* (LVT). Essentially, the LVT allows a banked design to behave like a true multi-ported design by directing reads to appropriate banks based on which bank holds the most recent write value. LVT is purely implemented in logic elements. Figure 3.5 show the general design.

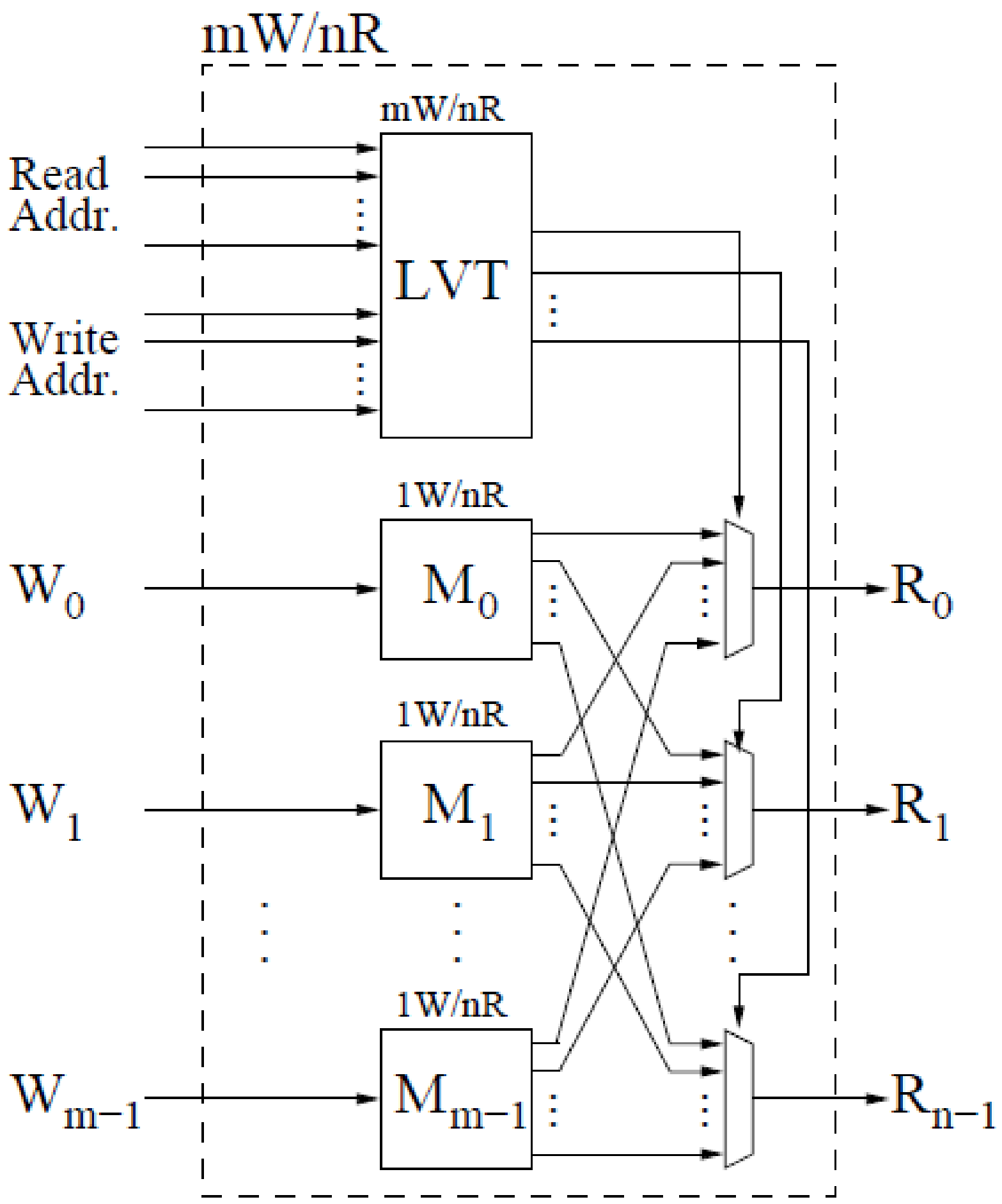

**Fig. 3.5** A generalized mW/nR memory implemented using a Live Value Table (LVT).

"Mn" Blocks uses replication technique in order to obtain a memory with 1 write port and n read ports, after they uses banking in order to increases the number of write ports, and finally with the LVT and multiplexors they can read the most recent write value. This design work a high frequency but operative frequency depends of the number of ports, the live value table could uses many logic elements (less than a complete memory with LE).

Laforest, Ming, Rapati and Steffan [24] proposed a new design based on the properties of XOR operation. This design is based in that XOR is commutative, associative and has the following properties:

$A \oplus 0 = A.$

$B \oplus B = 0.$

$A \oplus B \oplus B = A.$

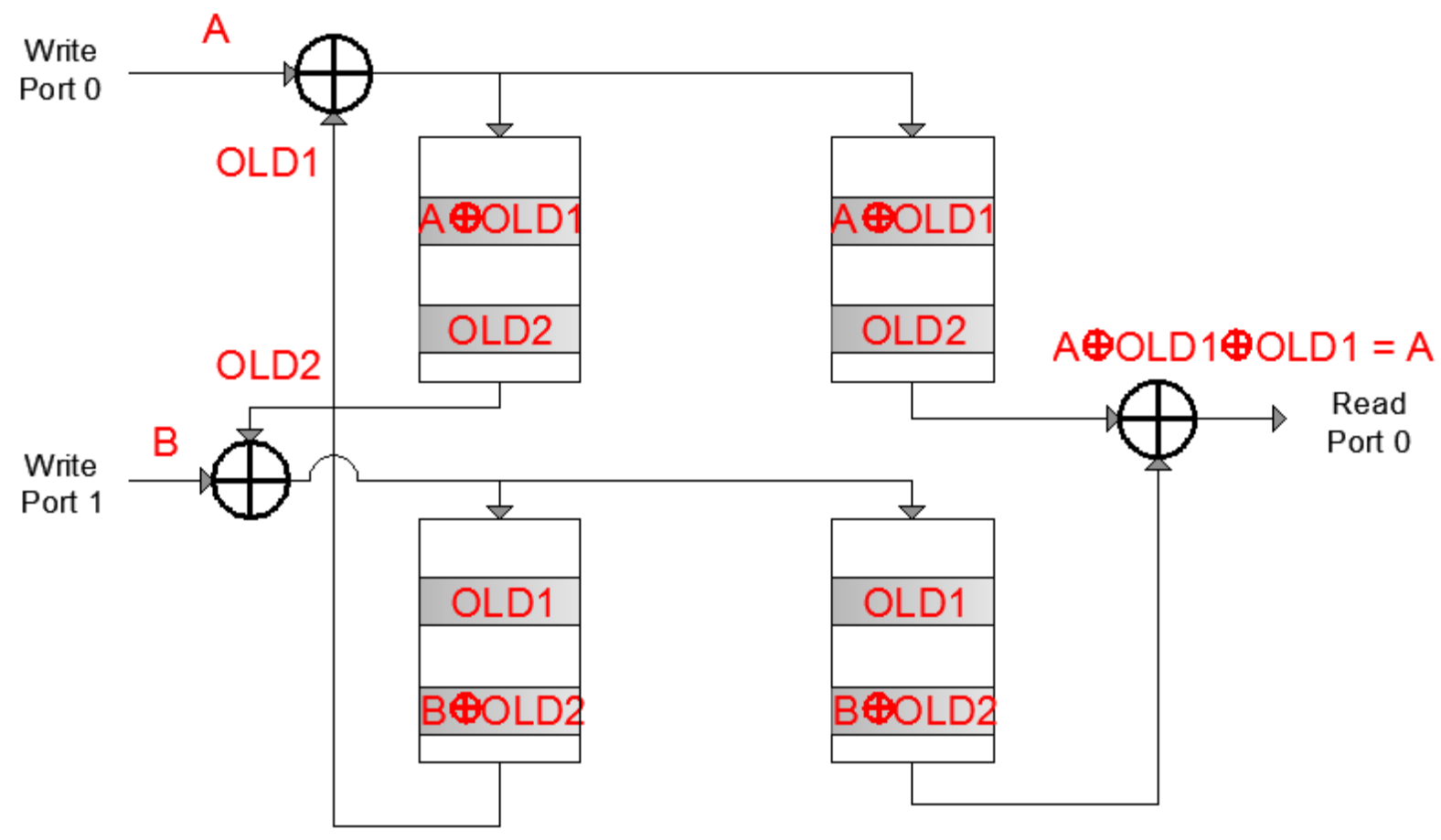

**Fig. 3.6** A 2W/1R memory implemented using XOR

Figure 3.6 shows a 2W/1R memory implemented using XOR design. In the example, it is required to save the value A using the W0 write port, thus we need read the value of the other write port and perform the XOR operation and save the result. In this case we save $A \oplus OLD1$, if we want to read the same address we need read all banking blocks and perform the XOR operation between them, in this case result as $A \oplus OLD1 \oplus OLD1 = A$, which is the most recent write value. Figure 3.7 shows a 2W/2R memory implemented using XOR design. This design requires m * (m-1+n) RAM Blocks to provide m writes ports and n reads ports as we can see in Figure 3.8.

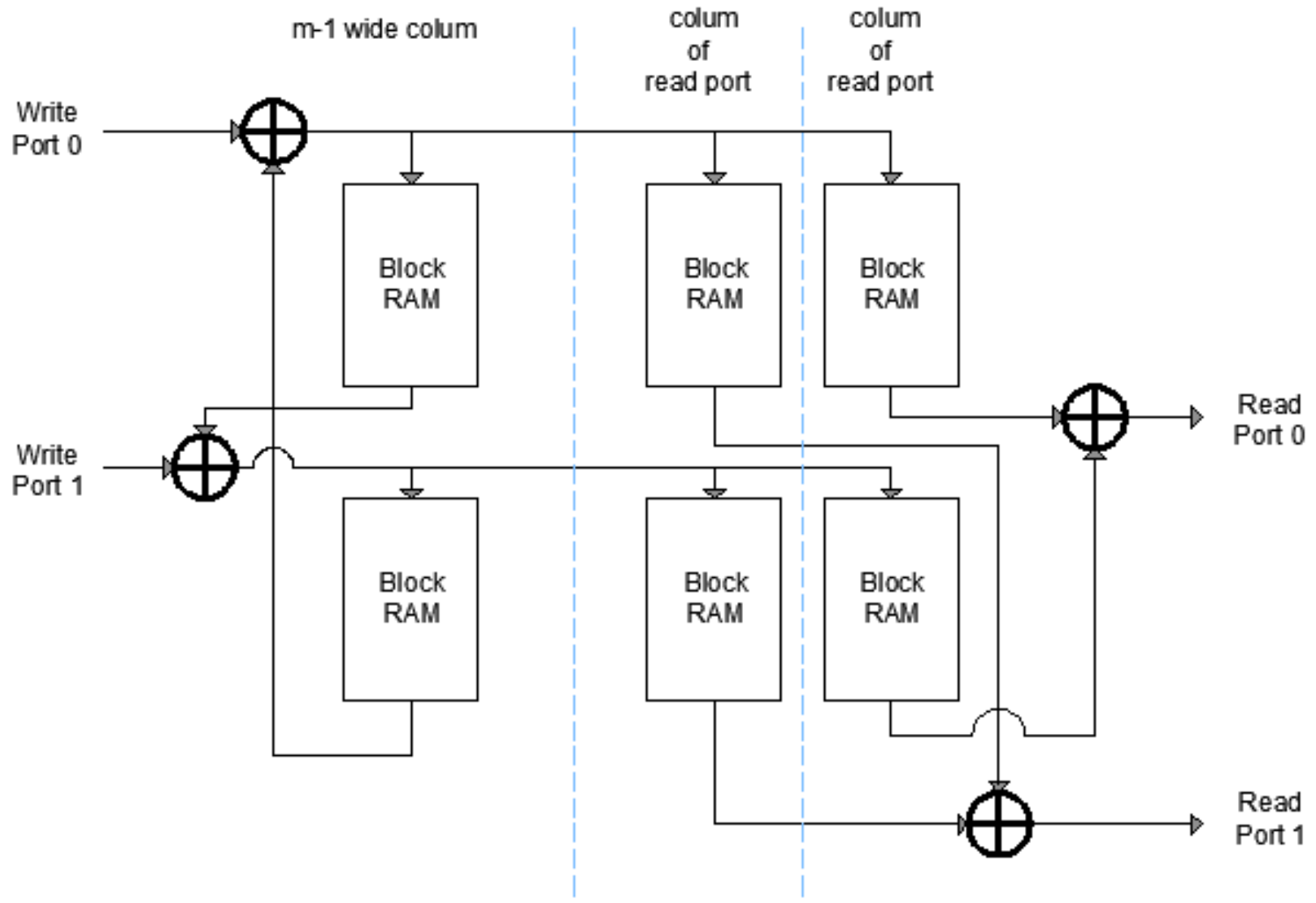

**Fig. 3.7** A 2W/2R memory implemented using XOR

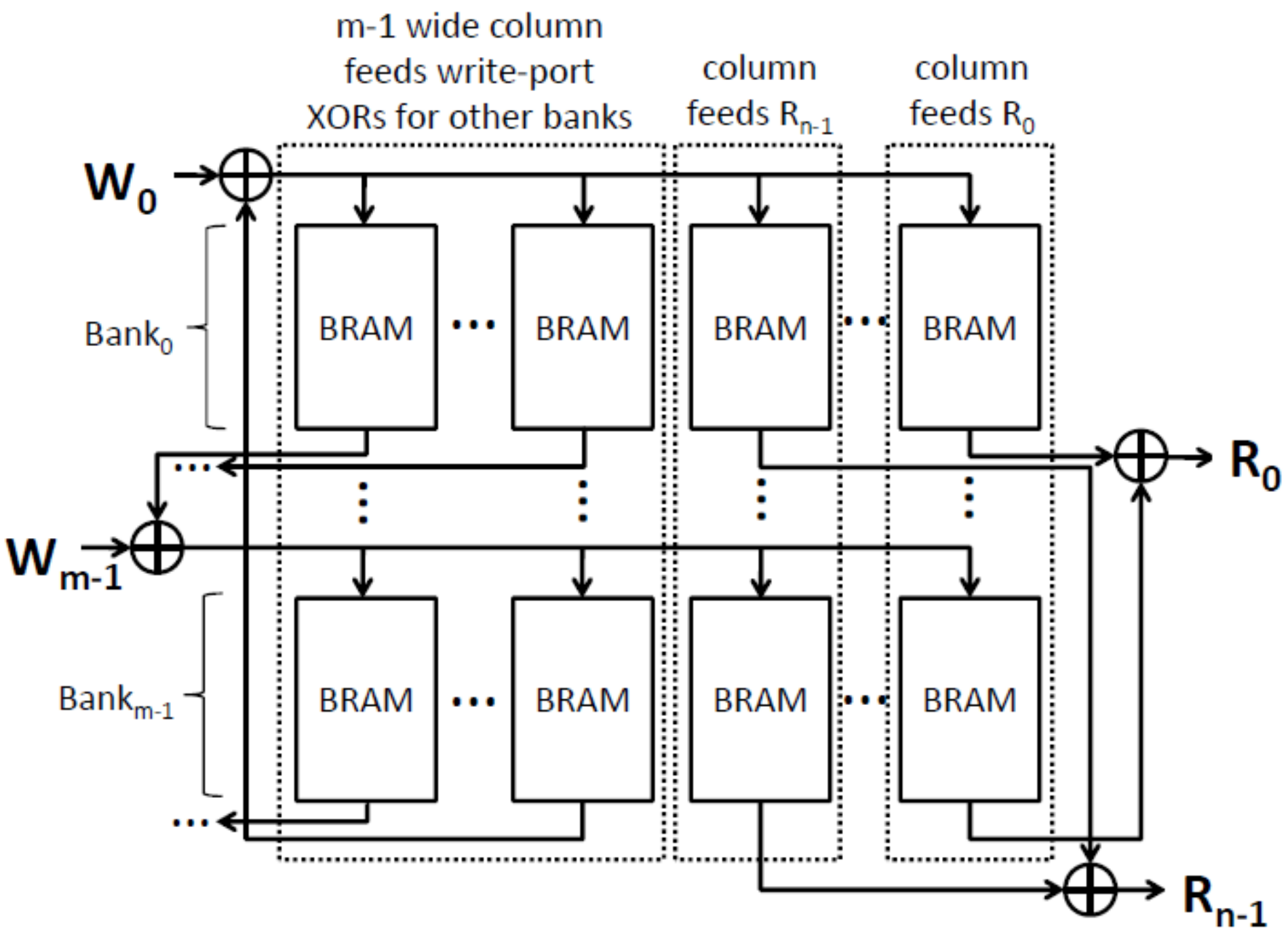

**Fig. 3.8** A generalized mW/nR memory implemented using XOR.

## 3.3. Execution Stage

The performance and area of a functional unit depend upon circuit style, logic implementation, and choice of algorithms. The three primary parameters in FP functional unit design are latency, frequency, and area. The functional unit latency is the time required to complete a computation, typically measured in machine cycles. Designs can be either Fixed Latency (FL) or Variable Latency (VL). Over the past two decades a lots of work has been dedicated to performance improvement of floating point computations, both at algorithmic level and implementation level. Several works also focused their implementation on FPGA platforms. In [25] we can found the basic algorithms for floating point operations like Adder/Subtractor, Multiplication, Division and Multiply-Accumulate with some improvements. Basically all current floating-point implementations are based in the basics algorithms with little modifications.

Following are described some proposals of Floating-point Adder/subtractor, Multiplication and Division. Furthermore, are presented the Floating Point LPM modules provided by Altera Corporation in the software Quartus II.

### 3.3.1. Floating Point Adder/Subtractor

Floating Point Adder/subtractor is one of the most frequent arithmetic operations in scientific computing. The design of FP adder/subtractor is relatively more complex than other FP arithmetic operations. The operations consist of three major task, pre-normalization, addition and post-normalization.
Pre-normalization consists of exponent difference and *right shift*. Post-normalization consists of a priority decoder to detect the leading zeros in a number after the addition and a *left-shift* operation.

Post-normalization quickly becomes part of the critical path due we need know the number of zeros to the left in the shortest possible time to after perform the shift and deliver the result. Exist many ways to obtain the leading zeros, two of the main techniques are called Leading Zero Counter (or Leading One Detector) and Leading Zero Anticipation.

Several works are available in the literature, for implementations of floating point adder/subtractor unit on FPGA. [26] Proposed a design of FP adder/subtractor that has optimized the individual complex component of the adder module like dynamic shifter and the leading one detector (LOD).

In [27] is presented a study on floating-point adders in FPGAs. They analyze the standard floating-point algorithm and the hardware modules designed as part of this algorithm. They compare algorithms that use a Leading One Detector (LOD) and Leading One predictor (LOP). Both Algorithms are shown in Figure 3.9.

Fist algorithm performs the pre-normalization, addition and post-normalization where use the Leading One Detector in order to obtain the number of zeros to the left and subsequently perform shift left to normalize the result. Second use LOP instead of LOD. The main function of the module is to predict the leading number of zeros in the operation result and this block is working in parallel with the adder. Also this specific algorithm was proposed by Bruguera and T. Lang [28] which detects the error concurrently with the leading one detection. Last one improves 6.5% in latency but with a cost of 38% more area expensive.

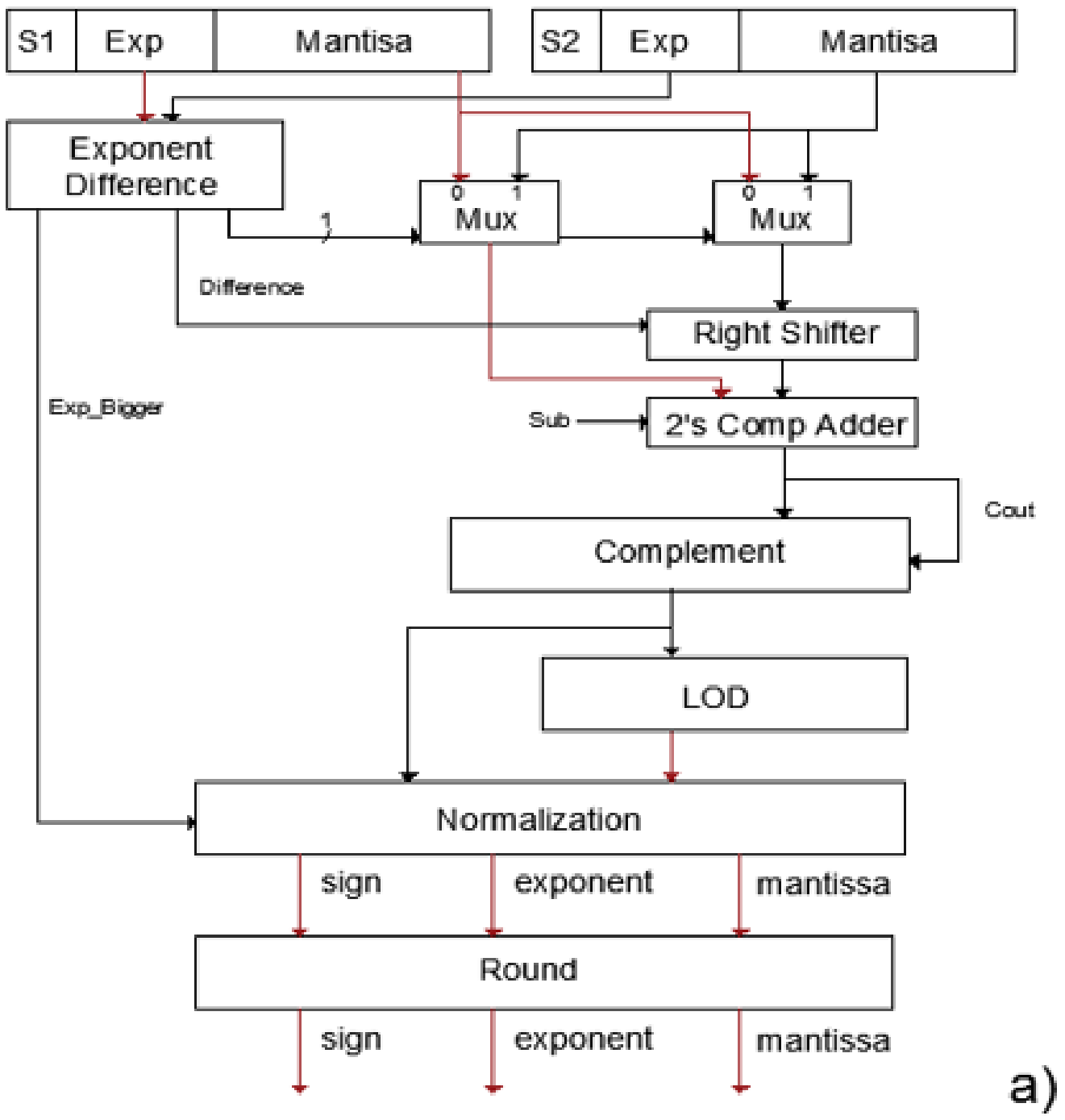

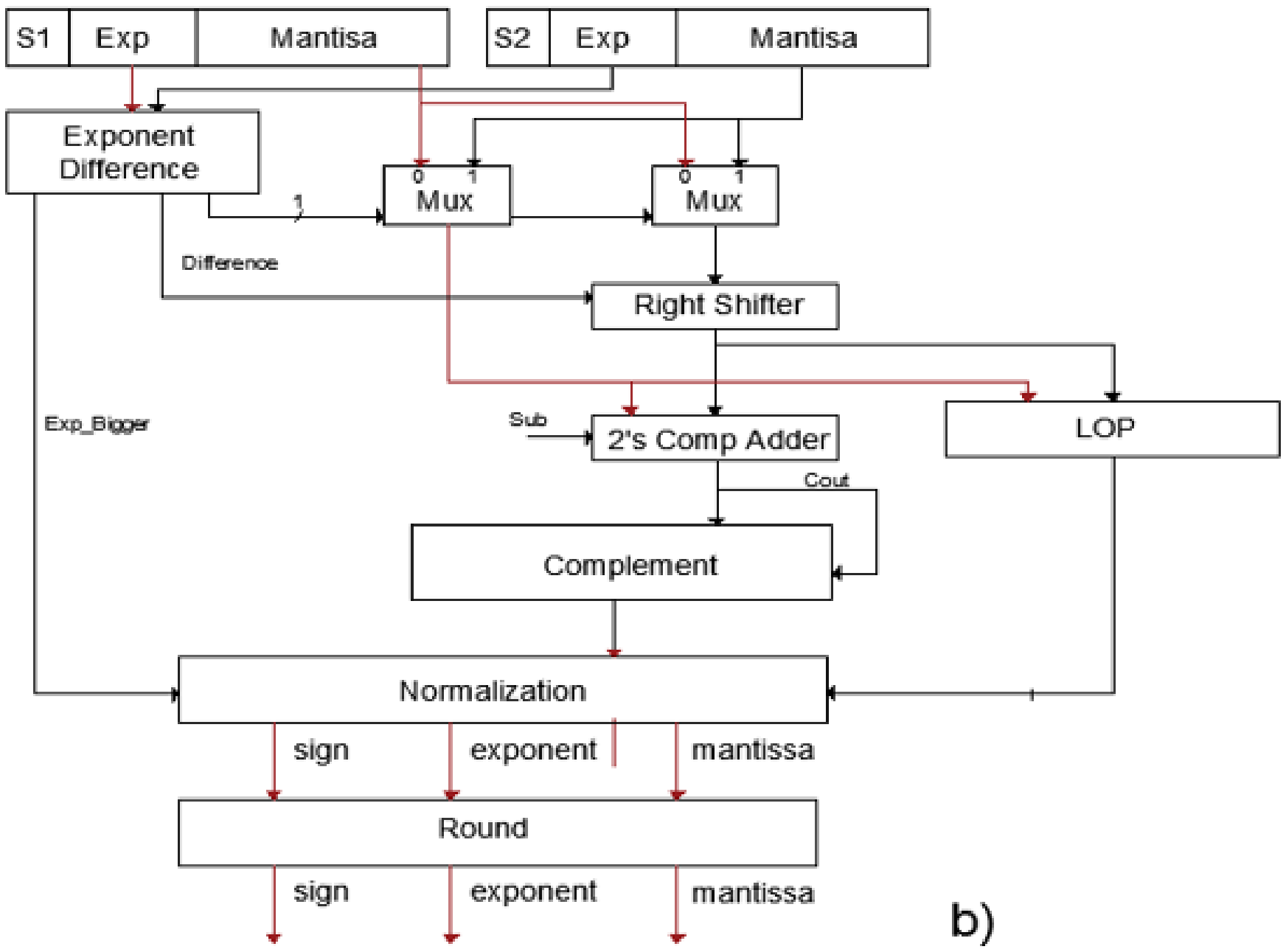

**Fig. 3.9** FP Adder Microarchitecture using a) LOD algorithm b) LOP algorithm

Dimitrakopoulos, Galanopoulos, Mavrokefalidis and Nikolos [29] proposed a new Low-Power Leading-Zero Counter for High-Speed Floating Point Units. Their computation is reduced using carry-lookahead techniques in a unified manner. They report that significant energy reductions are achieved by the proposed design compared to the most efficient previous implementations. Design is presented in Figure 3.10.

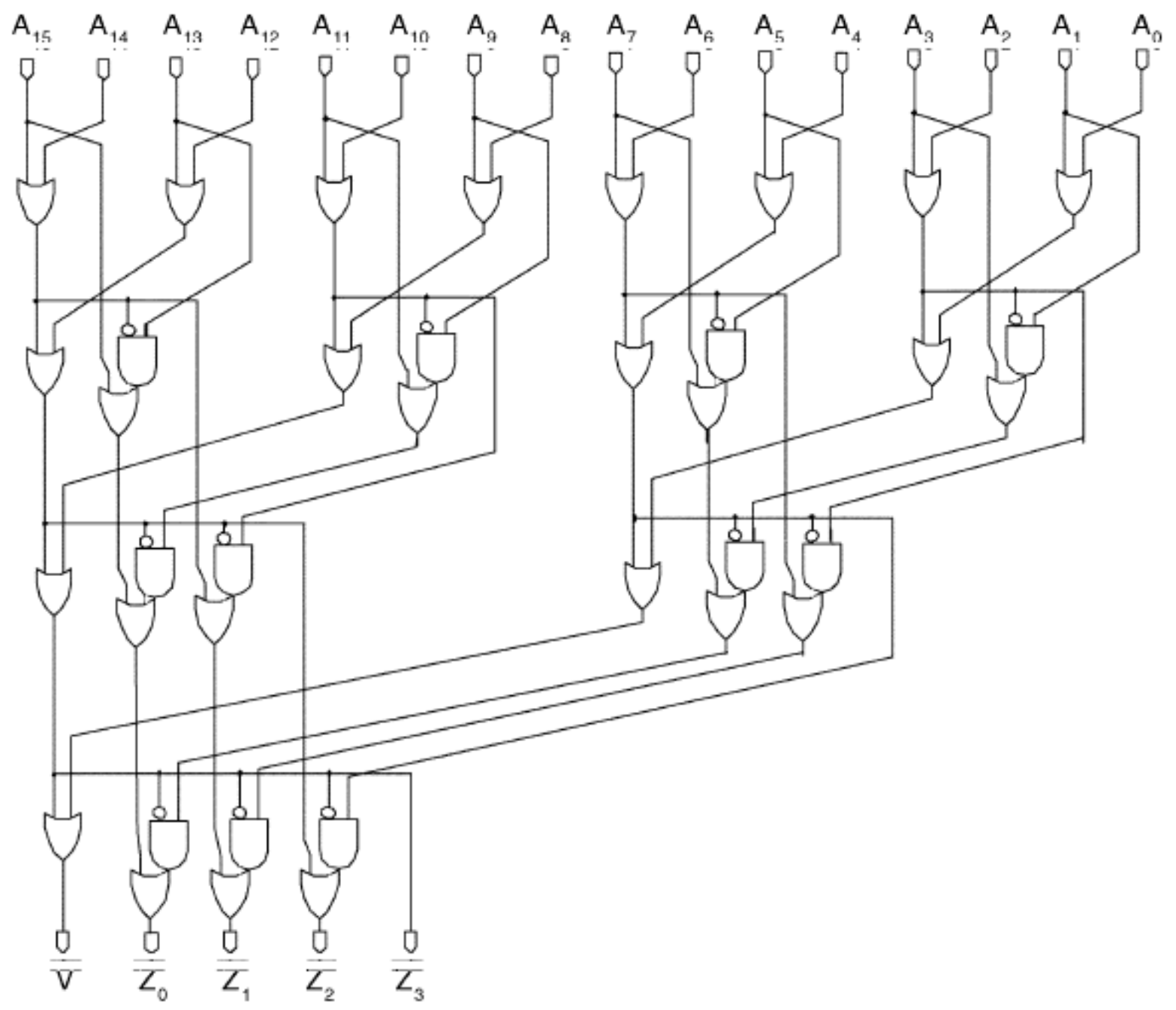

**Fig. 3.10** 16-bit LZC using the shared carry-propagate approach

### 3.3.2. Floating Point Multiplier

Floating Point multiplication is a core operation in many signal processing computations, and an efficient implementations of floating point multipliers is an important concern.
Multiplying two numbers in floating point format is performed in three main steps:

- Add the exponent of the two numbers and then subtracting the bias from their result.
- Multiply the significant of the two numbers.
- Calculate the sign by XOR operation of the two signs of the two numbers.

To multiply 2 numbers in double precision format, require the implementation of 53-bits x 53-bits multipliers in hardware, which is very expensive. This operation is relatively simple; proposals are based in how multiply the mantissa as fast as possible.

Manish Kumar and Chandrachoodan [30] propose an efficient implementation of IEEE double precision Floating-point Multiplier on FPGA; the proposed method is based on partial block multiplication. The main idea is divide the mantissa of the operands in small blocks and perform the multiplication using small size multipliers as is shown in Figure 3.11.

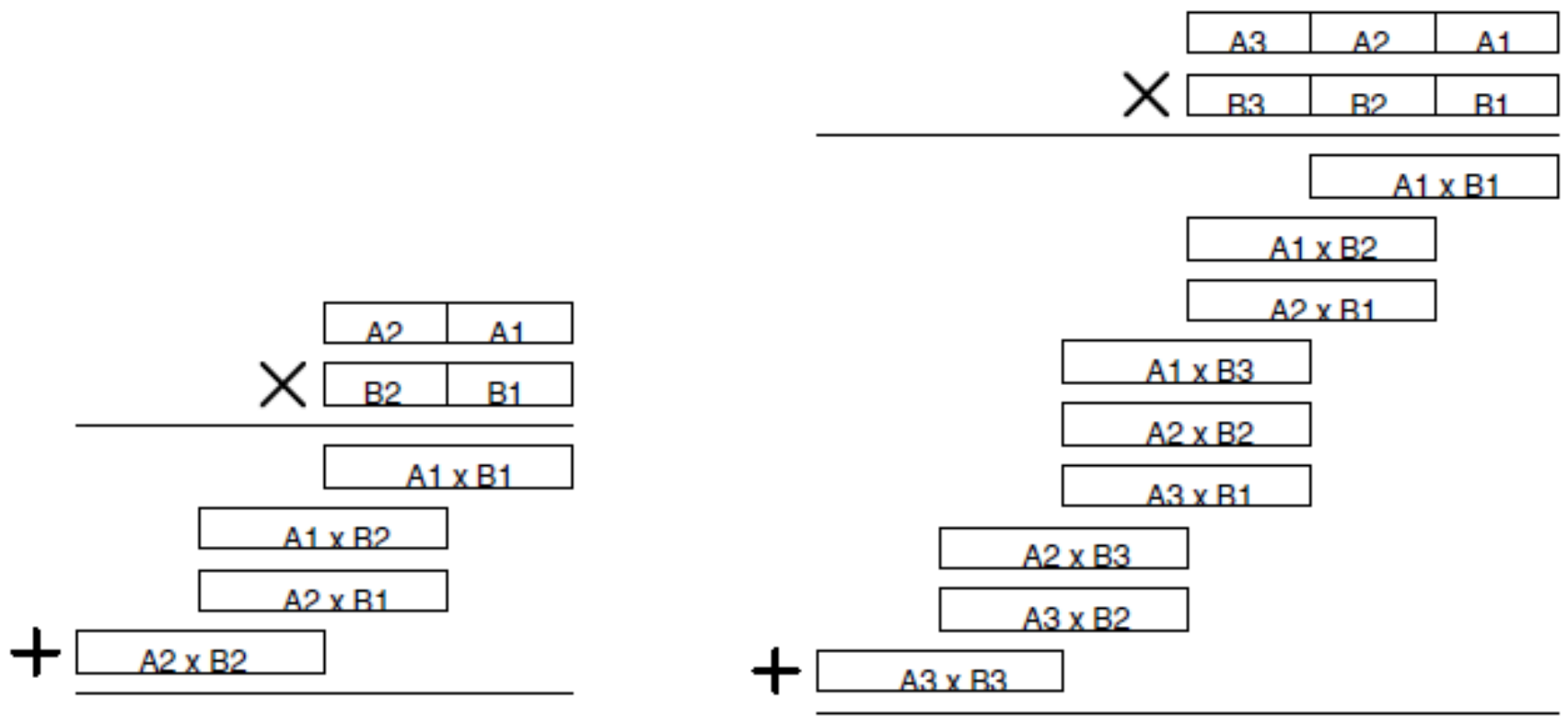

**Fig. 3.11** Block Multiplier for two and three blocks.

For implementing the module, they chose block size of 17-bit because Xilinx FPGAs provides a signed 18x18 multipliers. Figure 3.12 shows the partial division blocks. Partial products are arranged (varied for different latency) in suitable manner and added to get the result.

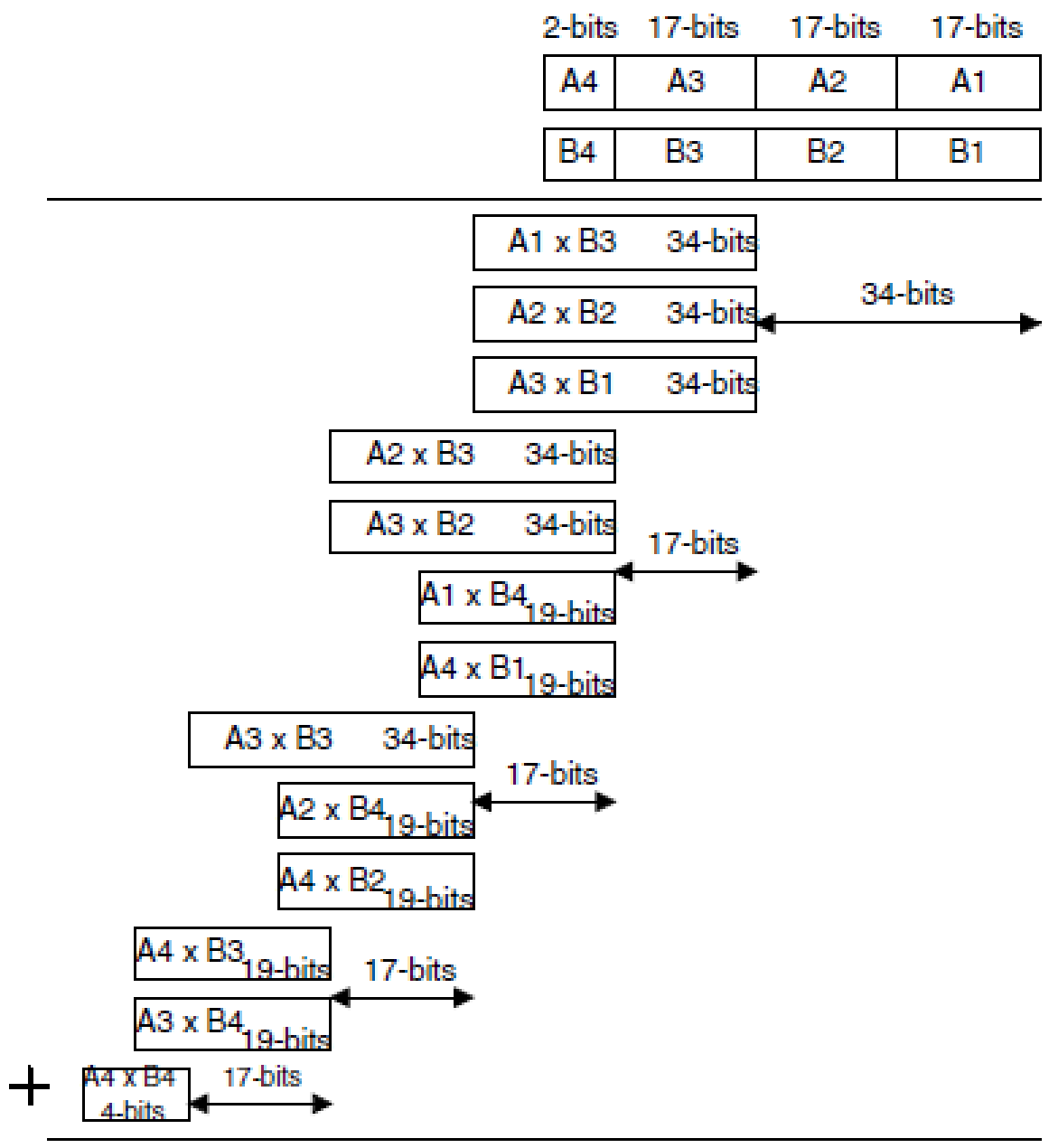

**Fig. 3.12** Partial Block multiplier for 53-bits

The cost of the design is an error when compared to the IEEE standard, of up to 1 unit in the last place when used with partial nearest value rounding, or up to 2 units in last place without rounding. Design is restricted to only normalized numbers.

### 3.3.3. Floating Point Divider

Floating Point divider needs many cycles to perform the division operation using an algorithm based on subtract and shift operations as a core of the functional unit. Among the arithmetic operations, the division is the operation that consumes more time, because the number of cycles used to determine quotient is proportional to the number of bits of the dividend and it is difficult to implement with pipeline due to the dependencies between the iterations.

***Floating Point Divider/Reciprocal***

Division operation can be expressed as $a = \frac{b}{c} = b\, x \frac{1}{c}$. Techniques such as Newton- Raphson and series expansion algorithms are usually used to compute the reciprocal for high-performance division.

A basic implementation of Newton-Raphson reciprocal for double precision is presented in [31]. This proposal begin with an initial approximation through a look-up table ($2^{10}x20\ bits\ ROM$) obtained using a Taylor series expansion. After that, uses two Newton-Raphson iterations. Complete algorithm is described below.

Obtaining the initial reciprocal approximation takes three clock cycles, which requires reading the look-up table to obtain the initial value to start, followed by multiplication and addition operations. In order to iterate the initial approximation, each Newton-Rapson iteration spend four clock cycles, which has two stages, and each stage consists of a multiplication and an addition.

This Unit perform the floating point reciprocal operation in only eleven cycles. The disadvantage of usesing this kind of method is that it does not guarantee the accuracy of the least significant bit. The design is presented in Figure 3.13.

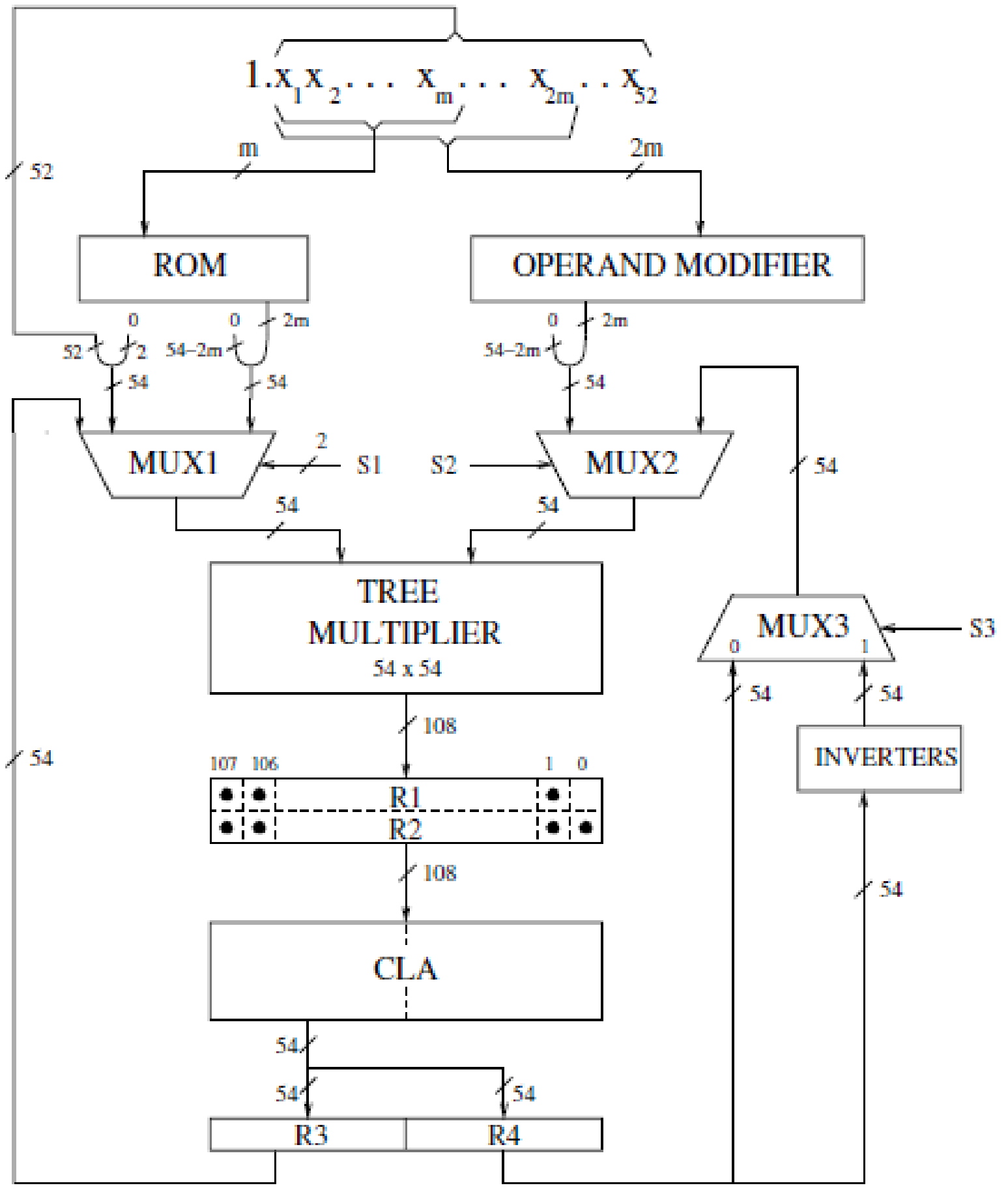

**Fig. 3.13** Reciprocal Unit implementation - first proposal

Other proposal is presented in [32] where an optimized design and its implementation of reciprocal unit is proposed, in which the initial approximation of the reciprocal is obtained using a look-up table and a multiplication. Also they describe in detail how to implement efficiently the look-up table. Their design utilizes a $2^7x16\ bits\ ROM$ followed by two Newton-Raphson iterations. Furthermore, this design spends 10 clock cycles to achieve the 52-bit of accuracy for double precision floating-point number. Design is presented in Figure 3.14.

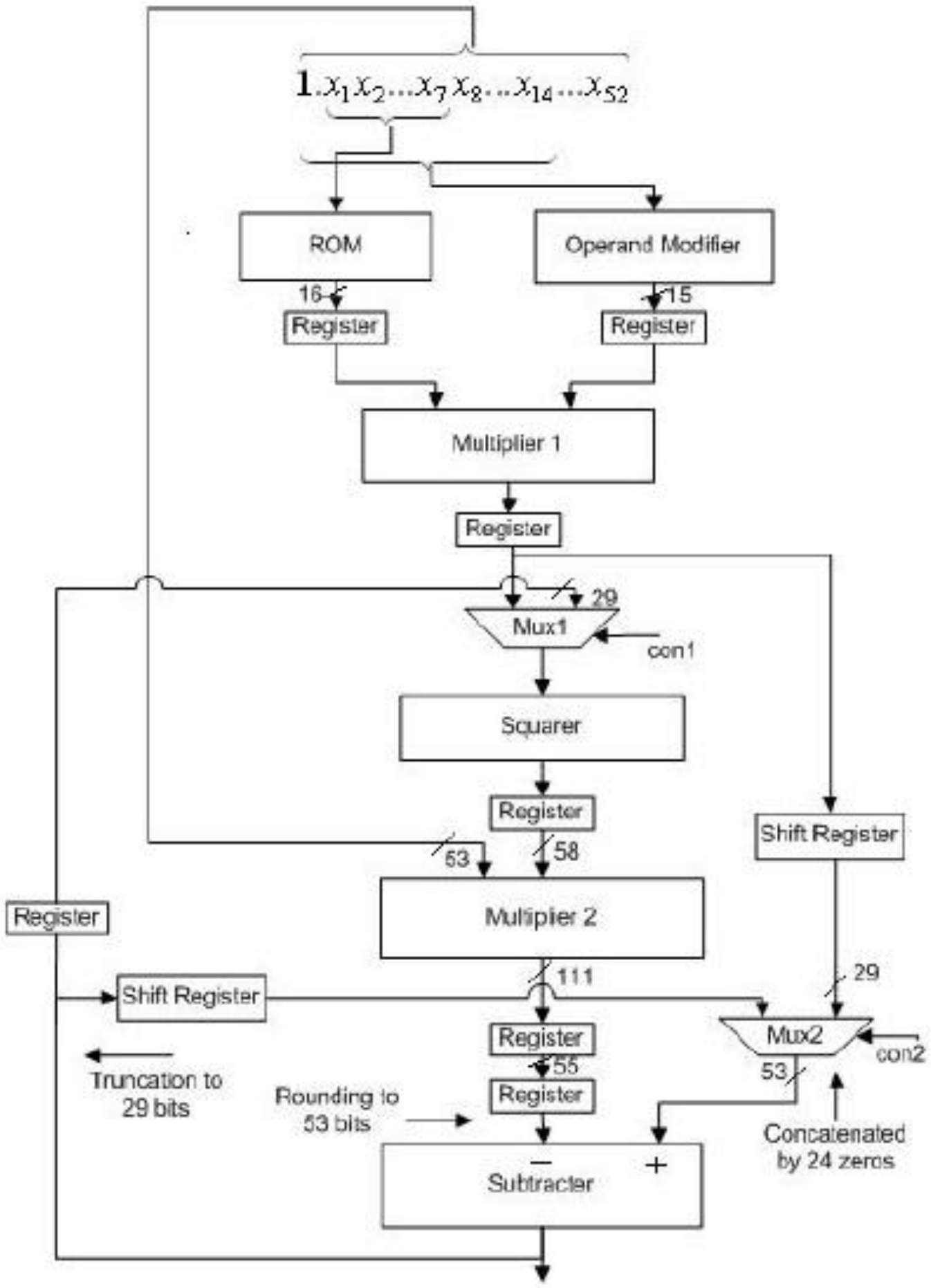

**Fig. 3.14** Reciprocal Unit implementation – second proposal

## Altera IP Cores

Altera provides many useful IP core functions for Floating point operations [33]. All Altera floating-point IP cores offer the following features:

- Support for floating-point formats.
- Input support for not-a-number (NaN), infinity, zero, and normal numbers.
- Optional asynchronous input ports including asynchronous clear (aclr) and clock enable (clk_en).
- Support for round-to-nearest-even rounding mode.
- Compute results of any mathematical operations according to the IEEE-754 standard compliance with a maximum of 1 unit in the last place (u.l.p.) error.

Altera floating-point IP cores do not support subnormal number inputs. If the input is a subnormal value, the IP core forces the value to zero and treats the value as a zero before going through any operation.

Following we describe only 3 IP cores (Adder/Subtract, Multiplier and Divider) in order to compare with our designs in Chapter 6.

***ALTFP_ADD_SUB – Floating Point Adder/Subtract IP core***

The ALTFP_ADD_SUB IP core offers the following features:

- Dynamically configurable adder and subtractor functions.
- Optional exception handling output ports such as zero, overflow, underflow, and NaN.
- Optimization of speed and area.
- Output latency available are 7, 8, 9,11,12,13 and 14 clock cycles.

Following table list the resource utilization and performance information for double precision floating point adder/subtractor for the Cyclone IV device family.

**Table 3.2** ALTFP_ADD_SUB Resource Utilization and Performance for the Cyclone Series Devices.

| Optimization | Output Latency | Total Logic Elements | Total Memory Bits | Embedded Multiplier 9-bit elements | Fmax (MHZ) |
|---|---|---|---|---|---|
| **Speed** | 8 | 1804 | 45 | 0 | 116.36 |
| | 14 | 2452 | 150 | 0 | 208.77 |
| **Area** | 8 | 1684 | 45 | 0 | 105.61 |
| | 14 | 2196 | 150 | 0 | 204.12 |

***ALTFP_MUL – Floating Point Multiplier IP core***

The ALTFP_MUL IP core offers the following features:

- Optional exception handling output ports such as zero, overflow, underflow, and NaN.
- Optional dedicated multiplier circuitries in Cyclone and Stratix Series.
- Output latency available are 5,6,10 and 11 clock cycles.

Following table list the resource utilization and performance information for double precision floating point Multiplier for the Cyclone IV device family.

**Table 3.3** ALTFP_MUL Resource Utilization and Performance for the Cyclone Series Devices with dedicated Multiplier circuitry.

| Optimization | Output Latency | Total Logic Elements | Total Memory Bits | Embedded Multiplier 9-bit elements | Fmax (MHZ) |
|---|---|---|---|---|---|
| - | 6 | 832 | 0 | 18 | 119.0 |
| - | 10 | 1041 | 110 | 18 | 132.59 |

***ALTFP_DIV – Floating Point Divider IP core***

The ALTFP_DIV IP core offers the following features:

- Optional exception handling output ports such as zero, division_by_zero, overflow, underflow, and NaN.
- Optimization of speed and area.
- Low latency option.
- Output latency available for double precision are 10, 24 and 61 clock cycles.

Following table list the resource utilization and performance information for double precision floating point divider for the Cyclone IV device family.

**Table 3.4** ALTFP_DIV Resource Utilization and Performance for the Cyclone Series Devices.

| Optimization | Output Latency | Total Logic Elements | Total Memory Bits | Embedded Multiplier 9-bit elements | Fmax (MHZ) |
|---|---|---|---|---|---|
| **Speed** | 24 | 1344 | 6441 | 44 | 117.91 |
| | 10 | 1325 | 4709 | 44 | 88.94 |
| **Area** | 24 | 1344 | 6441 | 44 | 117.91 |
| | 10 | 1325 | 4709 | 44 | 88.94 |

## 3.4. Intel Itanium Floating Point Architecture

Focusing on parallelism, the Intel Itanium processor was launched in 2001, followed by the Itanium 2 processor in 2002 and produced until 2010. Itanium 2 boasts of a particularly powerful floating-point architecture.

The Itanium floating-point architectures were designed to combine high performance and good accuracy. It has features such as floating point register set of 128 registers and the ability to execute multiple instructions per clock cycle. Furthermore, Itanium wanted to achieve the full IEEE-754 compliance.

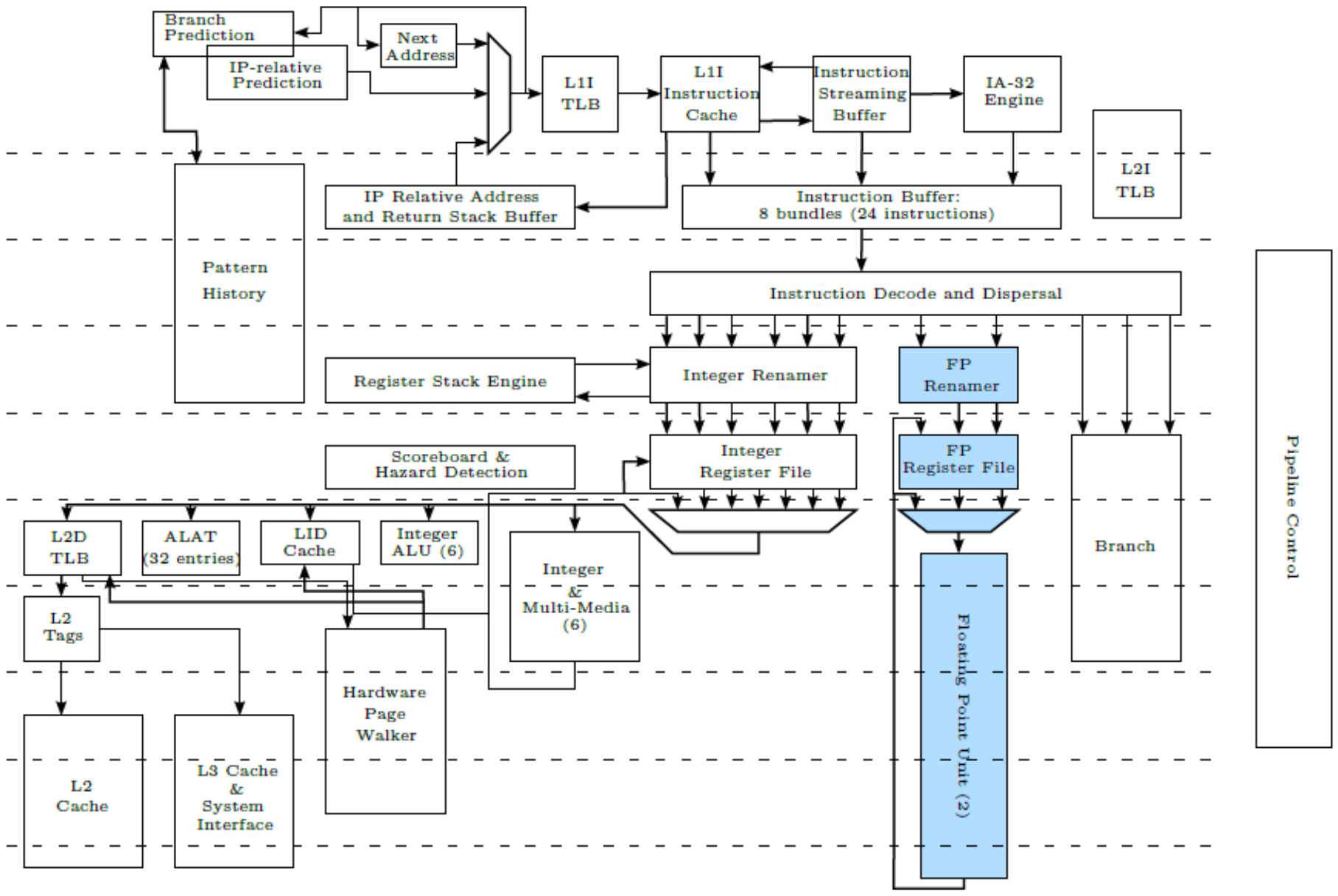

**Fig. 3.15** Intel Itanium Architecture

In most computer architectures, there are separate instructions for floating-point multiplication and floating point addition. Itanium include as a basic arithmetic operation the *floating-point multiply-add*, which allows higher accuracy and performance in many common algorithms. Addition and multiplication can easily be implemented as special cases of the fused multiply add (fma), for example $x + y = x.1 + y$ and $x.y = x.y + 0$.

Itanium processor support single, double and double-extended precision formats. All rounding modes have been implemented and all five exceptions in order to be fully compliant with the IEEE-754 standard. Also Intel define some specific exceptions for subnormal operands. [34]

## 3.5. AMD Bulldozer Architecture

AMD Bulldozer microarchitecture is used in the AMD CPUs since 2011. Bulldozer is the codename for the architecture, not for a specific processor.

The 15th AMD Processors family is aggressive, out-of-order, four-way superscalar AMD64 processors. They can theoretically fetch, decode and issue up to four AMD64 instructions per cycle. As shown in Figure 3.16, the two cores available in each Bulldozer module share the Fetch unit. The two cores also share the L1 instruction cache because it is an essential part of the fetch unit, but each CPU core has its own L1 data cache.
The AMD instruction set is complex (CISC). 15th AMD Processors family does not execute these complex instructions directly. Decode unit is in charge of converting the instructions provided by the compiler (macro-operations) into simpler fixed-length instructions called micro-operations [35]. The Bulldozer architecture has four decoders. The decoding of complex instructions takes several clock cycles to be completed, because they are converted into several microinstructions. Simple instructions, however, are usually converted in only one clock cycle because they are translated into a single microinstruction. After the instructions are decoded, they are sent to the appropriate scheduler, integer or floating-point. The Bulldozer architecture has only one floating-point unit, which is shared by two “cores” available. On the other hand, it has two completely independent integer units, the so-called “cores.”

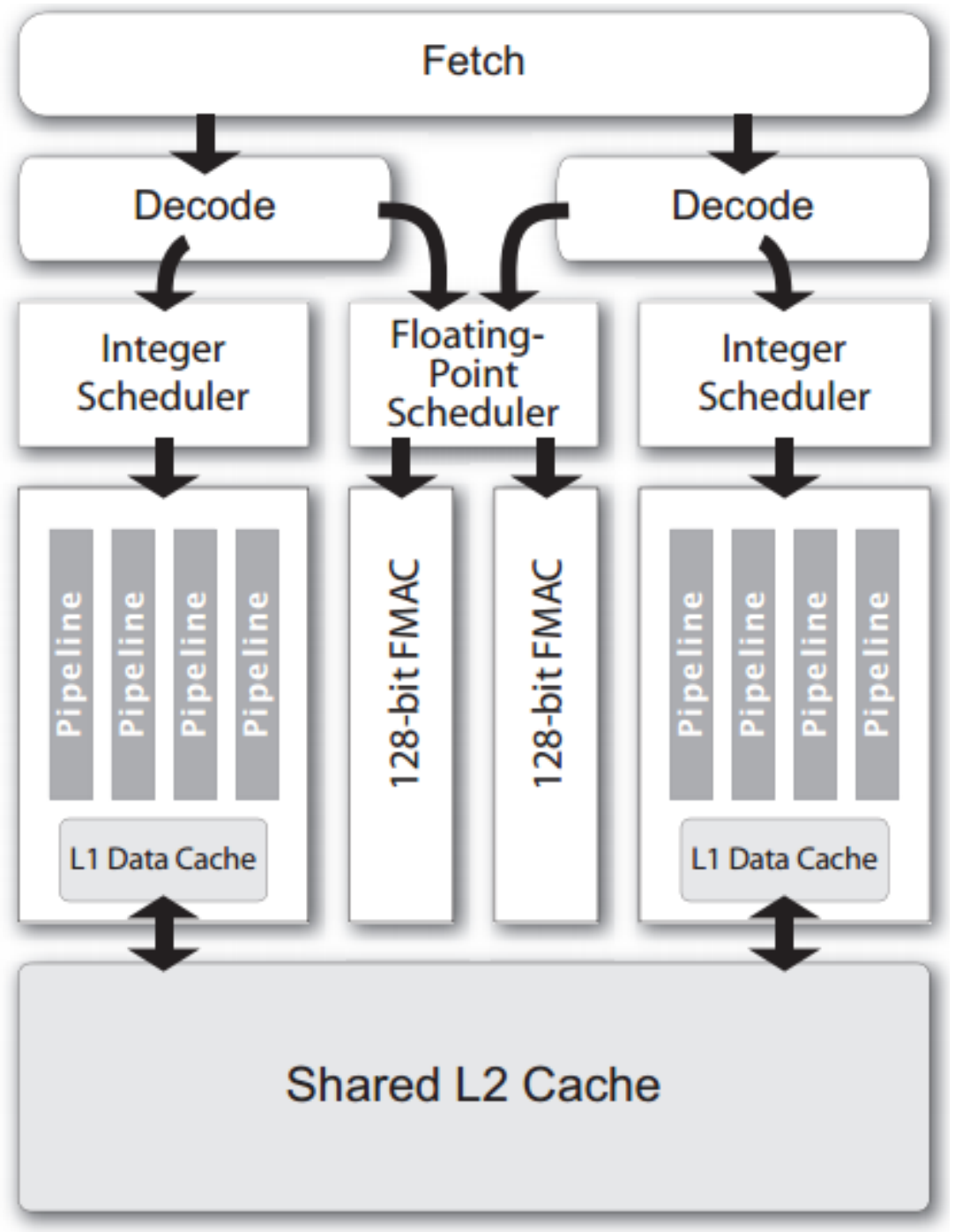

**Fig. 3.16** Bulldozer building block

The Bulldozer architecture uses an out-of-order execution engine, like AMD64 CPUs and Intel CPUs since the Pentium Pro (P6 architecture). After instructions are executed, perform commit in order as any out-of-order processor today.
The optimization comes from the fact that on a typical multi-core CPU several units inside the CPU remain idle, and these units could be combined in the Bulldozer architecture. And since the CPU will have less units, it can save area, register ports, save energy and reduce cost according AMD.

Each integer engine has four Execution units; it also has a Load/Store unit (“LD/ST”), which is in charge of getting from the memory or storing in the memory a data requested by an instruction.

Bulldozer architecture was designed to provide improved FADD and FMUL bandwidth over Opteron and Athlon 64 processors. It achieves this by means of two 128-bit fused multiply accumulate (FMAC) units which supports four single precision or two double precision operations. The FPU is a coprocessor model that is shared between the two cores. As such it contains its own scheduler, register files and rename units. In addition to the two FMACs, the FPU also contains two 128-bit integer units that perform arithmetic and logical operations on AVX, MMX and SSE packed integer data. Only one 256-bit operation can issue per cycle [35].

Users may notice differences in the results of program when using FMAC instead to perform a multiplication an addition. However, the combined result of the MUL and ADD is more precise, as is explained in chapter 4.3.

Bulldozer architecture includes support for Intel's Advanced Vector Extensions (AVX) instruction set, which supports and extended set of 128-bit (XMM) and 256-Bit (YMM) media registers [36]. The physical registers internally are 128-bits in size, equal to an XMM or half of a YMM register (it takes two internal registers to represent a YMM 256-bit register). To represent the Instruction Set Architected (ISA) registers it takes: 16 registers (YMM0-YMM15), or 32 (XMM0-XMM31) as shown in Figure 3.18.

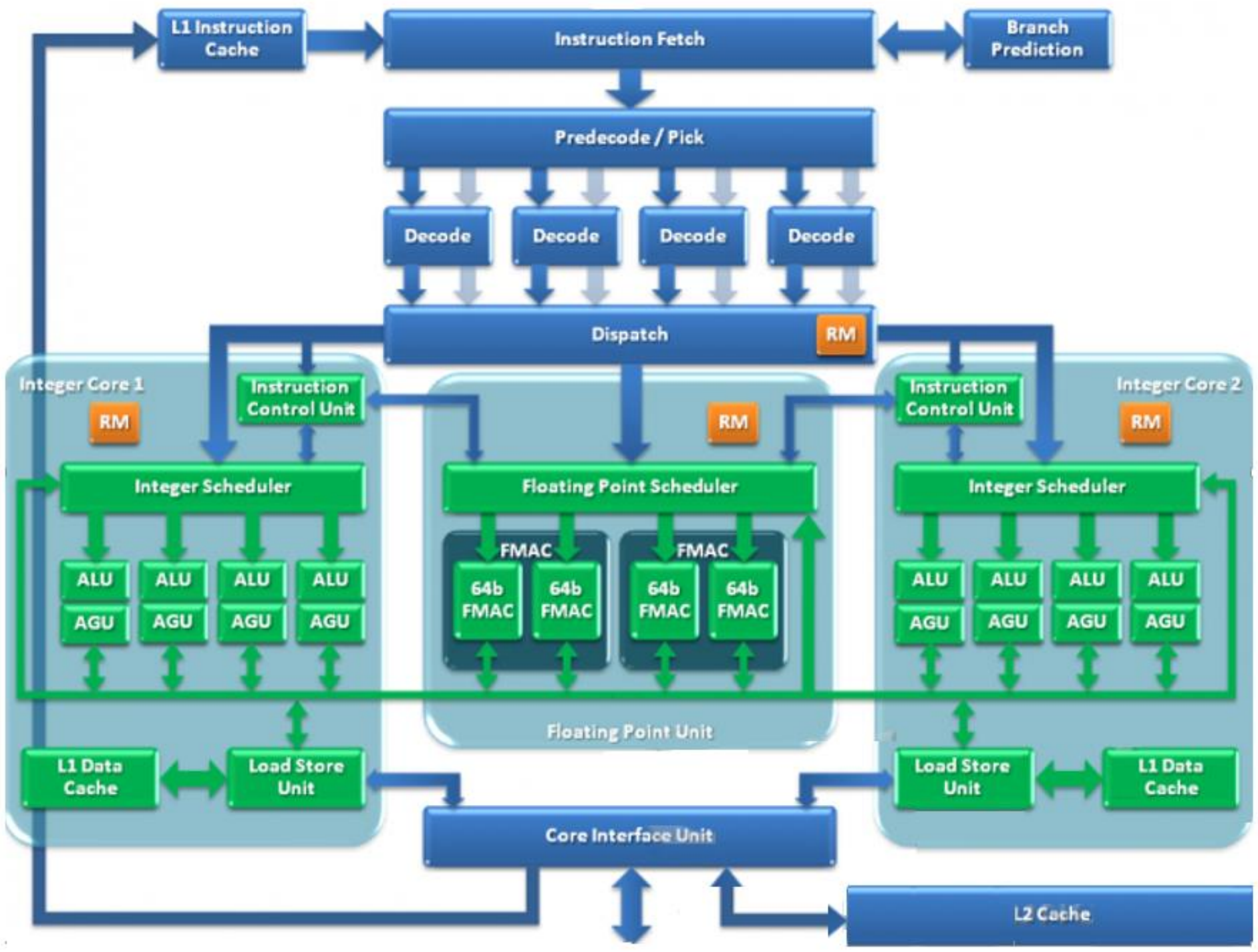

**Fig. 3.17** Inside Floating Point 128 FMAC

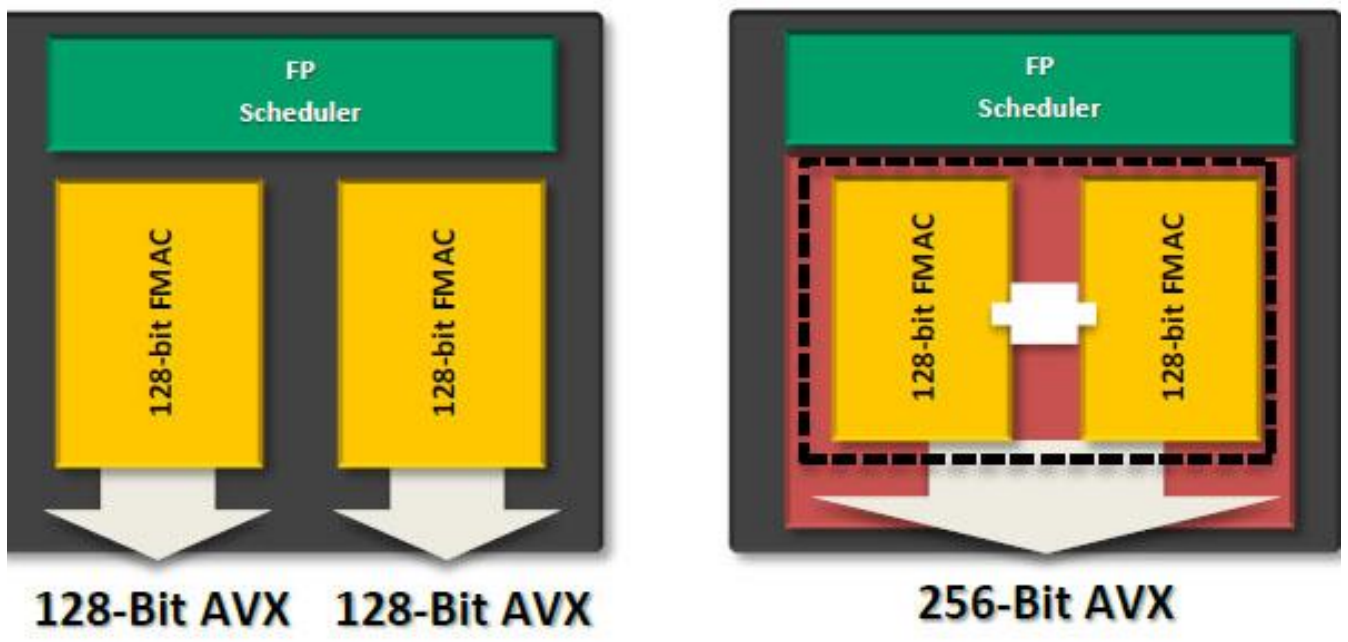

**Fig. 3.18** Execution of AVX instructions

Figure 3.19 shows the die of one bulldozer module in the AMD FX processors. The area consumed by the Floating-Point/SIMD Unit is bigger than each Integer datapath, also the benefits to share the same hardware between both units is huge due to usually the current processors can issue 2x64-bits or 4x64-bits FP scalar instructions per cycle, bulldozer module shares the FP hardware, then using 2x128-bits FMAC units can perform 1x256-bits or 2x128-bits SIMD operations or 4x64-bits FP scalar operations. The area of this units separated is almost the same, therefore share this hardware bring a huge benefit in terms of die area.

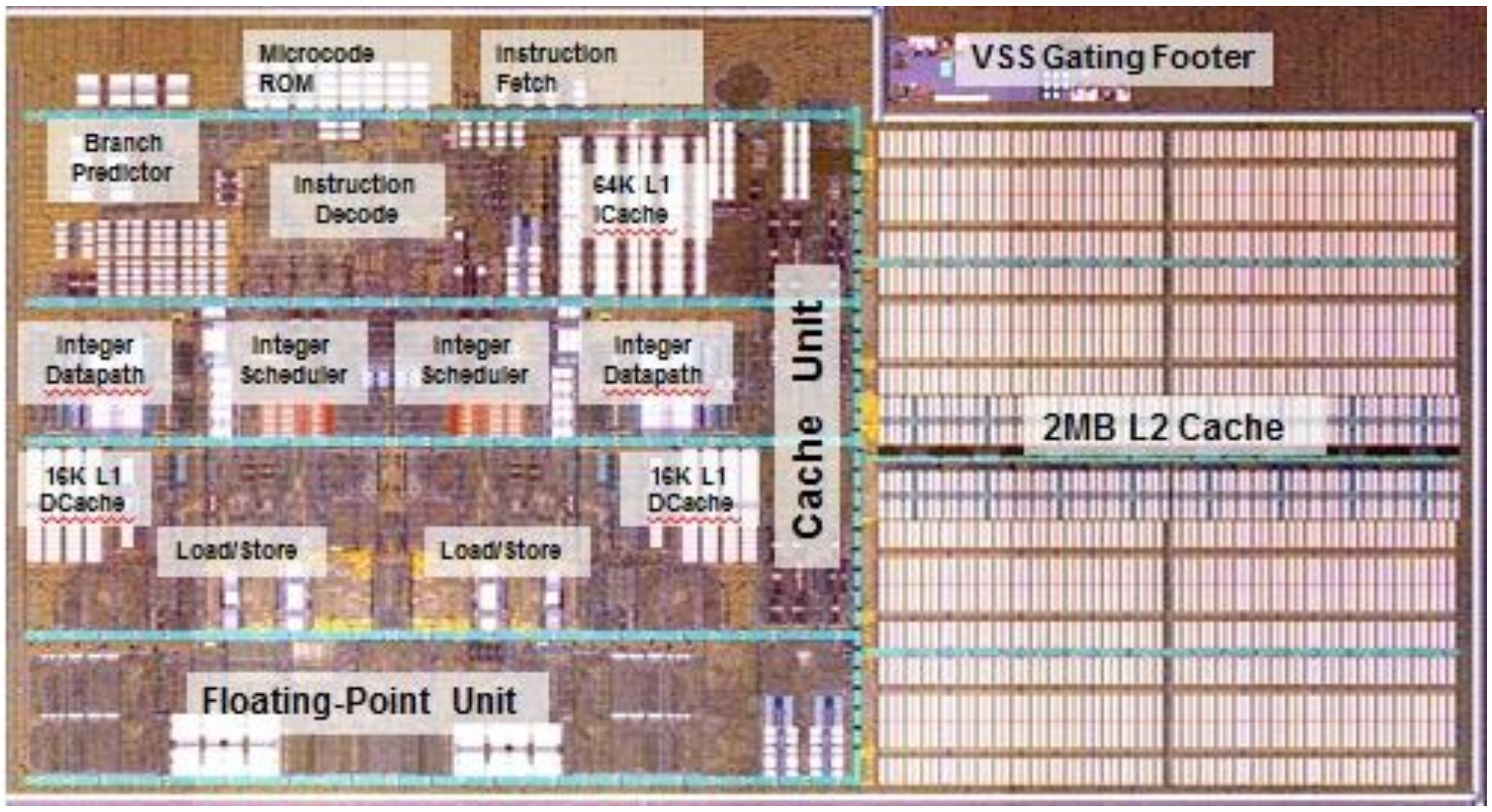

**Fig. 3.19** AMD Bulldozer Die (Fx Processors)

# Chapter 4

# 4. Design and implementation

Because exploiting instruction level parallelism (ILP), superscalar processors are capable of execute more than one instruction in a clock cycle. As we mention in Chapter 3, to do a dynamic scheduling we need:

- Fetch strategies that simultaneously fetching multiple instructions.
- Branch prediction strategies that predict the execution path of the instructions and fetching speculative code.
- Methods for determining true dependences involving register values, and mechanism for communicating these values to where they are need during execution.
- ***Methods for issue multiple instructions in parallel (Out of order execution).***
- ***Resources for parallel execution of many instructions.***
- Methods for committing the process state in correct order; these mechanisms maintain an outward appearance of sequential execution.

Lagarto II Processor has Instruction fetch strategies to fetch multiple instructions, while in same stage implement a 2 level branch predictor (GShare) in order to predict the branches. Afterward, decode stage identifies the main attributes of the instruction such as type and resources that it will require for their execution. The following stage performs a rename to delete the name dependences. Also, will execute instructions out of the original program order, then it has implemented a Reorder Buffer to preserve the original program order, also will need an out of order issue queue to send all possible ready instructions in a cycle which is part of the presented design in this work. Furthermore, Lagarto will have resources for parallel execution of many instructions (integer and floating point).

Processors that implement a dynamic scheduling exploit the instructions level parallelism but at the same time, these processors spend more energy than processors that implement a static scheduling.  This leads to a tradeoff between power consumption and high performance. To implement an efficient power-performance dynamic scheduling designer needs know about low power techniques as we said in last chapters, nowadays, power consumption is very important to obtain a large autonomy in mobile devices.

In this chapter we describe our implementations of each component of the general out of order execution engine, which include the Issue queue, the register File, execution units and the bypass logic. Was proposed two designs which are compared in Chapter 5.

## 4.1. First Proposal

### 4.1.1. Issue Queue

As we can see in previous chapters, issue queue design is an important component to exploit the instruction level parallelism, but in processors like Pentium 4 the issue logic is one of the main consumers of energy responsible for approximately 25% of the total energy consumption [15], then we need considerate some low power consumption techniques in order to get a low power consumption processor.

Also the width of fetch is an important parameter in the process design, a processor with large emission width it becomes more complex design and the complexity not produce performance necessarily, for example 2-wide processors has an average of 1.1 commit instructions per cycle, 4-wide processors has an average of 1.52 commit instructions per cycle, 6-wide processors has an average of 1.79 commit instructions per cycle and 8-wide processors has an average of 2 commit instructions per cycle, for integer benchmarks [37].

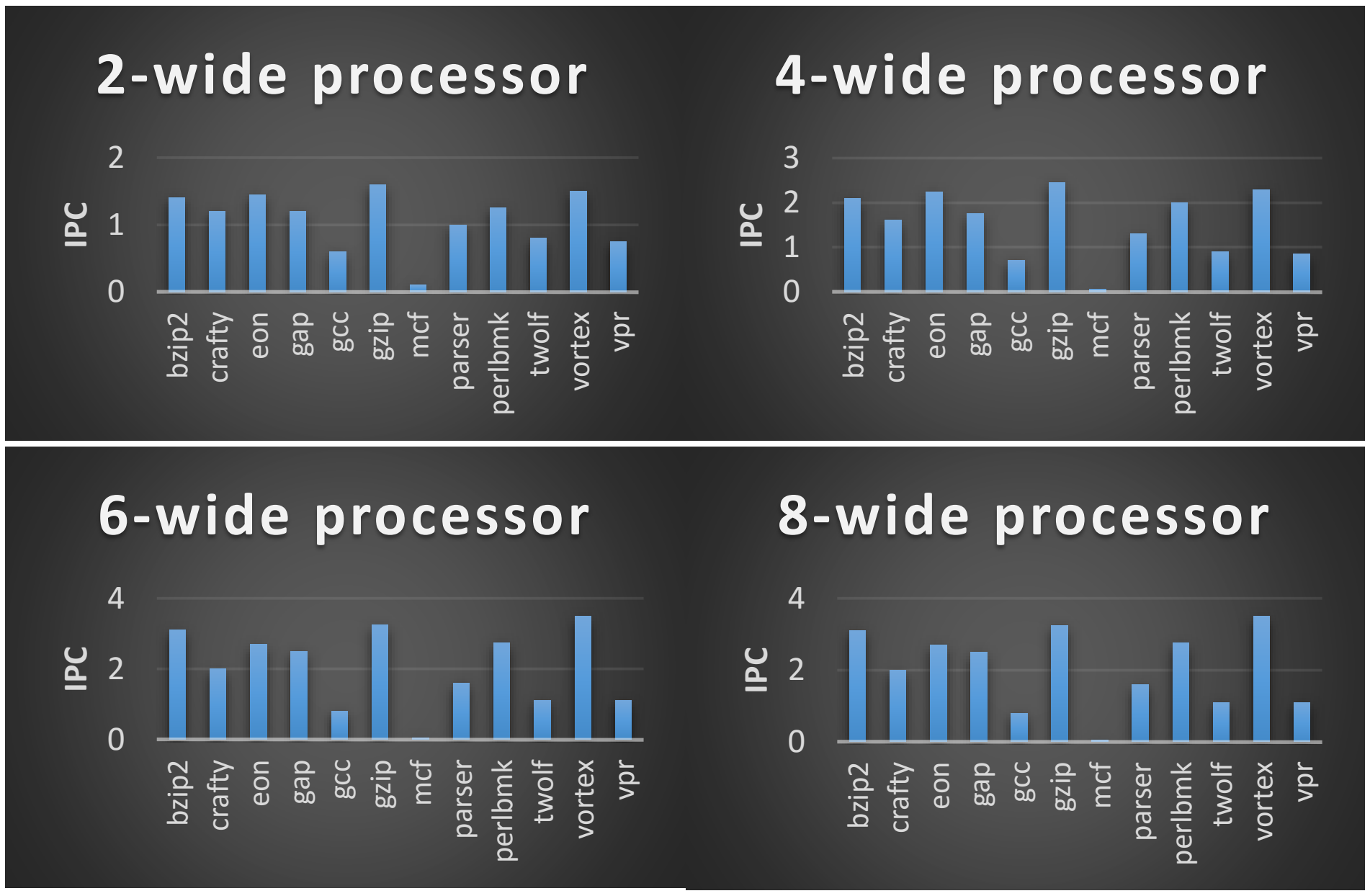

**Fig. 4.1** IPC for n-wide for a baseline processor

Lagarto II Architecture perform fetch, decode and dispatch up to 2 instructions per clock cycle to 3 different Buffers, Load/store Queue, Integer Queue and Floating Point Queue. Integer and Floating Point Queues can issue up to two instructions each one if instructions are ready, and Load/Store Queue can issue 1 instruction per clock cycle, these parameters were taken in order to reduce the number of ports in register bank and reduce the general complexity of the design because performance-complexity trend. The idea is that the design of the Instruction Issue Queue used in Lagarto II be a low power consumption architecture, and we decide take account the element proposed in [18] and add this element to our design which is described below.

**Instruction Queue Design**

Once the instructions have been decoded and renamed, they are allocated in a structure called Mapper, in this structure we read the operation vector (OPVEC) associated with each instructions to determinate the instruction queue in which should stay until their source operands are ready and execution unit required for their execution is available.

Lagarto II processor has 3 types of issue queues: Integer Instruction (Integer Instruction Issue Queue), floating point instructions (Floating Point Issue Queue) and Memory Access Instructions (Load / Store Queue). The wakeup instruction mechanism (Wakeup Logic) and selection (Selection Logic) are closely associated in an IQ, they determine the behavior of the instructions are stored in it.

Due to our design we will take account the Block Mapping Table presented in [18] the IQ design divide the queue in N blocks with M entries each one, then, according with the results presented in [18] where IQ is divided in 4 and 8 blocks with a similar performance (a little more in 8 blocks design), but later we will see that in the selection logic, the larger number of blocks becomes the selection logic a bit more complex. From this point 4 blocks are selected for this design.

***Allocation Logic***

Due to the CAM and RAM memories were divide in 4 blocks of 8 entries each one, an assignation algorithm was implemented which will define in which block the instruction should be allocate when one or two instructions arrive from the previous stage (MAPPER/DISPATCH) in the same clock cycle.

Allocation Logic receive a pair of signals called *Active Instructions* from the previous stage (MAPPER). It then passes them through inter-stage latch *DISPATCH/ISSUEQUEUE,* which signalize how many instructions are incoming to the floating point queue in this clock cycle. If the FP-Queue is full, Allocation Logic will send a signal (Full) in order to the *FETCH UNIT* not perform more fetch cycles.

An assignation algorithm was implemented based in the round-robin scheduler, in this scheme one instructions is allocated in each blocks. The algorithm can’t assign more than 1 instruction to the same block in the same clock cycle. Thus storage starts at the block B0 and ends in block

B3 to after that restarting. Figure 4.1 show the design of the Allocation Logic, showing the blocks Round robin, Data Assignation, FIFO B0, CAM B0 and PlayLoad RAM, the last three should be repeated 4 times.

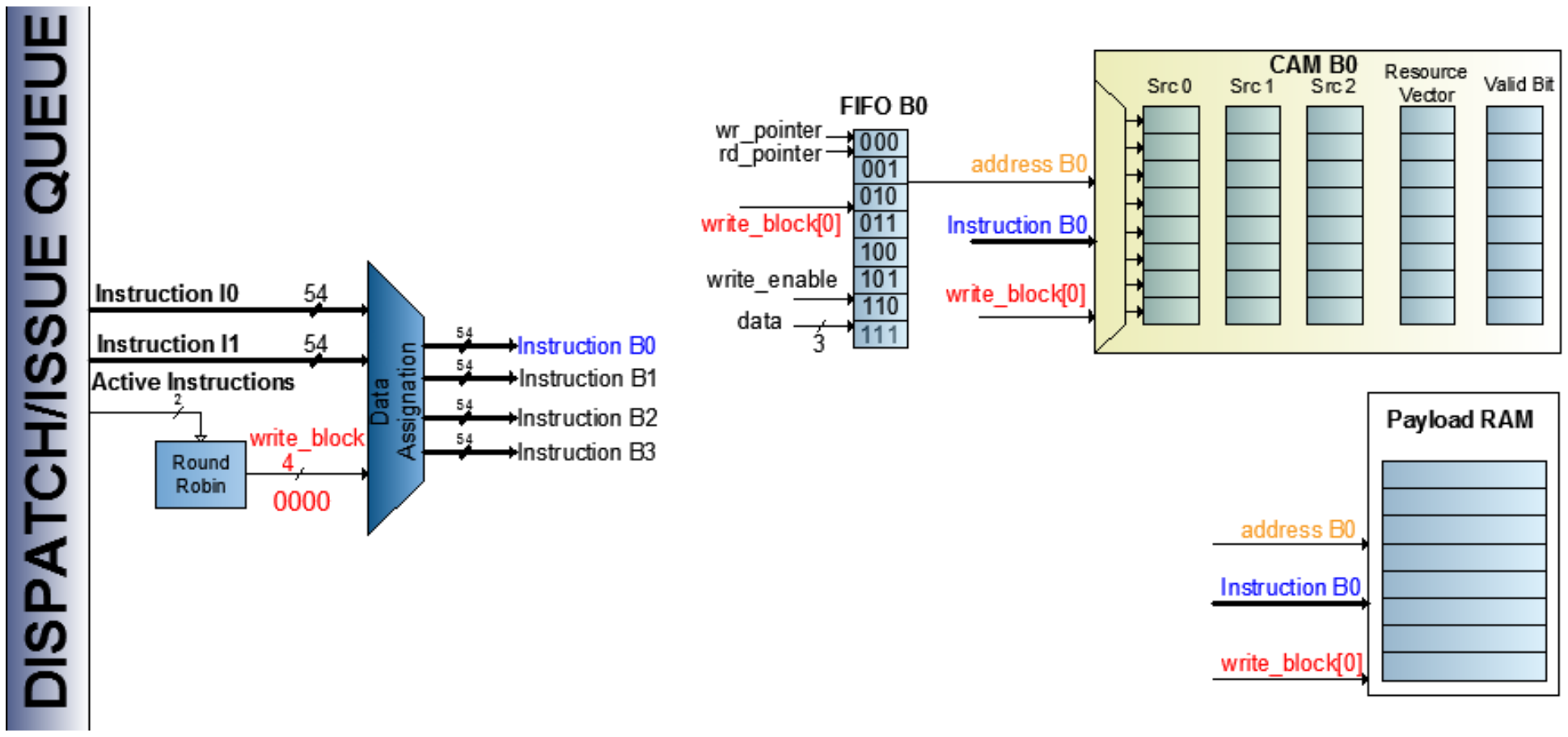

**Fig. 4.2** Block diagram of Allocation Logic

Data Assignation block only is a 2 to 4 Demultiplexor which put in the correct way the incoming instructions according to the signal *write_block* which is generated in the *Round Robin* block.

Round Robin block is a finite state machine with *Active instructions [1:0]* as entries and *write_block[3:0]* as outputs, depending of value of *Active instructions* this are distributed one by one in each IQ Blocks.

FIFO B0 is a little buffer of 8 locations of 3 bits each one, which contain the free locations in the IQ Block, which comprise both *CAM* and *PAYLOAD RAM Blocks.* When a new instruction arrives, FIFO Block select a free entry of each particular IQ Block appointed by the position of the *rd_pointer*. In the other hand when an instruction waiting in the IQ Block is issue to execution, the entry freely is recycled to FIFO block performing a write to entry appointed by *wr_pointer*.

***Low Power WakeUp Logic Mechanism***

As a part of dynamic scheduling, in the issue stage is required a wakeup mechanism for waking up instructions waiting in each IQ Block. The waking up are accomplished by associative comparisons of the destination register tag of the instructions computed each clock cycle with source register tags of instructions sleeping in the IQ Blocks while its operands become ready. One operand is ready when destination tag and source tag matched. This mechanism is power hungry because comparisons are always performed although it not produces operands ready.

One instruction is ready when all its operands become ready and the functional unit required for execution is free. Ready logic is responsible for signalize instruction ready to issue.

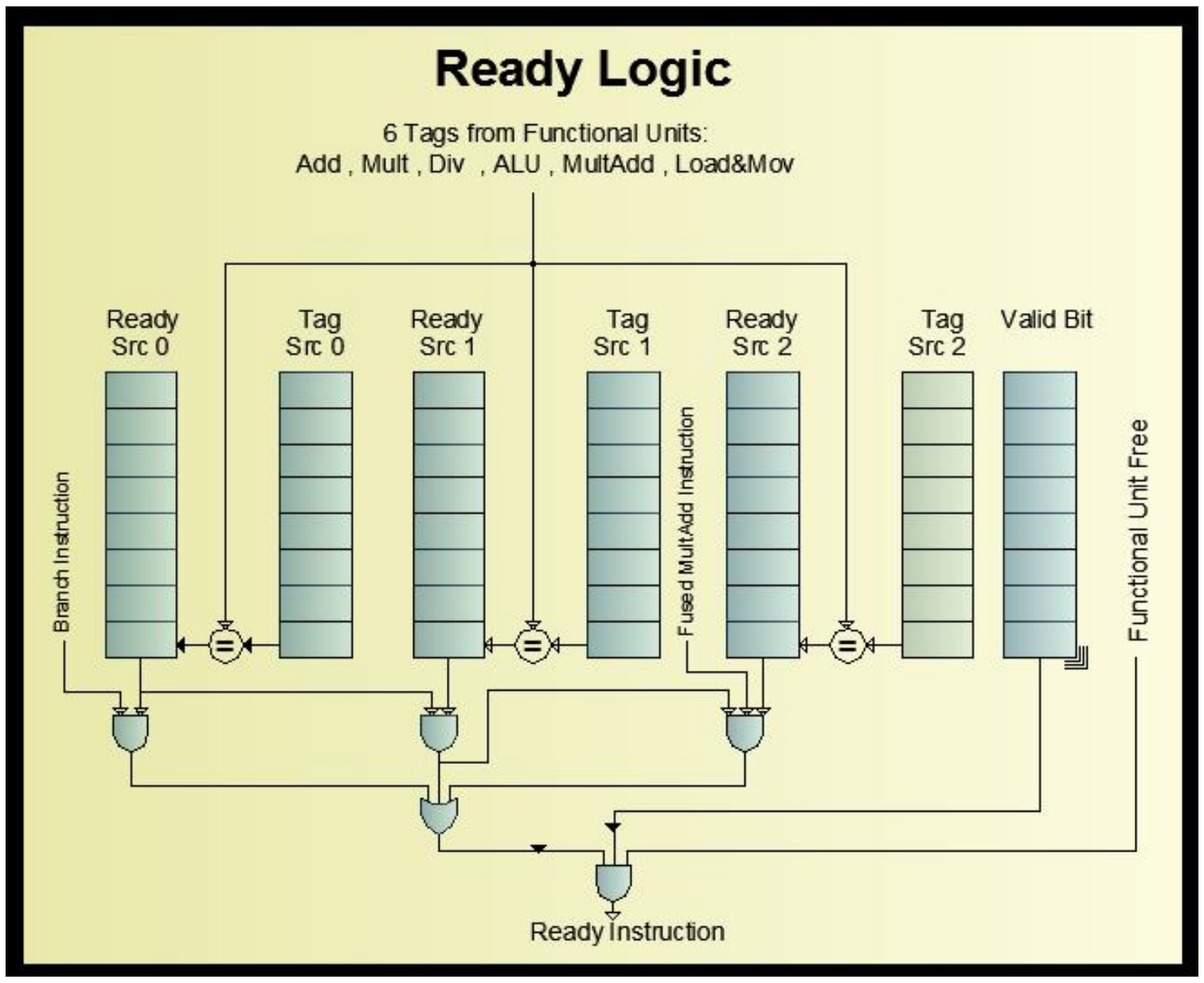

**Fig. 4.3** Ready Logic

The destination tag of current execution is broadcasted to all instruction in the queue N-Cycles before its execution will be completed. N is the number of cycles required for schedule the wakeup, selection, issue and read registers of consumer instructions. This Schedule must ensure that result's value is present in the bypass network at same cycle when consumer instruction arrives to the functional unit.

The destination tag in traditional CAM/RAM designs must be compare with all elements of the queue, 64 comparisons every clock cycle for a queue with 32-entries (two source operands for instruction). In the design also was added one source operand more in each location because *Lagarto II* Architecture execute instructions as Multiply Accumulate which uses three sources as shown in Figure 4.2, also in the design add an extra logic proposed in [18] called Block Mapping Table. In the following sequence illustrate the behavior of the design using the Block Mapping table.

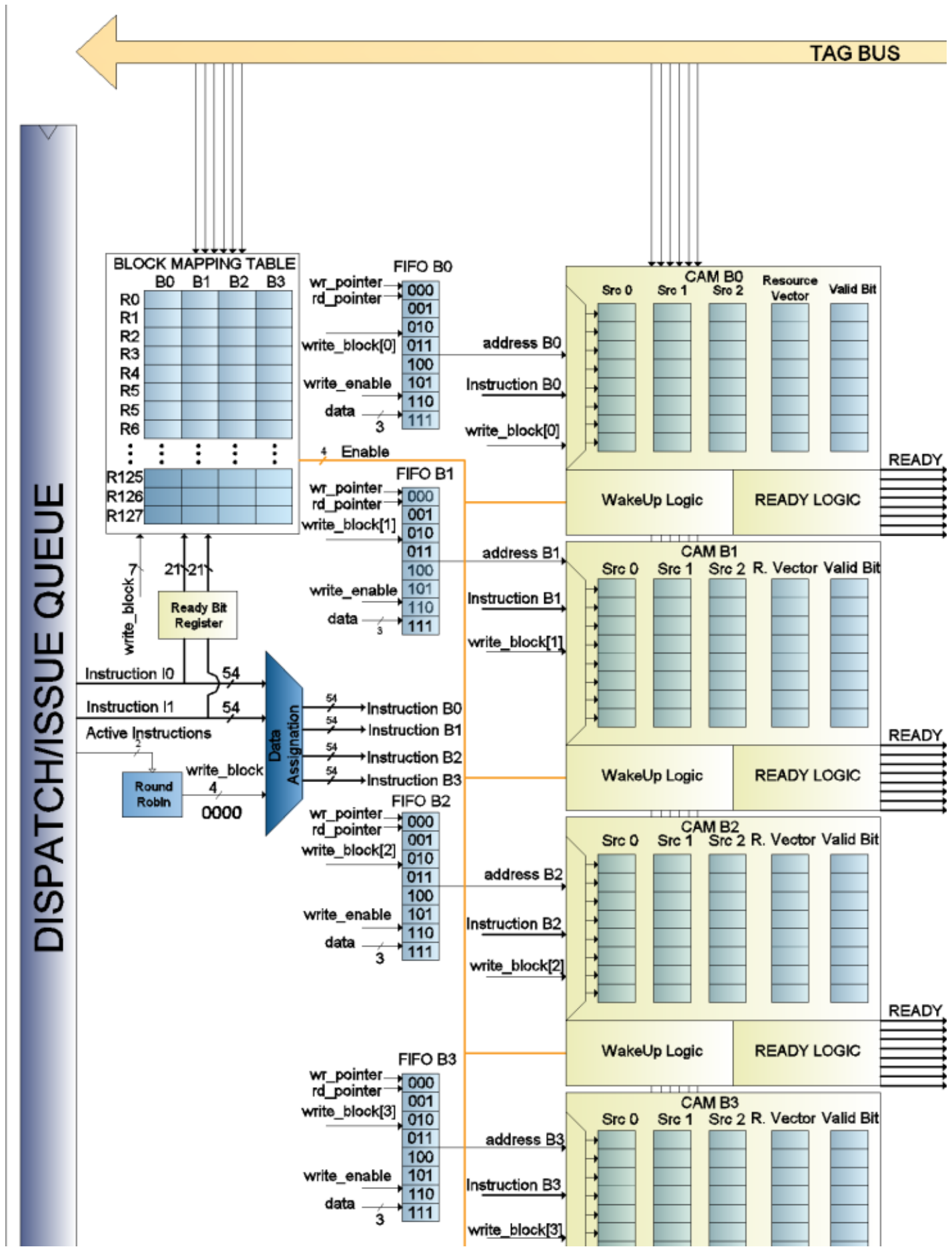

**Fig. 4.4** Behavior of the Wakeup using the Block Mapping Table

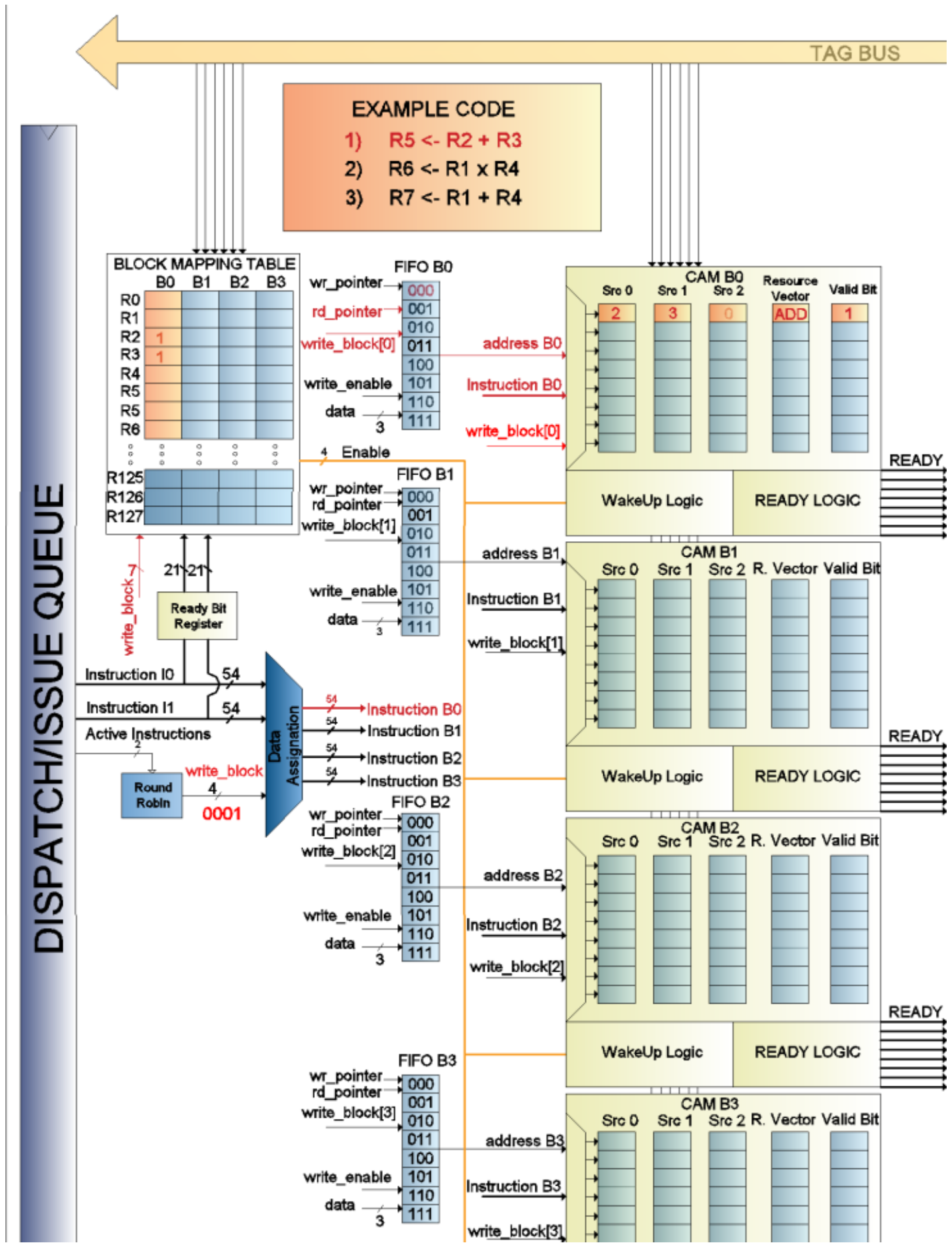

**Fig. 4.5** Behavior of the Wakeup using the Block Mapping Table

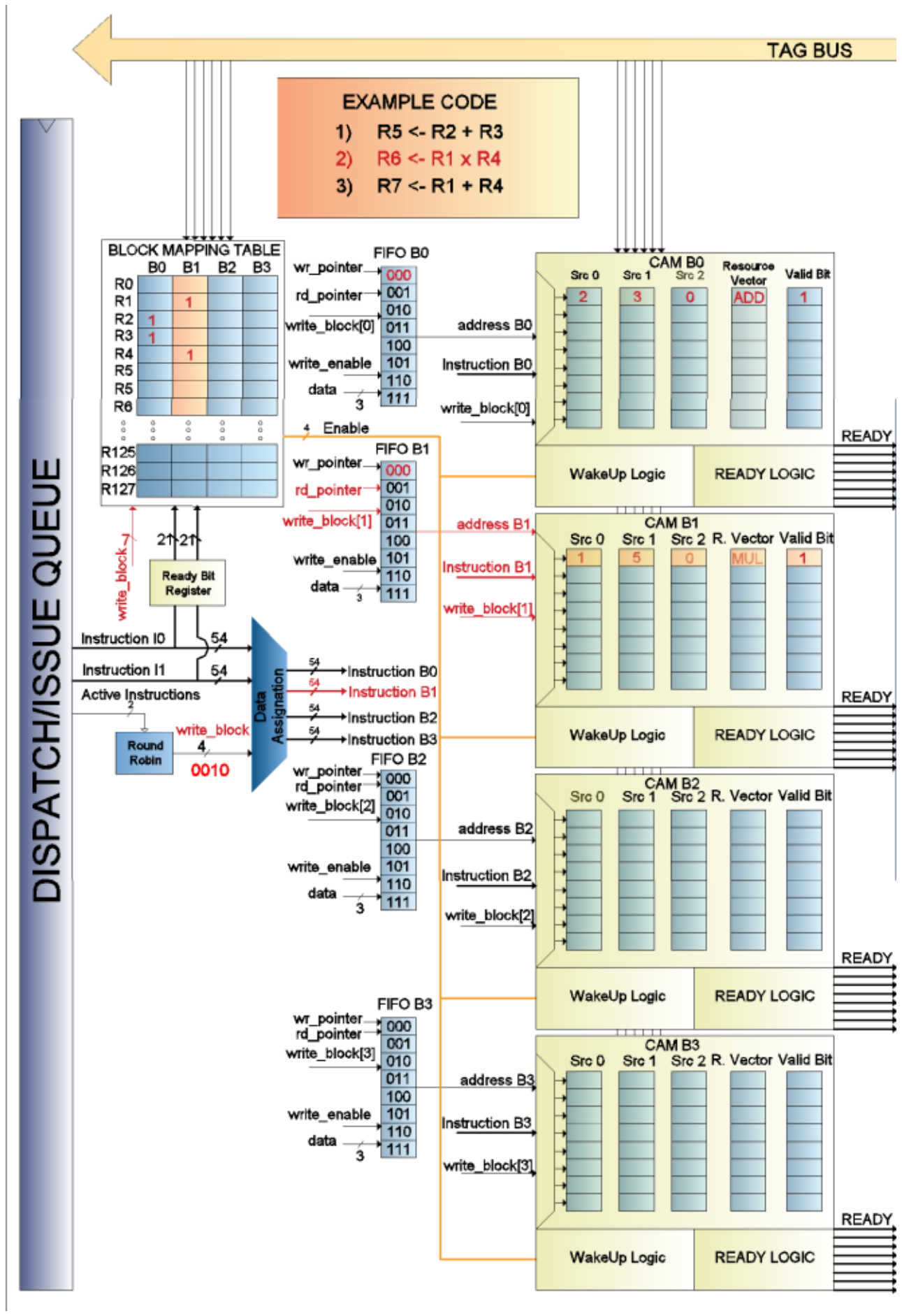

**Fig. 4.6** Behavior of the Wakeup using the Block Mapping Table

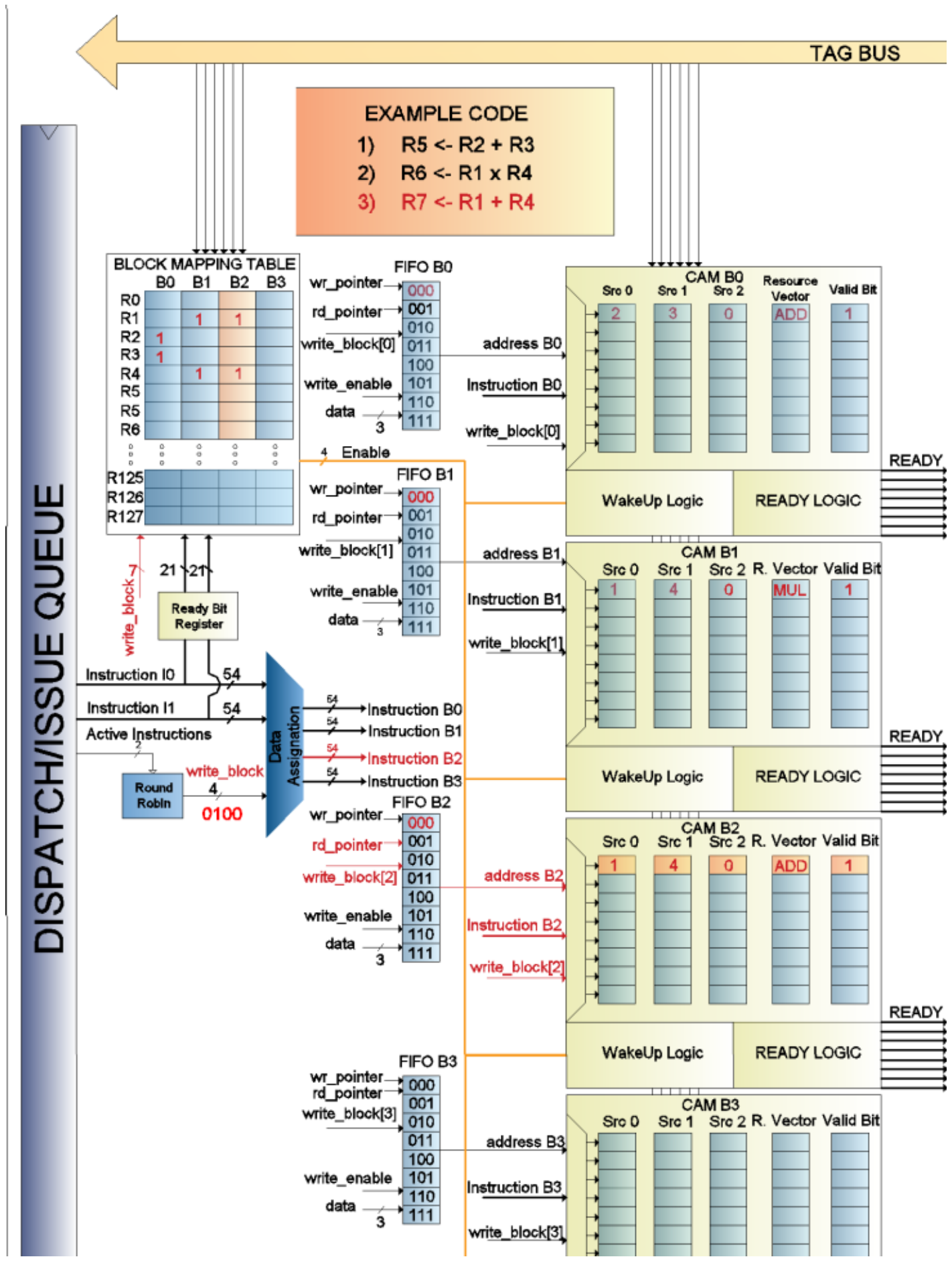

**Fig. 4.7** Behavior of the Wakeup using the Block Mapping Table

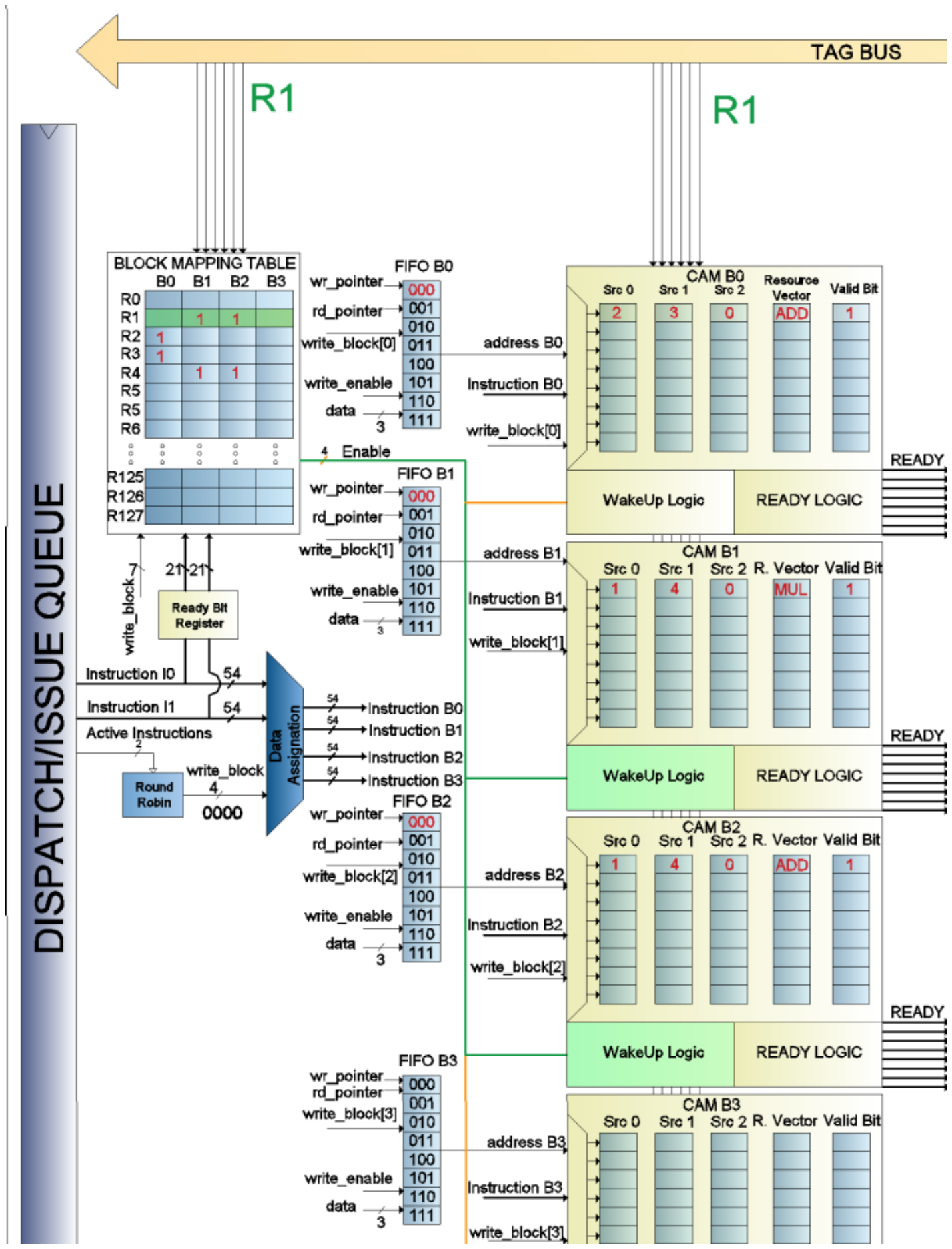

Fig. 4.8 Behavior of the Wakeup using the Block Mapping Table

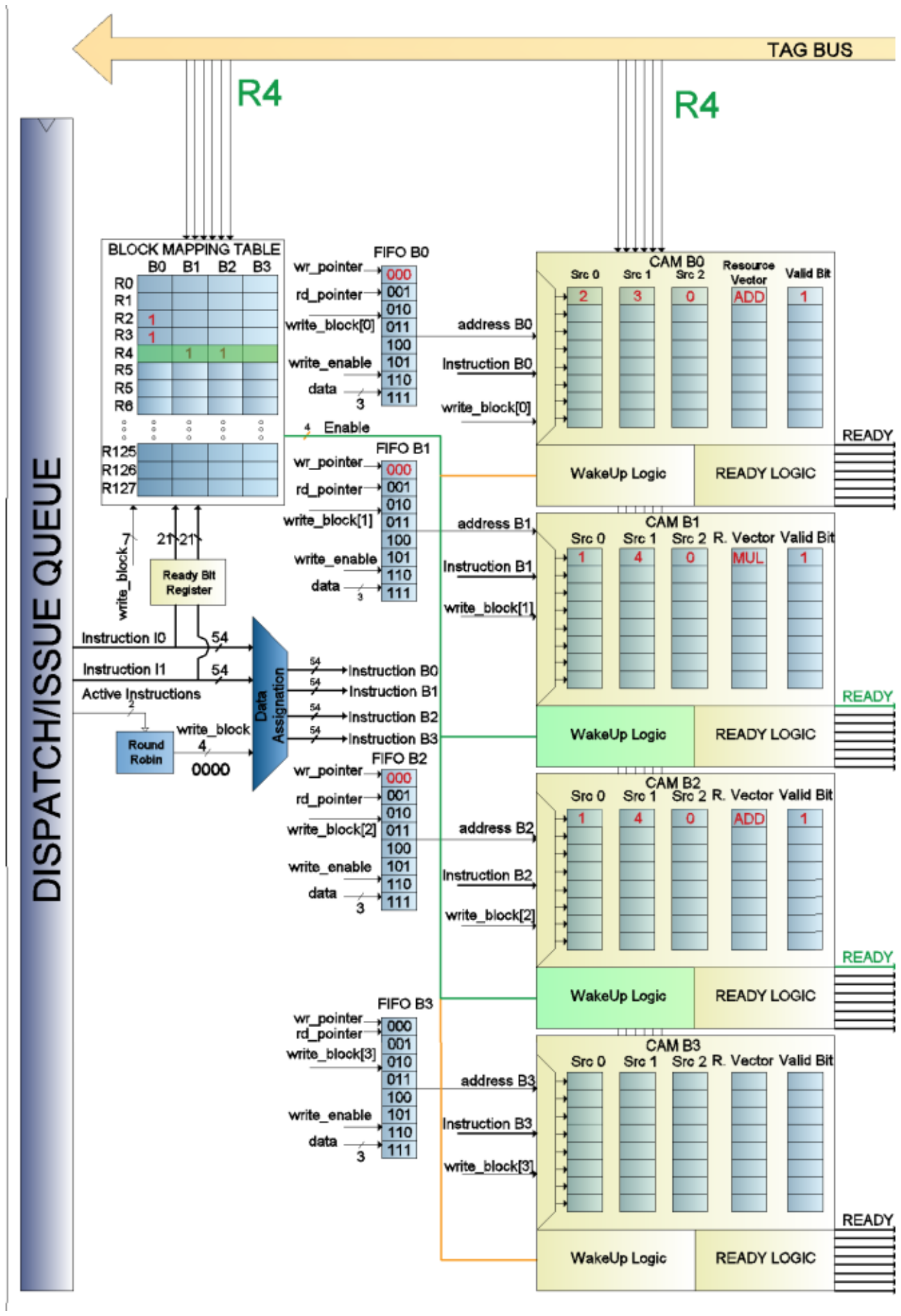

Fig. 4.9 Behavior of the Wakeup using the Block Mapping Table

In-flight instructions have only one identifier through all the time of its execution, specifically the destination register tag, then the *Block Mapping Table* is a structure associated to the register file to encoded IQ-blocks where instruction successor was allocated by the round robin logic. The length of the *Block Mapping Table* is same to the physical register file and the width is the number of blocks witch IQ was divided.

Figure 4.3 shows the first part of our design, the queue is divided in 4 blocks as we mentioned above and also the *Block Mapping Table* is included in the design.

Figure 4.4 show the first instruction (R5 <- R2+R3) arriving to the Queue, the Round *Robin* Block send the signal “0001” which means that only one instruction is arriving in this cycle and this instruction will be saved in the *CAM-B0* block, *FIFO-B0* block give the address in which the instruction will be allocated inside the *CAM-B0,* in this case, the address is “000”, at the same time the source operands read the *Registers ready bit vector* in order to know if its sources operands are ready or not, if one of the sources or both are ready, the ready flag are setting, otherwise the *Block Mapping Table* is set, indexed by the source operand tag that is not ready, in this case, the address two and three in the column B0 are set.

In the next cycle (Figure 4.5), a new instruction (R6 <- R1 + R4) is arriving, now the *Round Robin* block assigns the next CAM block (*CAM-B1*) to allocate the current instruction, also the *Block Mapping Table* is updating by setting the addresses one and four, in column B1 indicating that this operands R1 and R4 are not available.

In Figure 4.6 a new instruction is arriving (R7 <- R1 + R4), now *CAM-B2* is selected by *the Round Robin* Block to allocate this new instruction, also the *Block Mapping Table* is updated in the column B2 indexed by R1 and R4 designating that this sources are not available.

In Figure 4.7, functional units start the successor wakeup 3 cycles before that finalize the execution, sending the Tag destination register to read the *Block Mapping Table.* The data read is useful to enable the CAM blocks for comparison. Comparisons are not performed in CAM Blocks without successors. In this example *CAM-B1* and *CAM-B2* blocks only are enabled for comparisons.

At this moment any instruction can be issued to execute because still not comply with the condition that both source operands must be ready.

In the next cycle (Figure 4.8), Consider that R4 is at 3-Cycles to be computed, then the successors wakeup logic is started as was described above, reading the 4-entry of *Block Mapping Table* and enabling the blocks CAM-B1 and CAM-B2 for comparisons, resulting in two instructions ready for select and issue in the next cycle.

Adding the *Block Mapping table* to the design, wakeup logic can avoid perform comparisons in blocks without successors.

***Priority Arbiter***

To have more than one instruction ready to be issue in each one of the four *CAM Blocks* is possible, however only is possible perform the issue of two instructions per clock cycle to the execution units, it is important to minimize the number of ports in order to achieve a power-efficient design. For this reason, is necessary define a selection criteria.

*Priority Arbiter* is a structure implemented in two stages, which choose just 2 instructions to be issued to the execution. Both stages employ an aging policy for the selection of instructions, meaning, the instructions allocated first in the Payload-RAM will be the first to be issued.

The first stage consists of 4 modules of selection, one for each block. In each selection module the ready instructions signals are received. The second stage is capable of receiving up to 4 instructions lists, one for each block of the previous level and finally chooses only two instructions to be issued to execution.

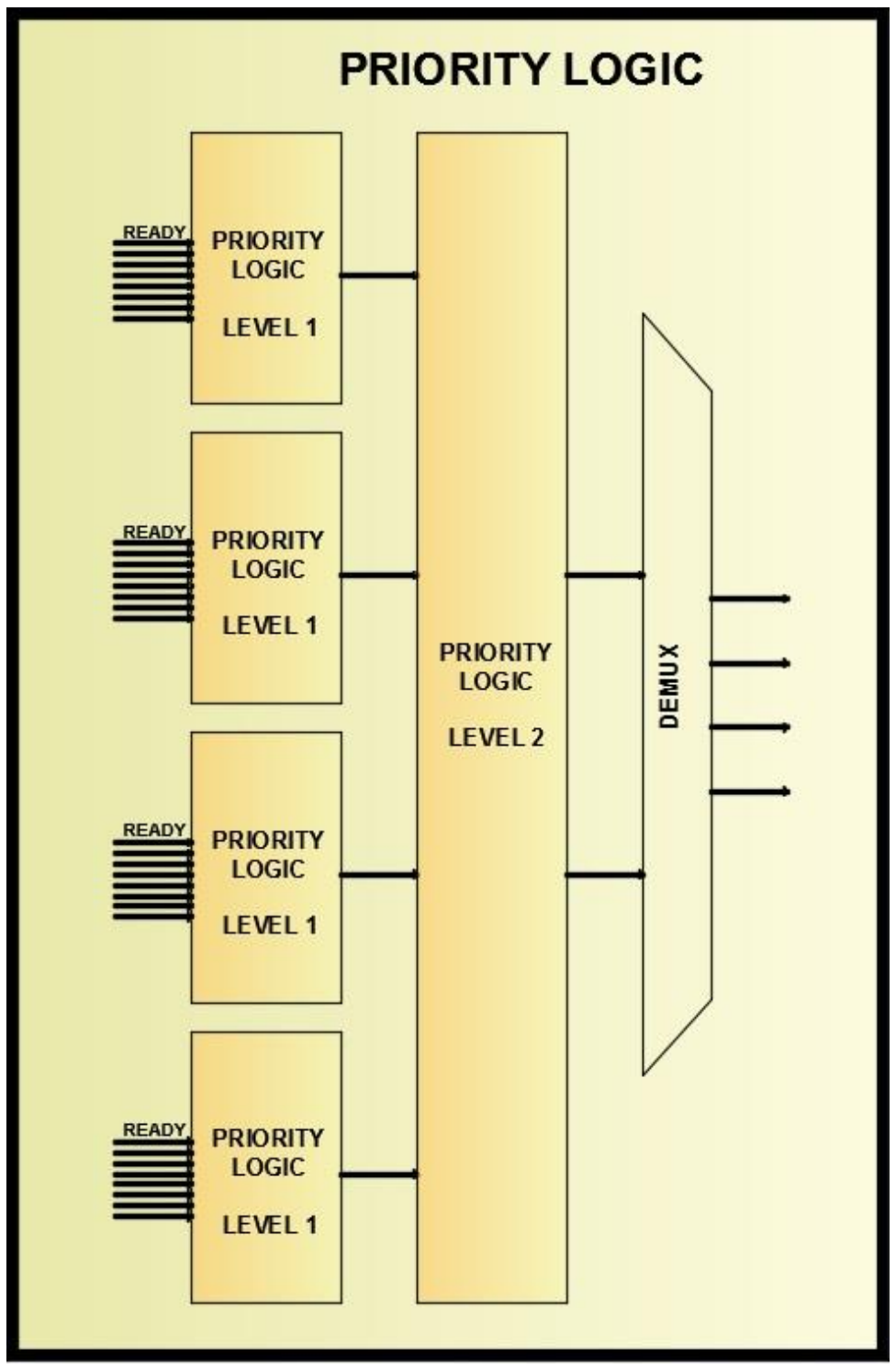

**Fig. 4.10** Priority Logic

***IQ Payload RAM***

IQ Payload RAM block is a set of 4 RAM blocks of 8-locations each one, where new instructions are allocated. Each instruction is composed for many fields:
*Format* which encode the instruction format (Single, Double, Word or Long).
*Source_0 , Source_1 and Source_2* which encode the operand address of the instructions. Note that each Source is composed of 7 bits to address 128 possible locations in the register file.
*Destination* which encode the physical register to save the result of the operation.
*Resource_Vector* which encode functional unit (Branch, MovToFrom, MulA, ALU, SQRT, DIV, MUL and ADD) needed to execute.
*Function* which encode what operation must be done.
*Dir_ROB* which encode the place of the instruction in the Reorder-Buffer.

| format | Source_0 | Source_1 | Source_2 | Destination | Resource_Vector | Function | Dir_ROB |
|---|---|---|---|---|---|---|---|
| [53:49] | [48:42] | [41:35] | [34:28] | [27:21] | [20:13] | [12:7] | [6:0] |

**Fig. 4.11** FP Instruction Format for the Issue Queue design

Each IQ Payload RAM Blocks has only one read port and one write port and only can write and read one instruction per cycle, because the module Round Robin require this behavior, so if two instructions are received in the same clock they are saved in two different Blocks RAMs. Similarly, when many instructions are ready can issue only two instructions per cycle and these instructions come strictly from different blocks. The design use a *PAYLOAD RAMs* splitting in blocks with 1 read and 1 write port each one.

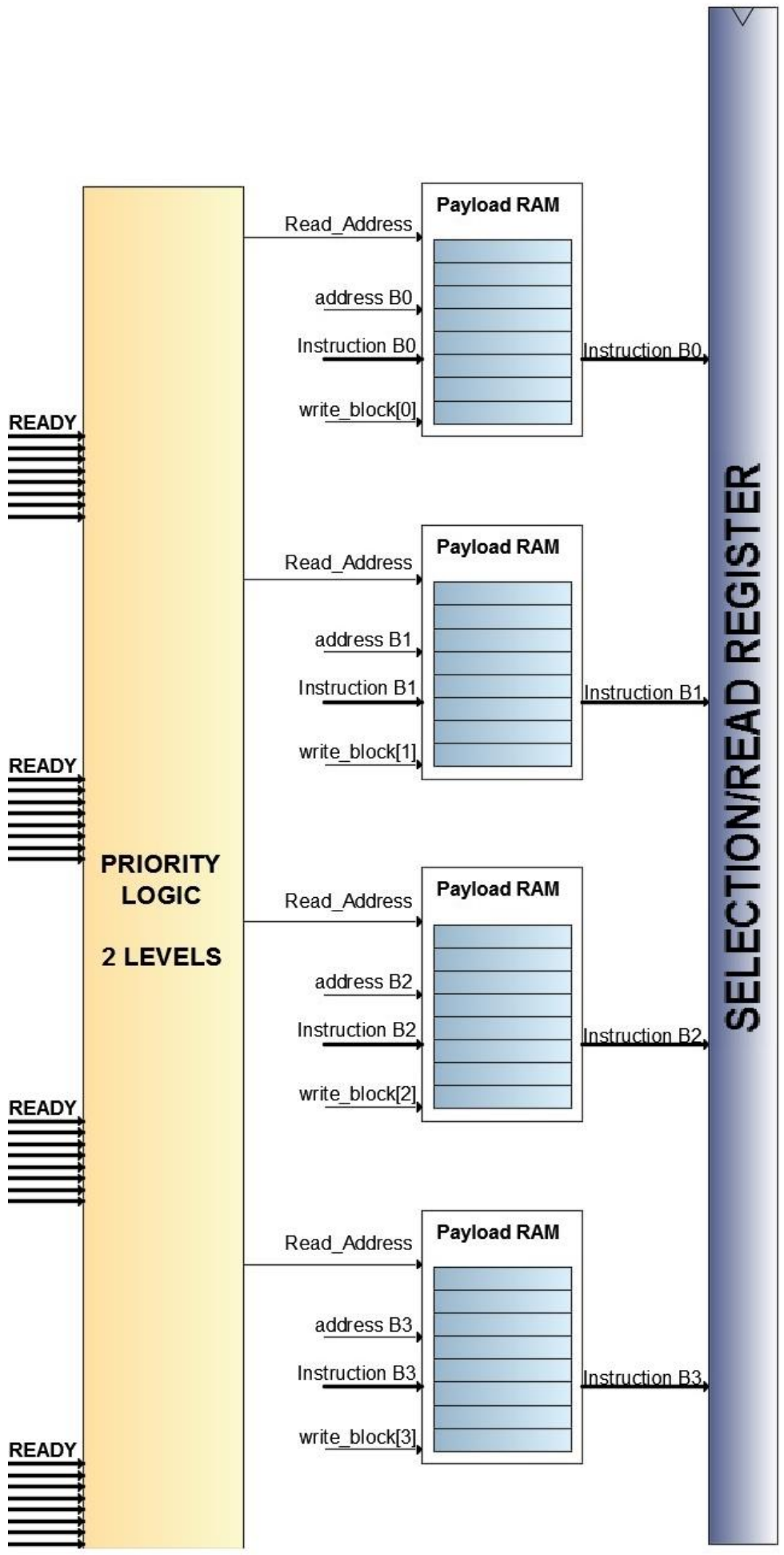

**Fig. 4.12** Payload RAMs

When an instruction is issue, the entry address used by the instruction is sent to the Allocation Mechanism in order to recycle it as a free entry in the corresponding FIFO, for future incoming instructions.

In Figure 4.13 the complete Low Power Issue Queue Design is show.

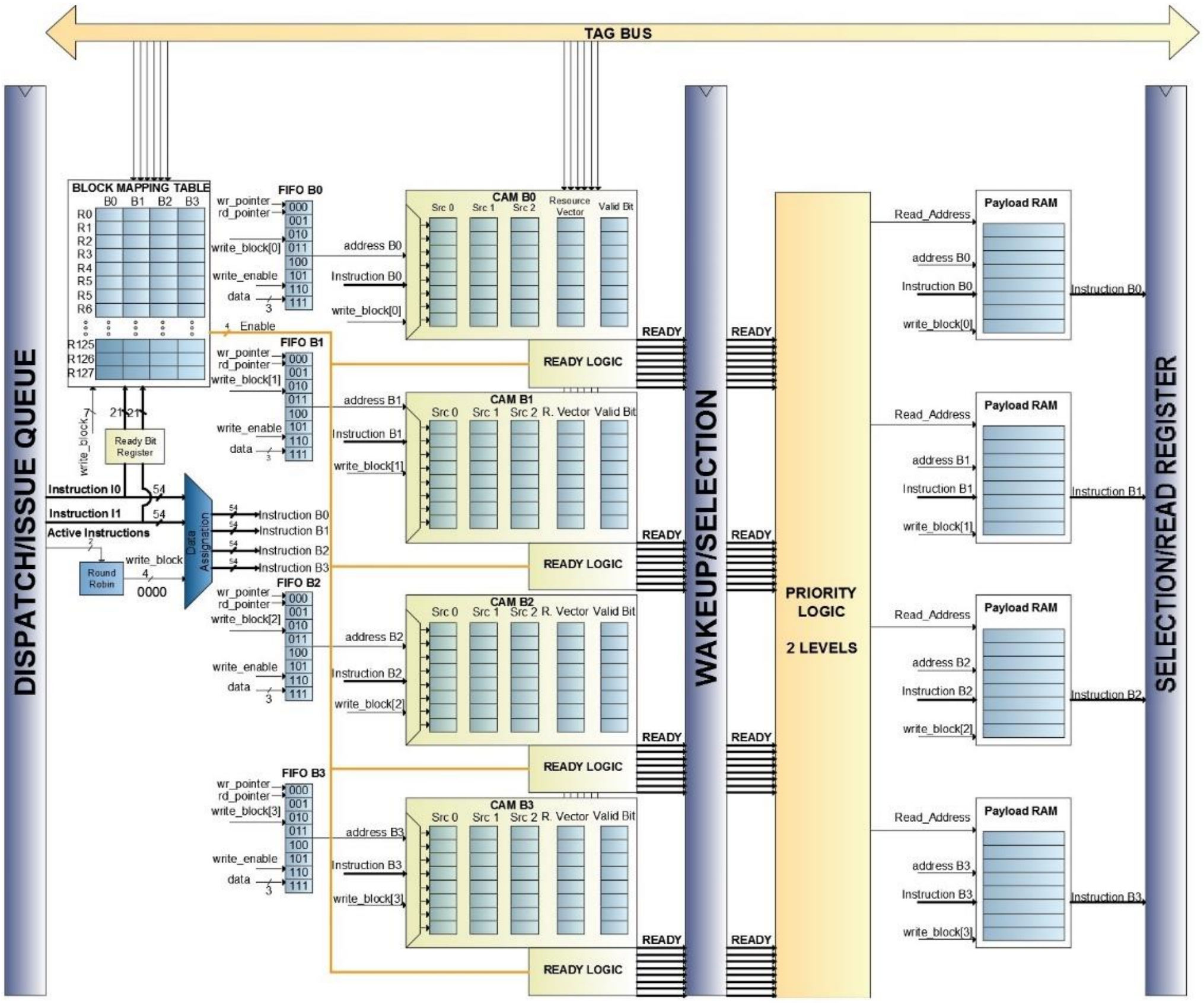

**Fig. 4.13** Complete Low Power Issue Queue Design

In following chapter, is detailed the results of this implementation.

### 4.1.2. Register Bank

**Design considerations**

The number of read and writes ports depend of the number of issue instructions and the number of functional units that have a dedicated write port. For this reason, was defined an issue width of only 2 instruction per clock cycle, in order to reduce the number of read ports in the register bank.

The Register Bank need 6 read and 6 write ports.
The six read ports are:

-SourceI0_0: read the Source 0 of the instruction 0
-SourceI0_1: read the Source 1 of the instruction 0
-SourceI0_2: read the Source 2 of the instruction 0
-SourceI1_0: read the Source 0 of the instruction 1
-SourceI1_1: read the Source 1 of the instruction 1
-Store : read the data to store in memory

Instruction 0 have 3 read ports, it is because Lagarto II processor can execute instructions as Fused Multiply Accumulate (FMAC) which need read 3 source operands, when one FMAC instruction is ready to be issue, is forced to leave for the port 0.

The six write ports are:

-Read_1: to write the result from the Add/sub functional unit.
-Read_2: to write the result from the Mul functional unit.
-Read_3: to write the result from the Div functional unit.
-Read_4: to write the result from the ALU functional unit.
-Read_5: to write the result from the MulAdd functional unit.
-Read_6: to write the result from the load or Move instructions

The implementation of the register file is based in the proposals [23] and [24].

**LVT design**

Basically to implement the proposal presented in [23] , first the replication technique is used in order to obtain a 1W/6R memory as is shown in Figure 4.14.

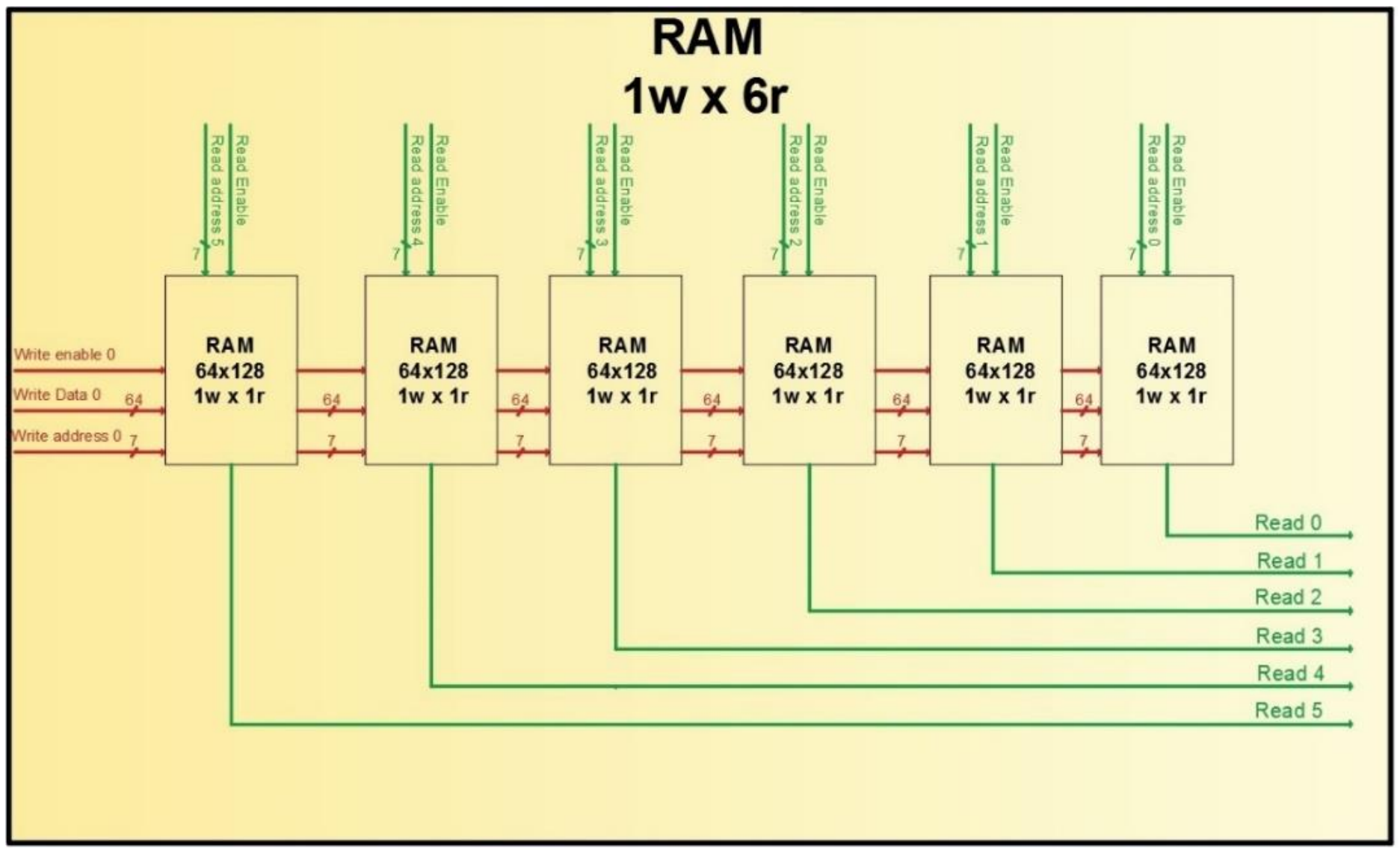

**Fig. 4.14** Replication technique to implement 1W/6R memory.

After that, banking technique is used in order to increase the number of write ports, and using the LVT table which select in the multiplexors the more recent write value to be read. The LVT table will use pure logic elements, but instead of build a memory of 64x128, only build a memory of 3-bits x128 locations with 6W/6R ports; therefore, the total logic elements will be dramatically reduced. The 3-bits are because with 3-bits is possible represent 8 possible combinations and although need 6 combinations.

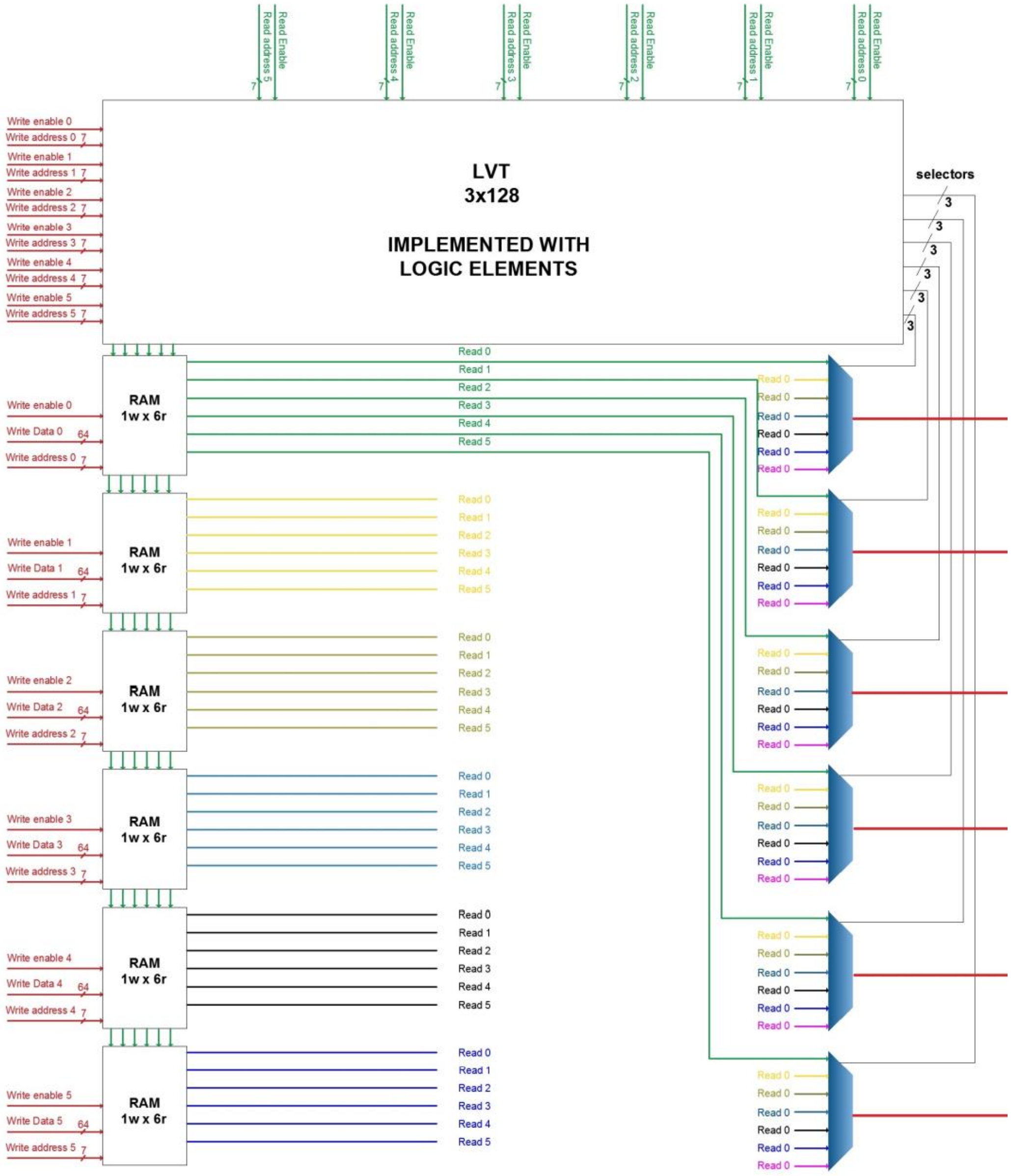

**Fig. 4.15** 6W/6R Memory using LVT Design

Implementing this proposal are used 36 memory blocks of 64-bit x128-locations, and the logic elements is reduced a lot, the exact numbers are presented in the following chapter.

**XOR design**

Furthermore, proposal presented in [24] was implemented for resources evaluation, which is based on the XOR operations.  The design requires m * (m-1+n) RAM Blocks to provide m writes and n reads ports, for the requirements presented before, are needed 6W/6R ports, then are needed 66 memory blocks of 64-bit x 128-locations, almost double that with using the LVT design, but the LVT table and its logic is removed, instead of latter, XOR gates are used. Because one XOR logic is added every new port, the performance of this design depends of the ports number.

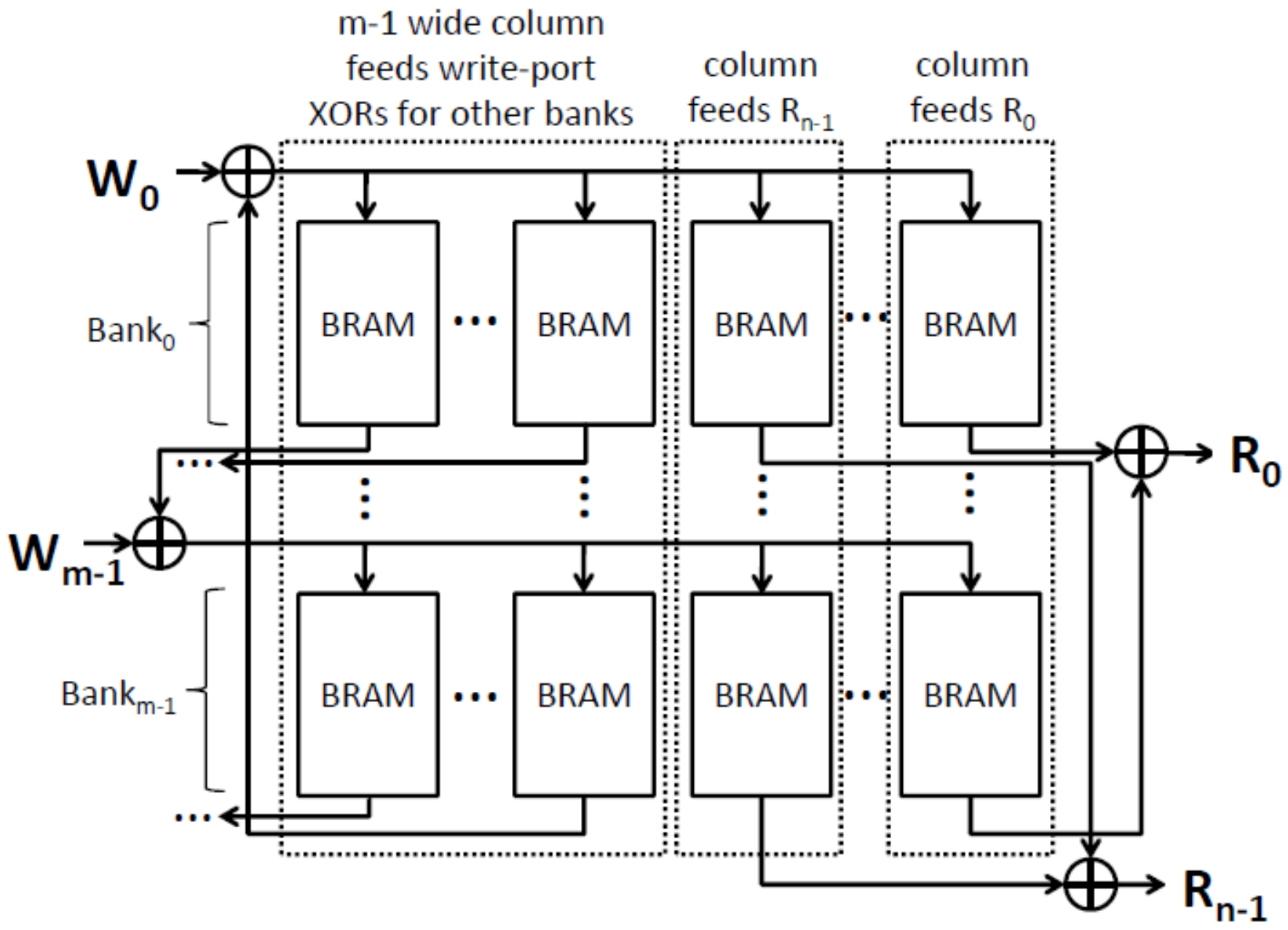

**Fig. 4.16** A generalized mW/nR memory implemented using XOR.

In following chapter are compared both designs and both designs have a good result, depending of the needs of the final design requirements, but both designs reduce a lot of logic elements used compared with multiported registers array implementation.

### 4.1.3. Execution Stage

In this sections are presented several designs of FP functional units based on the fundamental algorithms with some modifications taken account the proposals of the state of the art, which were mentioned before. Designs for the basics operation as Add/subtract, multiply, and Divide/Reciprocal are presented, also the Fused Multiply-Accumulate unit was implemented which first multiply two operands, then accumulate the result and finally add to a third operand to produce the result; Furthermore, an ALU unit is built in order to execute comparisons, moves and others instructions, and finally a branch unit to compute the conditional branch instructions. Last units were performed in order to try to execute all possible instruction set of the MIPS 64 R6 for evaluation purpose.

**FP Add/Subtract Unit**

In contrast to the integer arithmetic units, FP addition and subtraction are more complicated than multiplication and division. As mention in Chapter 3.3.1, three major task are presented in this operation, pre-normalization, addition and post-normalization. The integer adder is a crucial part of the design, but, also is needed a quick Leading Zero Detection which becomes part of the critical path due to is needed known the number of zeros to the left in a word in the shortest possible time and release the result; Furthermore it is possible predict the leading zeros in parallel with the integer addition operation, but this method may be erroneous by one position [27], if this is the case, then it can be fixed by shifting to the right one position. Also other techniques exist to produce an exact result, however these techniques are very expensive in terms of area [28].

The proposed design follows the basic algorithm but adding elements from other proposals [28] [29] and own proposals, also the design will support subnormal numbers which increases the complexity of the design compared with all presented in the state of the art including the IP Cores provided by Altera. This unit will execute two instructions that are shown in following table.

**Table 4.1** Add/subtract instructions

| Instruction | Description |
|---|---|
| ADD.fmt | Floating-Point Add |
| SUB.fmt | Floating-Point Subtract |

Figure 4.17 show adder/subtractor input signals where *Enable* (1 bit) which encodes a valid operation, *Source 1*(64 bits) and *Source 2*(64 bits) are the source operands and *Operation* (1 bit) encodes if the current operation is an addition or subtraction. In the other hand, the output signals are: *Ready* that indicates that current operation is complete, *Result* give the final result of the operation. 4 exceptions are contemplated according to the IEEE 754 standard, which are *Invalid Operation, Overflow, Underflow* and *Inexact*; Also MIPS 64 R6 specify Not *Implemented Operation* exception which for this case not apply.

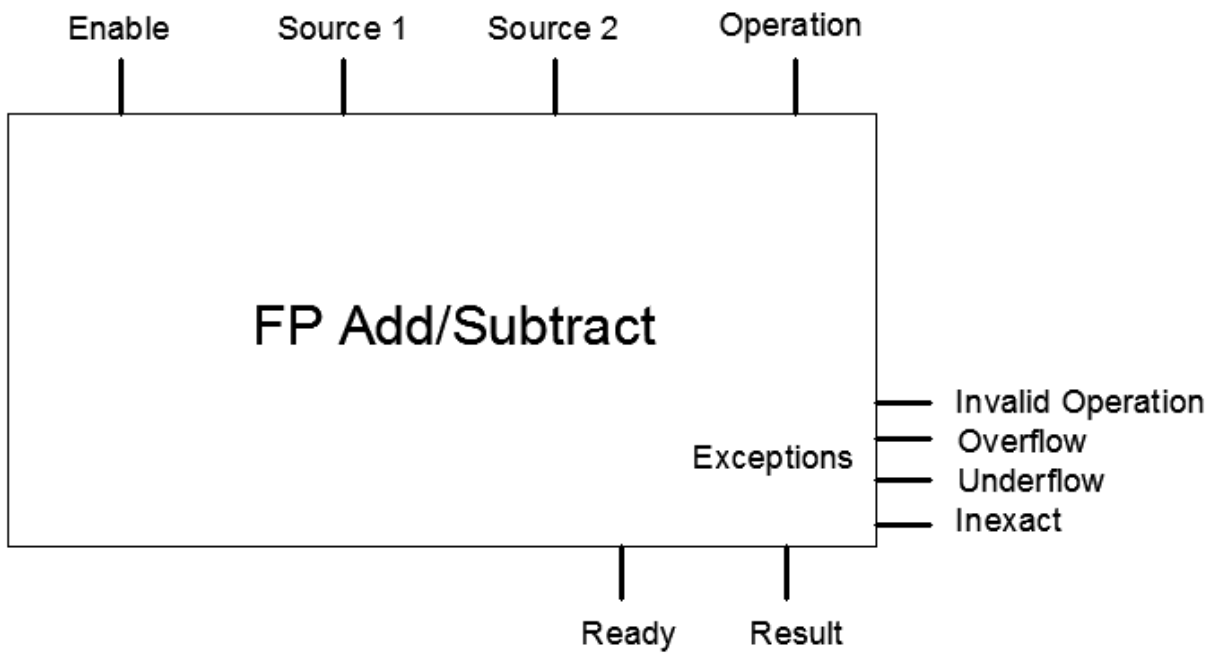

**Fig. 4.17** FP Add/Subtract Inputs/Outputs.

Figure 4.22 show the complete design of FP Add/Subtract Unit, which is divided in 8 stages.

First Stage perform 5 main activities:

- In block *Initial Conditions* is monitored if some source is *Infinity, Zero, QNaN* or *SNaN*, if someone of this condition is true, means that the operation can finalize because the result is known and the operation can skip the following steps.
- Identify the bigger number (absolute value) to in following stage perform a prenormalization. This is doing by obtain the exponents difference between both exponent, if the operation results negative minds that Source 2 is bigger and using the MUX block is possible change the path of this Sources. If result is positive, can be for 2 reasons, *Source 1* is bigger than *Source 2* or both have the same exponent and then is needed check which mantissa is bigger in order to obtain the bigger and smaller numbers.
- Is necessary check if smallest number is subnormal in order to adjust the exponent difference, if smallest number is denormalized, subtract operation is performed to the previous difference calculated (previous difference calculated -1).
- Is necessary build the complete format of the mantissas which is given by 55-bits instead of the original 52-bits. The format is presented in Figure 4.18, basically are 55-bits vector where in the two less significant bits are added two zeros which help to save some digits after pre-normalization and to perform the rounding in future stages; next are concatenate the 52-bits of the mantissa and finally 1 bit for the signal normalized which for normalized numbers is equal to 1, for subnormal numbers is equal to 0.

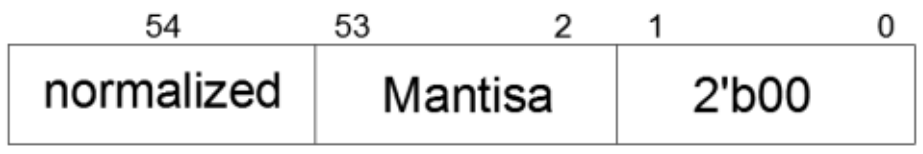

**Fig. 4.18** Complete 55-bits mantissa format

- And finally, in block *Sign* is readjust the operation and the sign bits in order to always obtain a positive result, doing this step, 2-complement after addition is don't needed because ensure that result ever will be positive, also is performed in parallel with the other activities. Figure 4.19 shows the eight combinations, for “000” and “011” cases, don’t perform any change. Basically the final sign is the sign of the biggest number.

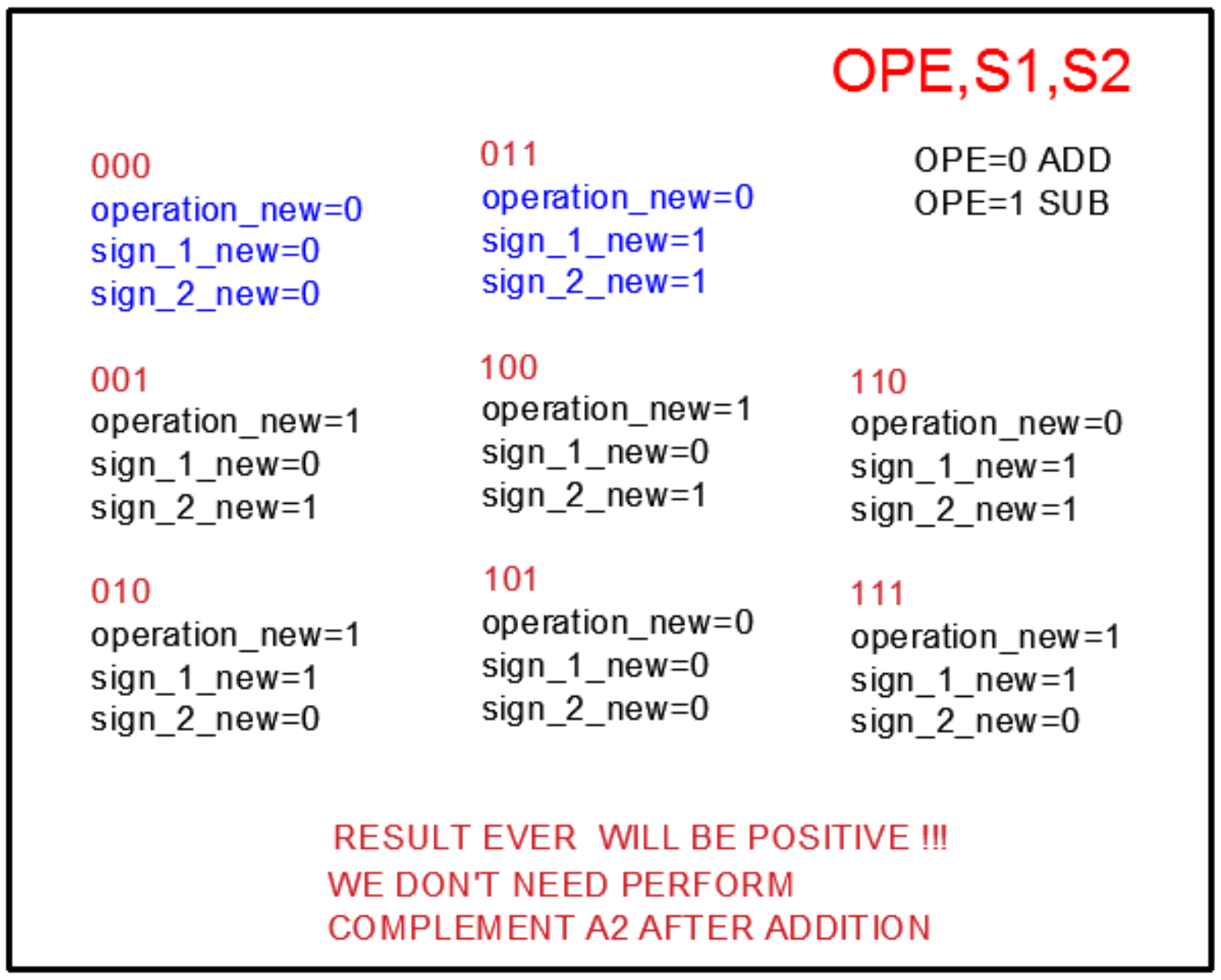

**Fig. 4.19** Readjustment of the Add/Subtract operation

Second stage basically performs 2 activities, pre-normalize and 1-complement if it is necessary:

- Pre-normalize is done by Shift to the right the smaller mantissa using the difference obtained between the two exponents.
- Complement a1 is performed if the final operation re-defined in the last stage is subtraction. Finally, in the following step in Carry_in signal is added “1” in order to complete the 2-complement.

Third stage perform only the addition of modified mantissas, in this block a 55-bits KoggeStone Adder is used in order to perform the addition as faster as possible.
Result is given in a 56-bits vector, where bit 56 indicates if some overflow occurs during the operation, in order to normalize the result in future cycles.

Fourth stage is in charge of count the zeros to the left after addition. Figure 4.20 show an example where is necessary to obtain as faster as possible the number of zeros to the left after a subtraction or can be an addition of two subnormal numbers that need to be shift to the left one position. For this stage in the design the proposal presented in [29] is used, but adapted to 55-bits. Bit 56 as mentioned before, only say that overflow has occurred, then is not taken account to the Leading Zero Counter.

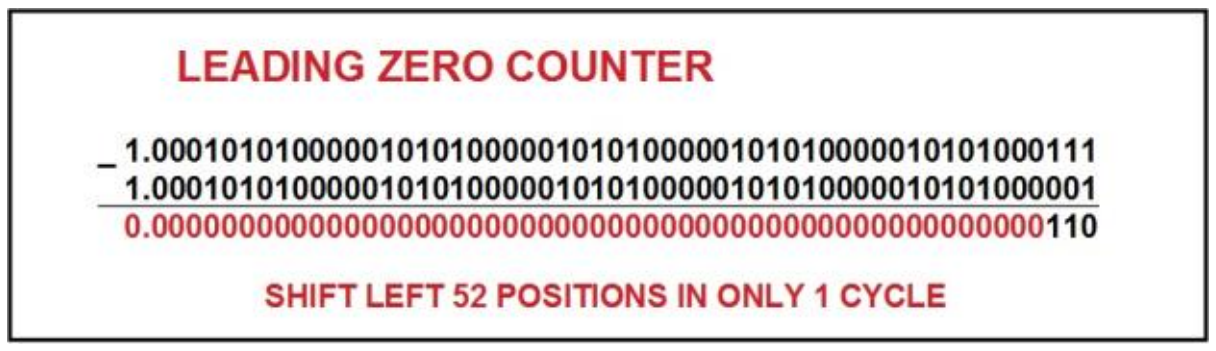

**Fig. 4.20** Example of result that needs shift to the left n number of bits.

Taking into account the previous example and considering that both number have an exponent with the value “1010” binary (not shown in figure 4.20), the Leading Zero Count algorithm only give the number of zeros to the left in order to perform the Normalization (in this case 52 position to the left). However, it can’t shift all this position because when 1 shift to the left is performed, the exponent field decrease by one, then if the current exponent is 10 decimal, only can perform shift to the left 10 positions and decrease the exponent to 0,  then the final result will be:

“0.0000000000000000000000000000000000000001100000000000”

With exponent “0”, then in this stage taking advantage that the Leading Zero Count (LZC) is so faster, one extra operation is performed, it consists of comparing the LZC result with the exponent in order to send the shift final value, and if LZC is less than the Exponent the result is given by the Leading Zero Count in other case by the Exponent.

Fifth stage performs the normalization, basically check next cases: if the operation is a subtraction, a shift to the left some positions is needed. If overflow occur a shift to the right one position is perform. If both numbers are subnormal and a normalized number is obtained as result, then the exponent need increased by one.

The normalized results often need to be rounded because most of them cannot be represented exactly in floating point representation. Next stage is in charge to round the result according to the rounding mode configured in FCRS (FP Control/Status register); In order to support all possible, the *IEEE 754 Standard* the Round module was designed to support the four rounding modes provided by the Standard which are *Round to zero, Round to infinity, Round to minus infinity and Round to nearest.*

Basically with 3 bits all floating-point arithmetic can be rounded as if it was computed with infinite precision. These three bits are called *Guard, Round* and *Sticky* bits, Figure 4.21 show the position of this bits.

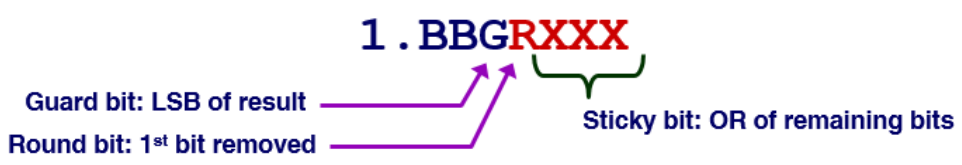

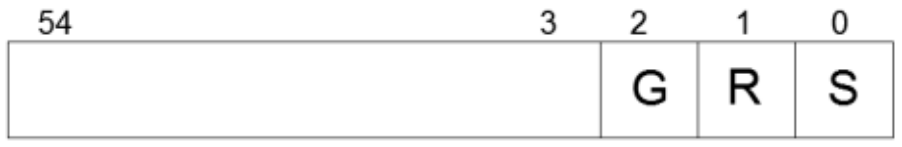

**Fig. 4.21** Guard, Round and Sticky bits

Sticky bit is required to guarantee correct rounding in the final stage of floating point arithmetic. The purpose of sticky bit is to indicate that the unrounded result is inexact.

*Round to Nearest*

Perhaps Round to nearest is the most common used in IEEE floating point arithmetic. In this rounding model all numbers are rounding to the nearest representation. Following is presented the algorithm implemented to Round to Nearest.

**Algorithm IEEE-754 round to nearest**

**Input: G-guard bit, R-round bit, S-sticky bit**

**Output:**
**If G = 0 then**
$Significant_{rounded} = Significant_{normalized}$
**else if ( R = 1 or S=1) then**
$Significant_{rounded} = Significant_{normalized} + 1$
**else**
$Significant_{rounded} = Significant_{normalized}$

*Round Toward Zero*

This is the simplest rounding mode. Basically is a truncation regardless of the state of the guard, round and sticky-bit. Because this mode is so simple, does not require any additional hardware to implement. Some high-speed floating-point units choose to support only this rounding mode as the Cell Processor.

*Round Towards Plus Infinity*

This mode rounds the result to the value closest to but not less than the result.

**Algorithm IEEE-754 round towards plus infinity**

**Input: G-guard bit, R-round bit, S-sticky bit, sign bit**

**Output:**
**If (G = 1 or R = 1 or S=1) and sign bit=0 then**
$Significant_{rounded} = Significant_{normalized} + 1$
**else**
$Significant_{rounded} = Significant_{normalized}\ (truncate)$

*Round Towards Minus Infinity*

This mode rounds the result to the value closest to but not greater than the result.

**Algorithm IEEE-754 round towards minus infinity**

**Input: G-guard bit, R-round bit, S-sticky bit, sign_bit**

**Output:**
**If (G = 1 or R = 1 or S=1) and sign_bit=1 then**
$Significant_{rounded} = Significant_{normalized} + 1$
**else**
$Significant_{rounded} = Significant_{normalized}\ (truncate)$

Seventh stage is in charge to identify if some exception occurs during execution. Floating Point adder/subtractor can produce four exceptions: *Invalid Operation, Overflow, Underflow* and *Inexact Operation.*

Eighth stage is in charge to check if some overflow exception occurs in seventh stage, then taking account the current rounding mode put the correct representation as +infinity, -infinity, largest representable positive number or the largest representable negative number and finally use a MUX in order to select if the final result come from seventh stage or first stage.

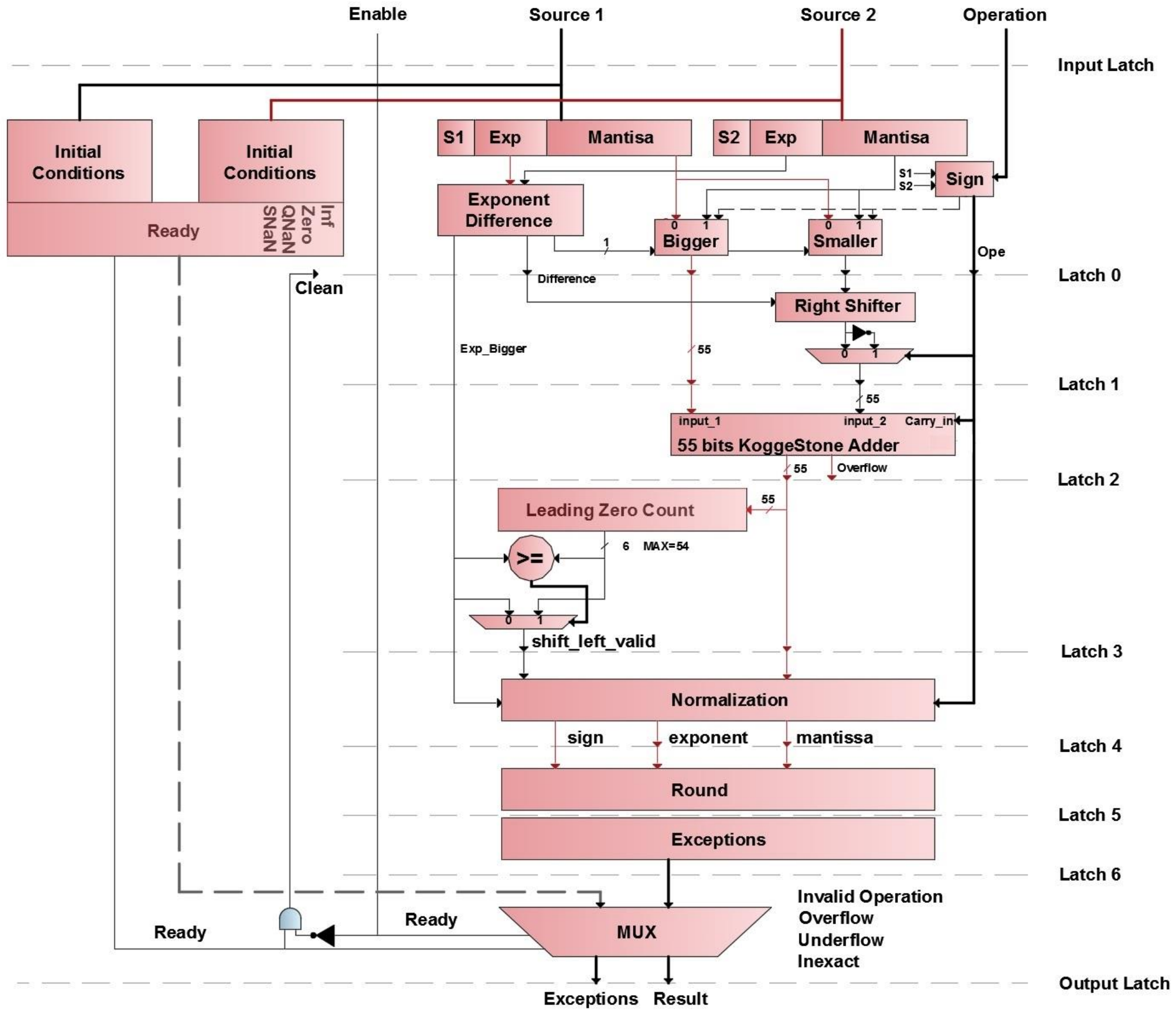

**Fig. 4.22** Design of FP Add/Subtract Unit

**FP Multiply unit**

Floating Point multiplication is a core operation in many kernels application, and efficient implementation of floating point multipliers is important concern.

Multiplying two numbers in floating point format is perform in three main steps:

- Add the exponent of the two numbers and then subtracting the bias from their result.
- Multiply the significant of the two numbers.
- Calculate the sign by XOR operation of the two signs of the two numbers.

To multiply 2 numbers in double precision format, require implementation of 53x53 multipliers in hardware, which is very area and power expensive.

This unit will execute only one instruction, which is shown in following table.

**Table 4.2** Multiply instruction

| Instruction | Description |
|---|---|
| MUL.fmt | Floating-Point Multiply |

Following image show Multiplier input signals where: Enable (1 bit) say that this is a valid operation, Source 1(64 bits) and Source 2(64 bits) are the operands. In the other hand, the output signals are Ready that indicates that the operation is complete. Result give the final result of the multiply and can to expose 4 exceptions according to the IEEE 754 standard which are: Invalid Operation, Overflow, Underflow and Inexact; Also MIPS 64 R6 specify Not Implemented Operation exception which for this case not apply.

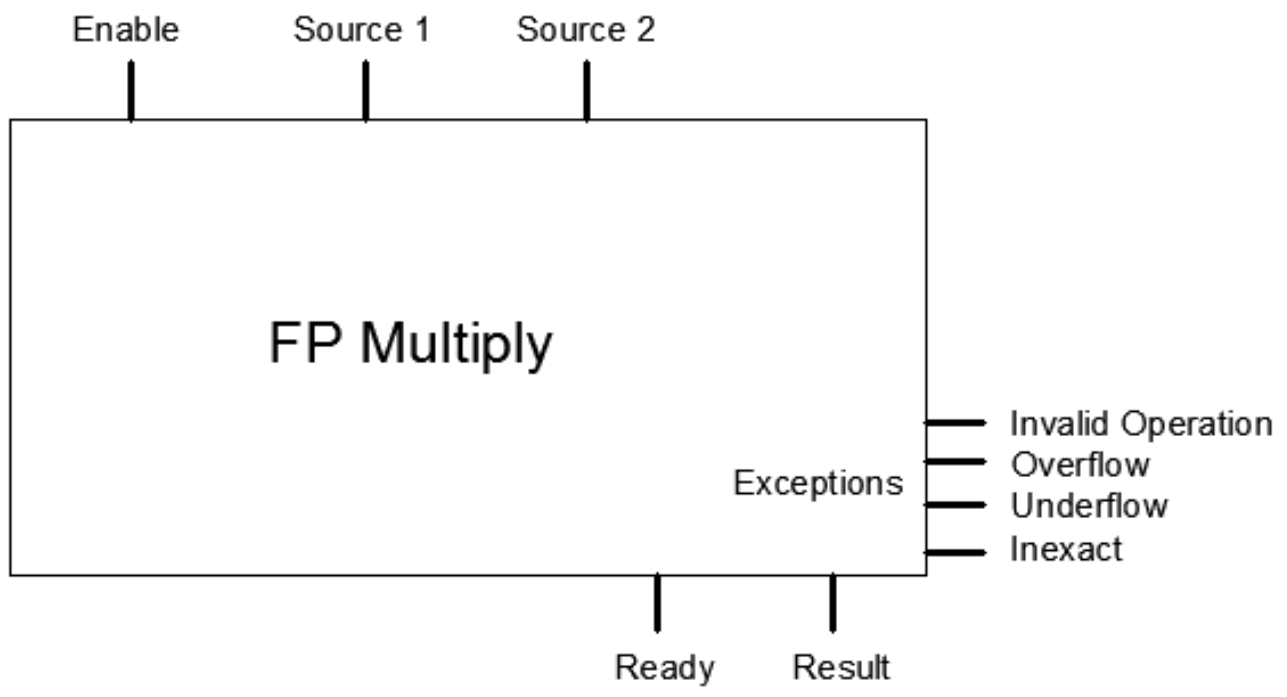

**Fig. 4.23** FP Multiply Inputs/Outputs.

Figure 4.24 shows the complete design of 7 stages FP multiply and in following lines are described each stage.

First stage basically performs 4 main activities:

- In block *Initial Conditions* is monitored if some source is Infinity, Zero, QNaN, SNaN, both subnormal or overflow (obtaining the final exponent), if someone of this condition is true means that the operation can finalize because the result is known and the following steps can be skip. This module accepts subnormal numbers, but if both are subnormal means that the result will be Zero.
- Add the exponent of the two numbers and then subtracting the bias from their result.
- Calculate the sign by XOR operation of the two signs of the two numbers.
- And Start the Multiplication of the mantissas.

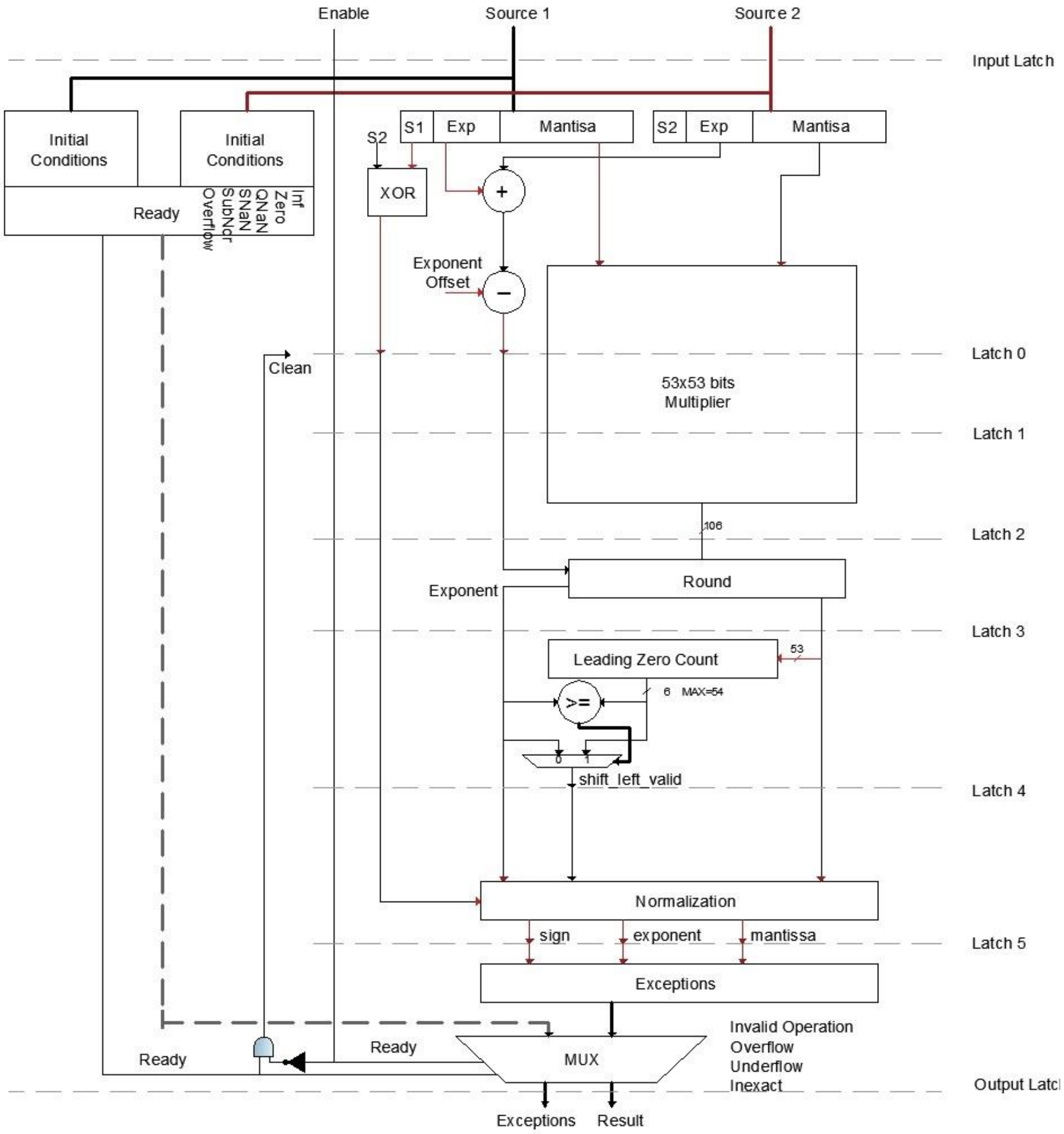

**Fig. 4.24** Design of FP multiply unit

Figure 4.24 shown the complete design, 53x53 Multiplier takes 3 cycles, the design of this multiplier is based in the proposal of Manish Kumar [30], where the proposal is perform the multiplication using small size multipliers, also in Cyclone IV Device Handbook [38] specify that for Altera FPGA Cyclone IV which is the FPGA on which the implementation took place, provide of dedicated 9x9 or 18x18 bits multipliers configurations and also propose the design presented in Figure 4.27 which is basically the same idea presented in [30].

This idea is presented in Figure 4.25, where to multiply two numbers of 54 bits each one, the mantissa is divided in three small blocks of 18 bits each one, just the size of multipliers provided by Altera Cyclone IV FPGA.

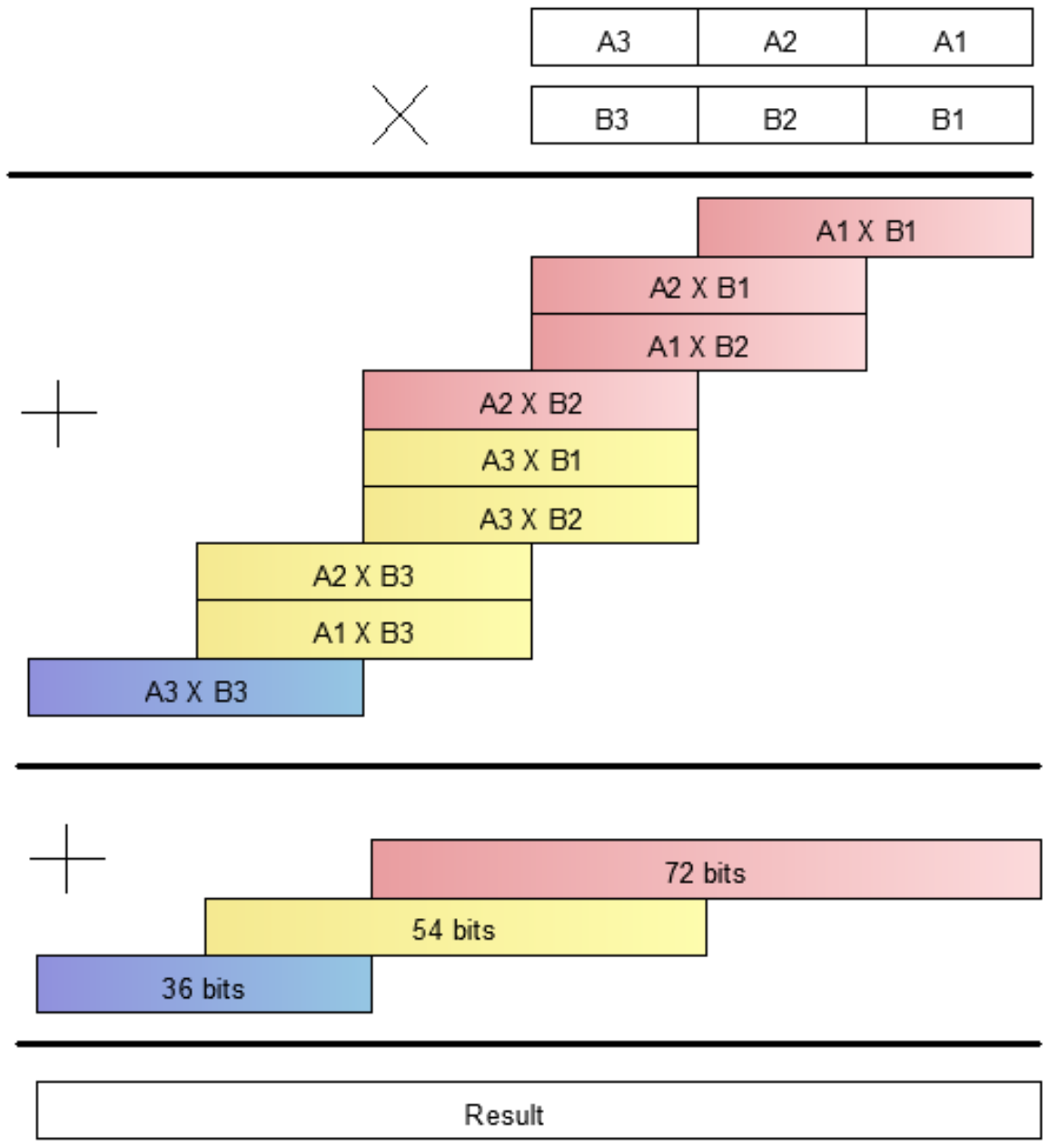

**Fig. 4.25** Multiplication using small size multipliers

Basically if in one stage are performed 9 18x18-bits multiplications in parallel and perform the addition in pairs of results, in a second stage the addition of the previous results in pairs is performed, the 36x36-bits multiplier could be designed with two stages as is show in Figure 4.26.

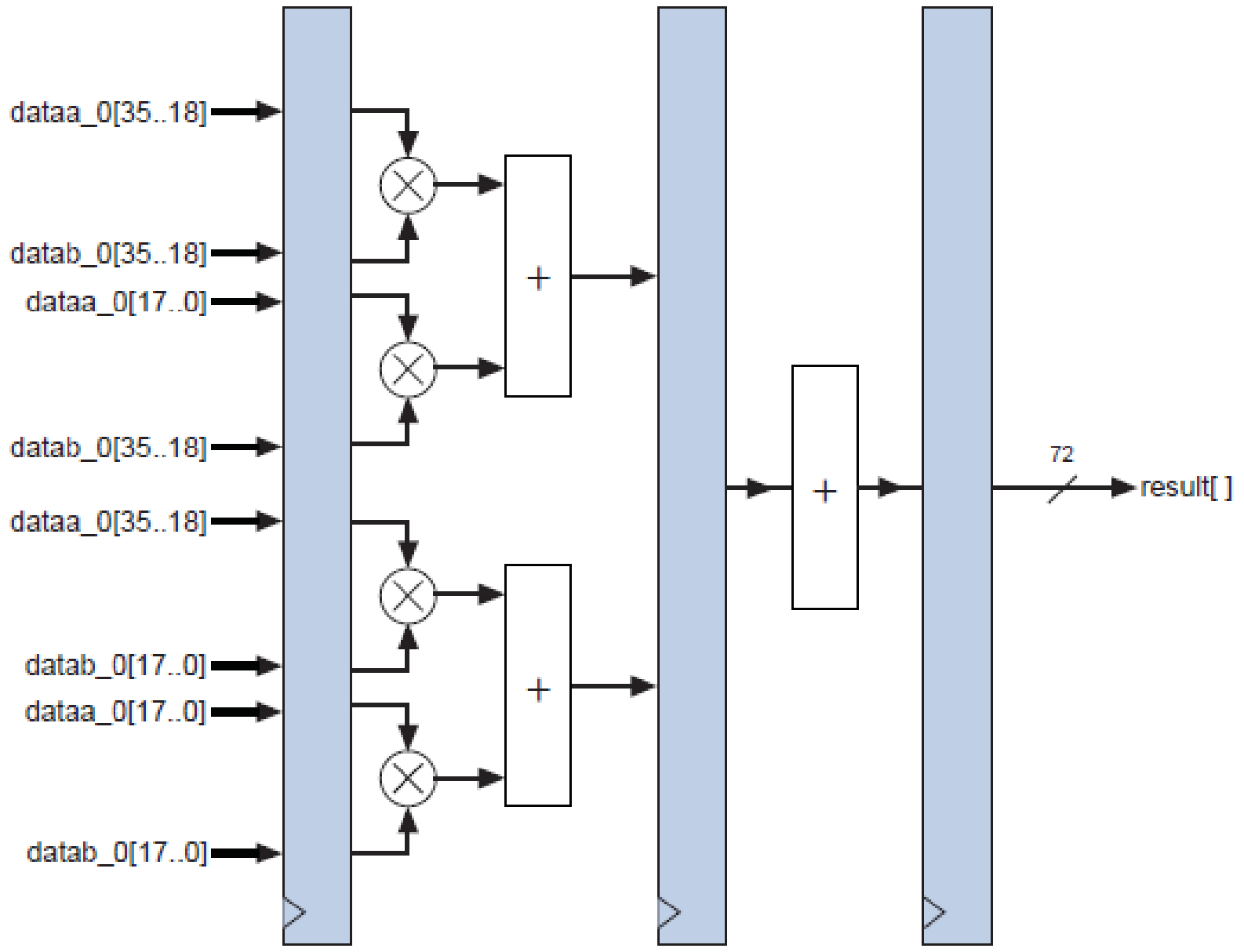

**Fig. 4.26** Two stage 36x36 bits multiplier

Third stage performs the final addition to obtain a 108-bit result. Following stage perform a rounding to reduce the result to 53 bits check. Next stage checks for subnormal case, if the Leading Zero Count give a value bigger than the exponent, signalize to the next stage to start the normalization, to only perform a shift of the value of the exponent, obtaining a subnormal number.

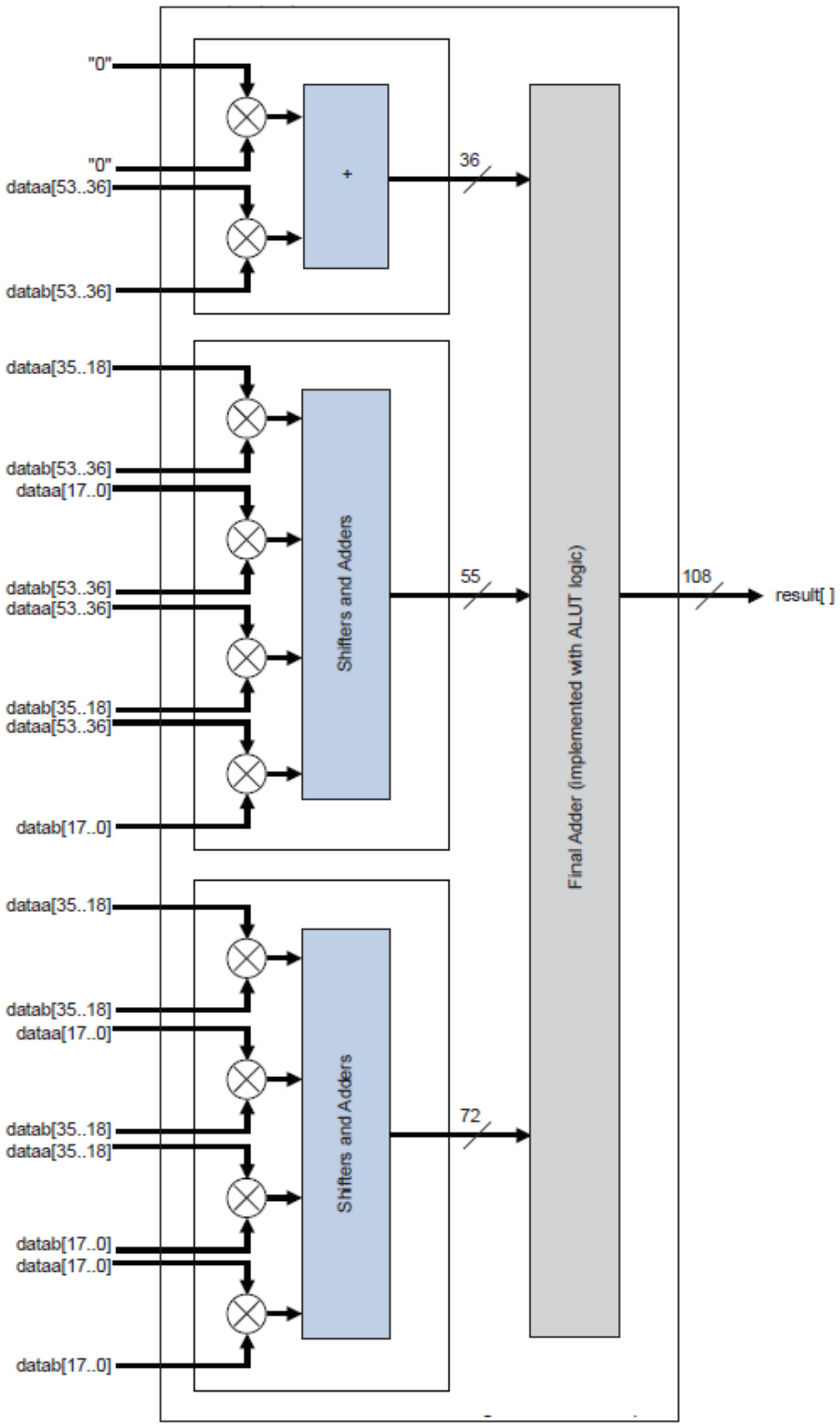

**Fig. 4.27** Three stages 54x54 bits multiplier

Five stage is in charge to normalize the multiplier result, if someone of the operands was subnormal, then is needed perform a shift to the left according the result of Leading Zero counter of the previous step, if after rounding stage detect an overflow then perform one shift to the right and increase the exponent by 1.

```
NORMALIZATION
SUBNORMAL INPUT
        MANTISA_OUT = MANTISA << shift_left_valid
        EXPONENT_OUT =  EXP_BIGGER - shift_left_valid
OVERFLOW (BIT 106)
        MANTISA_OUT = MANTISA >> 1
        EXPONENT_OUT =  EXP_BIGGER + 1
ANY CASE
        MANTISA_OUT = MANTISA
        EXPONENT_OUT =  EXP_BIGGER
```

**Fig. 4.28** Normalization

Exceptions are similar to the module presented in PF Adder/Subtractor.

**FP Divide unit**

Floating Point Divide was implemented using a Reciprocal unit presented in [32] and after that performing a multiplication using the previous FP Multiply design.

This unit will execute two instructions, which are shown in following table.

**Table 4.3** Multiply instruction

| Instruction | Description |
|---|---|
| DIV.fmt | Floating-Point Divide |
| RECIP.fmt | Floating-Point Reciprocal Approximation |

Following image show Divider/Reciprocal input signals where: Enable (1 bit) say that this is a valid operation, Source 1(64 bits) and Source 2(64 bits) are the operands. In the other hand, the output signals are: Ready that indicates that some operation is complete, Signal *Result* give the final result of the operation, and implements 5 exceptions according to the IEEE 754 standard which *are Invalid Operation, Overflow, Underflow, Inexact* and *Division by Zero*; Also MIPS 64 R6 specified "Not Implemented Operation" exception which for this case is not apply.

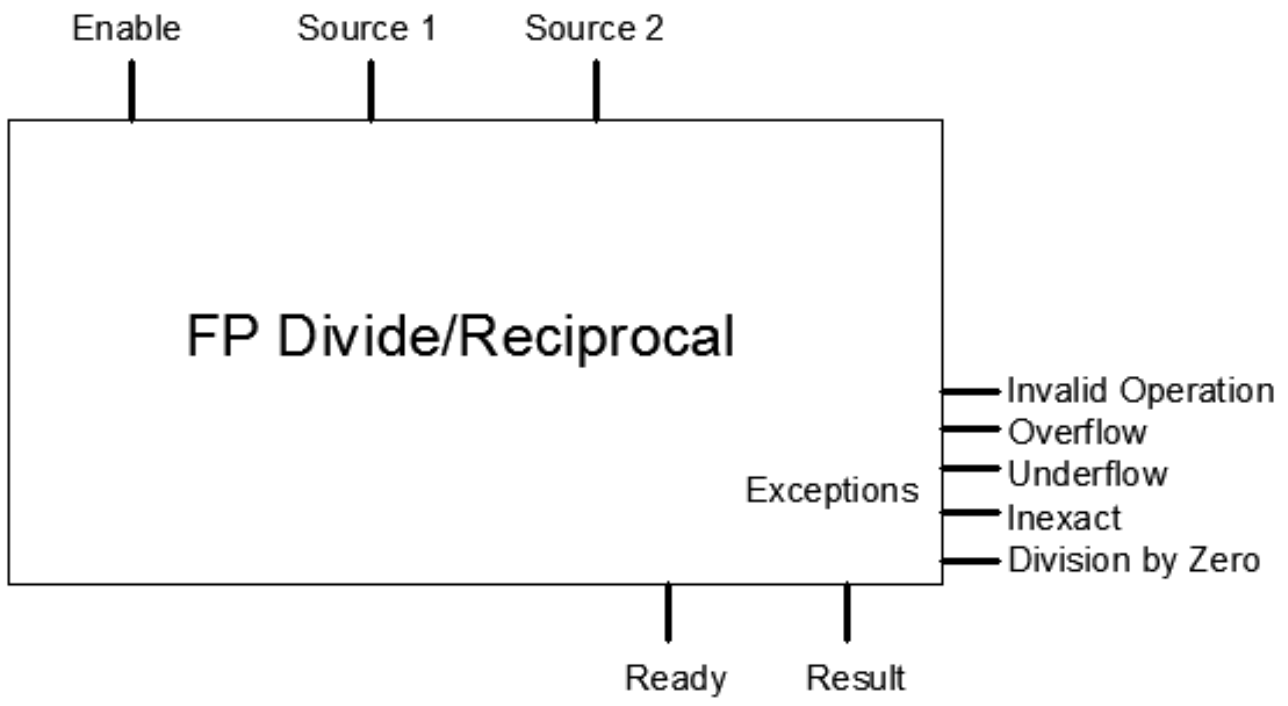

**Fig. 4.29** FP Divide/Reciprocal Inputs/Outputs.

The first step performs the initial approximation of the reciprocal which is obtained reading a value of 16 bits from the look-up table using the 7 most significant bits of the mantissa without the leading 1, and after that a multiplication between the read value and the 15 most significant bits modified previously. Next, the design is described in detail, including the design of the look-up table.

First approximation is based on Taylor series expansion taking until the first derivative term, as is presented in formula 1.

$$f(x_i + 1) = f(x_i) + f'(x_i)f(x_{i+1} - x_i) \quad (1)$$

And the 53-bits mantissa is represented as:

$$X_{mantisa} = [1.x_1x_2x_3 \dots x_{52}] \quad (2)$$

To represent $X^{-1}$ by Taylor series expansion, operand $X$ can be split into two parts as in formula 3 and 4.

$$X_{m1} = [1.x_1x_2x_3 \dots x_m] \quad (3)$$
$$X_{m2} = [0.x_{m+1}x_{m+2}x_{m+3} \dots x_{52}]x2^{-m} \quad (4)$$
$$X_{mantisa} = X_{m1} + X_{m2} \quad (5)$$

The initial reciprocal approximation $X^{-1}$ is computed by following equation:

$$X^{-1} = (X_{m1} + 2^{-m-1})^{-1} - (X_{m1} + 2^{-m-1})^{-2}(X_{m2} - 2^{-m-1}) \quad (6)$$

And can be rewriting as:

$$X^{-1} = (X_{m1} + 2^{-m-1})^{-2}[(X_{m1} + 2^{-m-1}) - (X_{m2} - 2^{-m-1})] \quad (7)$$

Where the first term $(X_{m1} + 2^{-m-1})^{-2}$ is read from ROM and the remaining term $(X_{m1} + 2^{-m-1}) - (X_{m2} - 2^{-m-1})$ will be formed with the operand modifier module. Basically the operand modifier module performs an inversion of the bits from $(m + 1)^{th}\ to\ 2m^{th}$ bits.

In order to obtain a ROM with reasonable size, in [32] perform many test with different values of m, first they do a test with m=6 because theoretically 52-bit accuracy can be achieved with only 2 iterations, but in their simulations more than half of the result not achieve this accuracy, then performed more proves with m = 7, 8 and 9 and results presented shown that with this three values the results are very similar, and the accuracy was better. Finally m=7 then needs a ROM of $2^7x16\ bits$. Table 4.3 shows some of the first and last locations of the ROM.

**Table 4.4** Some calculated values for the ROM memory using the method above explained.

| Address bits (7bits) | ROM values (16 bits) | Address bits (7bits) | ROM values (16 bits) |
|---|---|---|---|
| 0000000 | 1111111000000010 | 0010000 | 1100100011011111 |
| 0000001 | 1111101000011010 | 0010001 | 1100011000011111 |
| 0000010 | 1111011001001001 | 0010010 | 1100001101101101 |
| 0000011 | 1111001010001101 | … | … |
| 0000100 | 1110111011101000 | 1110100 | 0100000001000000 |
| 0000101 | 1110101101010111 | 1110101 | 0100010110010111 |
| 0000110 | 1110011111011010 | 1110110 | 0100010100000111 |
| 0000111 | 1110010001110001 | 1110111 | 0100010001111000 |
| 0001000 | 1110000100011100 | 1111000 | 0100001111101011 |
| 0001001 | 1101110111011000 | 1111001 | 0100001101100000 |
| 0001010 | 1101101010100111 | 1111010 | 0100001011010111 |
| 0001011 | 1101011110001000 | 1111011 | 0100001001001111 |
| 0001100 | 1101010001111001 | 1111100 | 0100000111001001 |
| 0001101 | 1101000101111011 | 1111101 | 0100000101000100 |
| 0001110 | 1100111010001101 | 1111110 | 0100000011000001 |
| 0001111 | 1100101110101111 | 1111111 | 0100011000101001 |

Operand modifier will read $1.x_1x_2x_3x_4x_5x_6x_7x_8x_9x_{10}x_{11}x_{12}x_{13}x_{14}$ and perform the inversion of the bits from $x_8$ $to$ $x_{14}$.

In the second stage, the result of operand modifier and the value obtained from the ROM will be multiplied in order to obtain an initial approximation. The result is truncated to 16 bits and concatenated by 13 bits zeros.

Following steps are in charge to perform two Newton-Raphson iterations. Newton-Raphson is a sophisticated method that needs less number of iterations to reach convergence than other iteration methods as Gauss-Seidel, which is one of the common iterative methods. Newton-Raphson takes less computation time.

Iterations will be given by:

$$x_{i+1} = x_0 - \frac{f(x_0)}{f'(x_0)} \tag{8}$$

Where $x_0$the initial approximation is calculated cycles before and

$$f(x_0) = \frac{1}{x} - X \tag{9}$$

$$f'(x_0) = -\frac{1}{x^2} \tag{10}$$

Substituting the equations (9) and (10) into equation (11) is obtained:

$$x_{i+1} = x_i(2 - Xx_i) = 2x_i - Xx_i^2 \tag{11}$$

Is possible to implement the first term $2x_i$ with only a left shifter of one position (one shift to the left is equal that multiply by two). Second $Xx_i^2$ term is implemented using a squarer to obtain $x_i^2$ and a 53x53 multiplier to obtain the product which must be rounded to 53 bits, and to finalize the iteration perform a subtraction of the two terms. This process will be doing 2 times (2 iterations to obtain a 52-bits of accuracy). Design is presented in Figure 4.30.

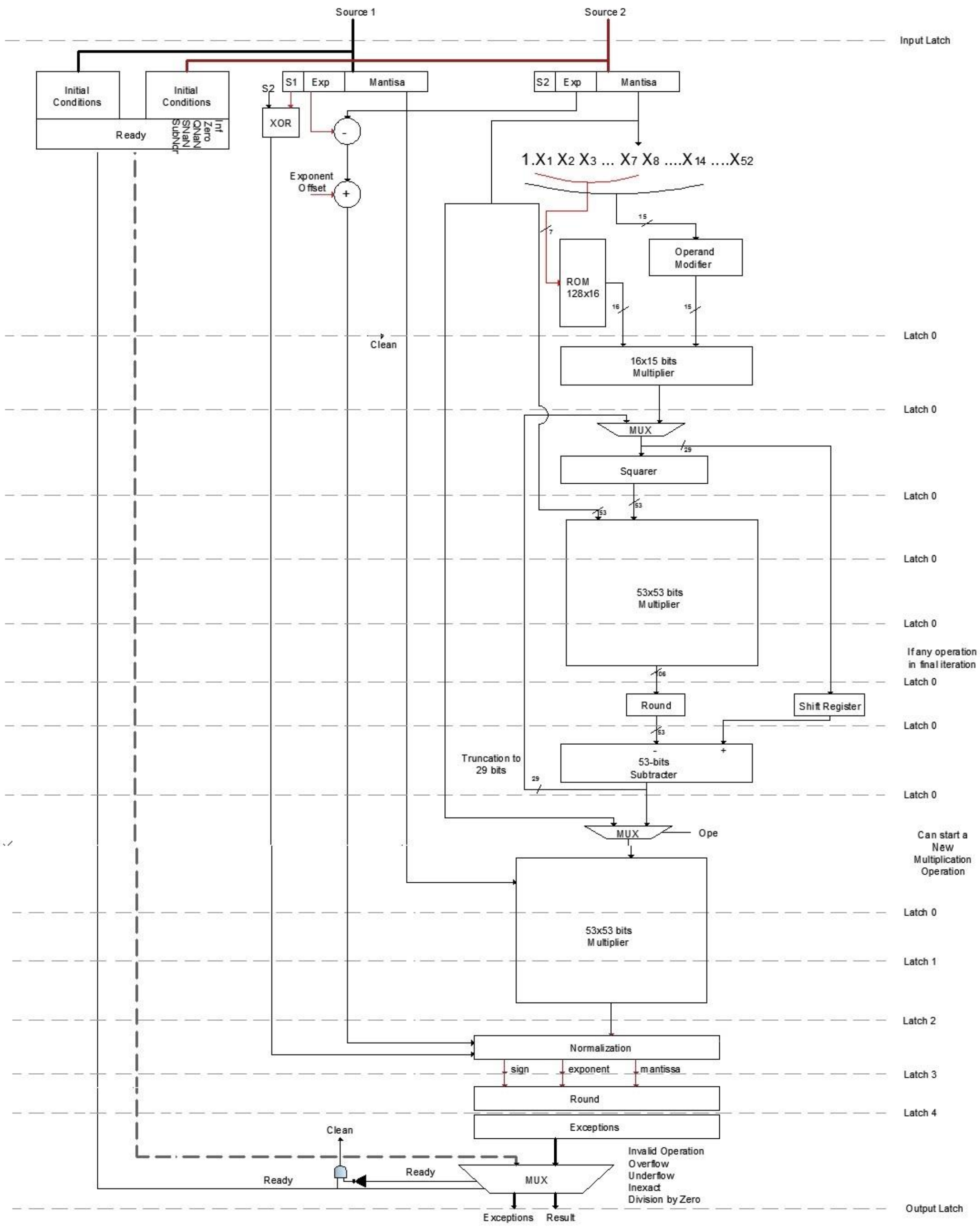

**Fig. 4.30** Design of FP Divider/Reciprocal unit

**FP Fused Multiply Accumulate Unit**

Many floating-point units can actually be thought of as collection of datapaths, one for each operation. In some modern microprocessors designs the FMAC unit entirely replaces all independent execution units. Good examples are presented in Itanium Processors [34] and more recent in AMD bulldozer microarchitecture [35].

**Table 4.5** Fused Multiply Accumulate Instructions and compatibles with this unit

| Instruction | Description |
|---|---|
| MADDF.fmt | Fused Floating Point Multiply Add |
| MSUBF.fmt | Fused Floating Point Multiply Subtract |
| ADD.fmt | Floating Point Add |
| SUB.fmt | Floating Point Subtract |
| MUL.fmt | Floating Point Multiply |

Following image show the Fused Multiply Accumulate input signals where Enable (1 bit) which encode if it is a valid operation, *Source 1(64 bits), Source 2(64 bits) and Source 3(64 bits)* are the operands and *Operation (3 bits)* encode the type of operation that will be execute for the functional unit. This unit can execute 5 instructions (Table4.5). In the other hand, the output signals are Ready that indicates that some operation is complete, Signal *Result* give the final result of the operation, and give 4 exceptions according to the IEEE 754 standard which are *Invalid Operation, Overflow, Underflow* and *Inexact*; Also MIPS 64 R6 specify Not Implemented Operation exception which for this case is not apply.

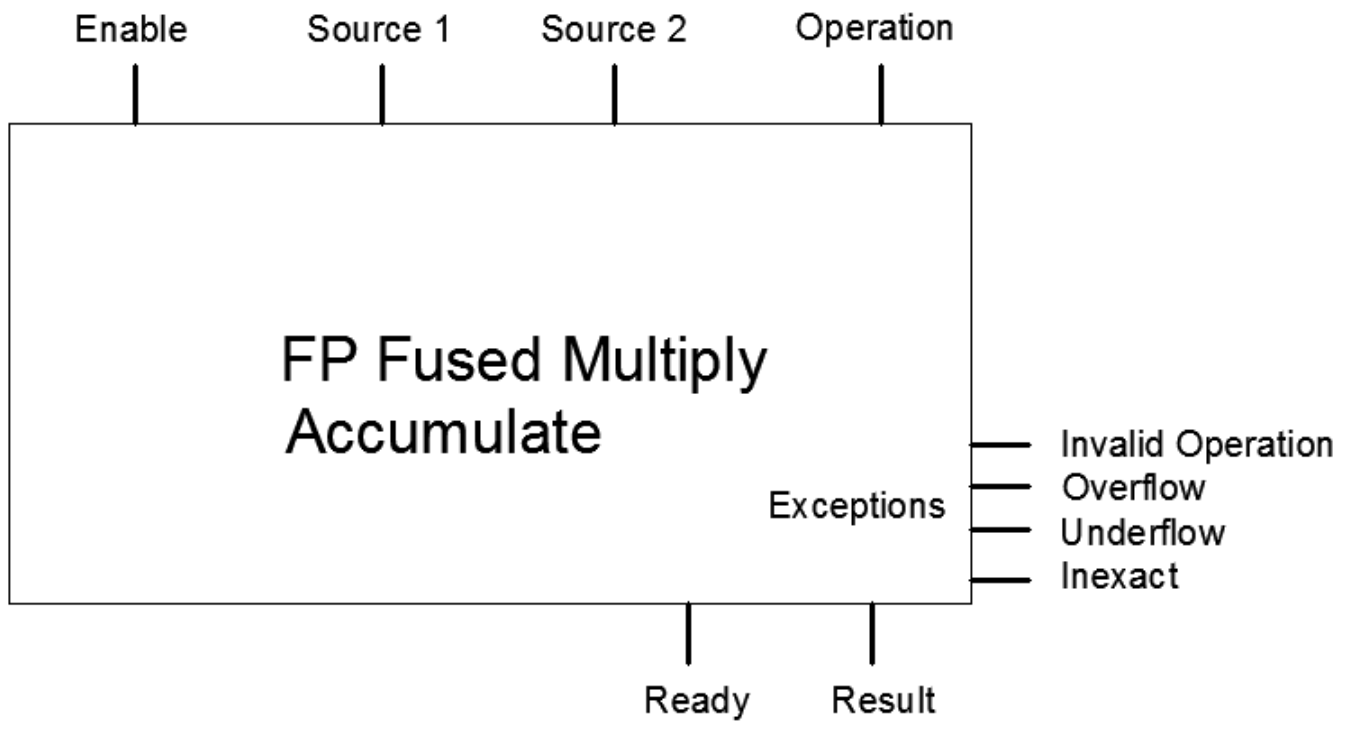

**Fig. 4.31** FP Divide/Reciprocal Inputs/Outputs.

Fused Multiply Accumulate performs a Multiplication of A (Source1) and B (Source2) and the result if added to C (Source3) as is shown in Figure 4.32. Also not only offers improved performance, the precision also increases due to the elimination of a rounding operation after the first operation (multiply).

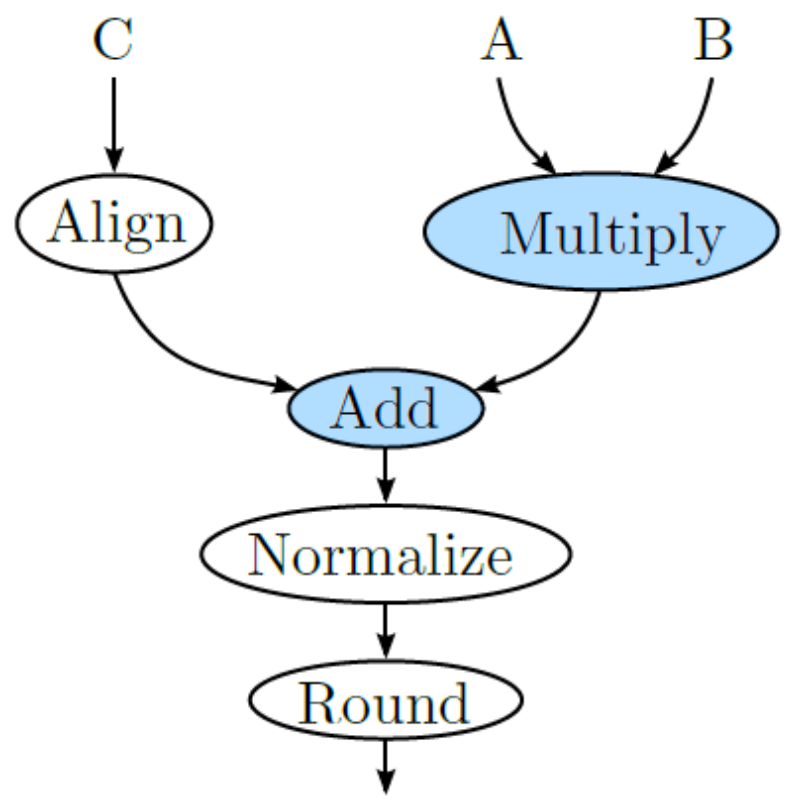

**Fig. 4.32** Fused Floating Point Multiply-Add

Figure 4.33 shown the complete design of the Fused Multiply Add unit, basically take the previous individual designs of ADD/SUB and MUL and join with some little modifications as the elimination of the round after multiplication, and sticky bit take an important role. The most significant 53 bits plus Round- bit and Sticky- bit are kept, also sticky bit perform a OR operation between the ten most significant reminder bits in order to give to the addition operation a more precise value, not only the 53 bits as in traditional operation.
Fused Multiply Accumulate takes only 13 cycles to perform the operations instead of perform first the multiplication with latency of 7 cycles and after an addition with latency of 8 cycles with a latency total of 15 cycles in the best case, because when multiplication send the tag to the FP Queue to perform wakeup of the next operation (ADD), other instruction can take this cycle to be issue to execute delaying the accumulation.

Furthermore, other important feature of this unit, is that was designed to perform individual operations as addition, subtraction or multiplication.

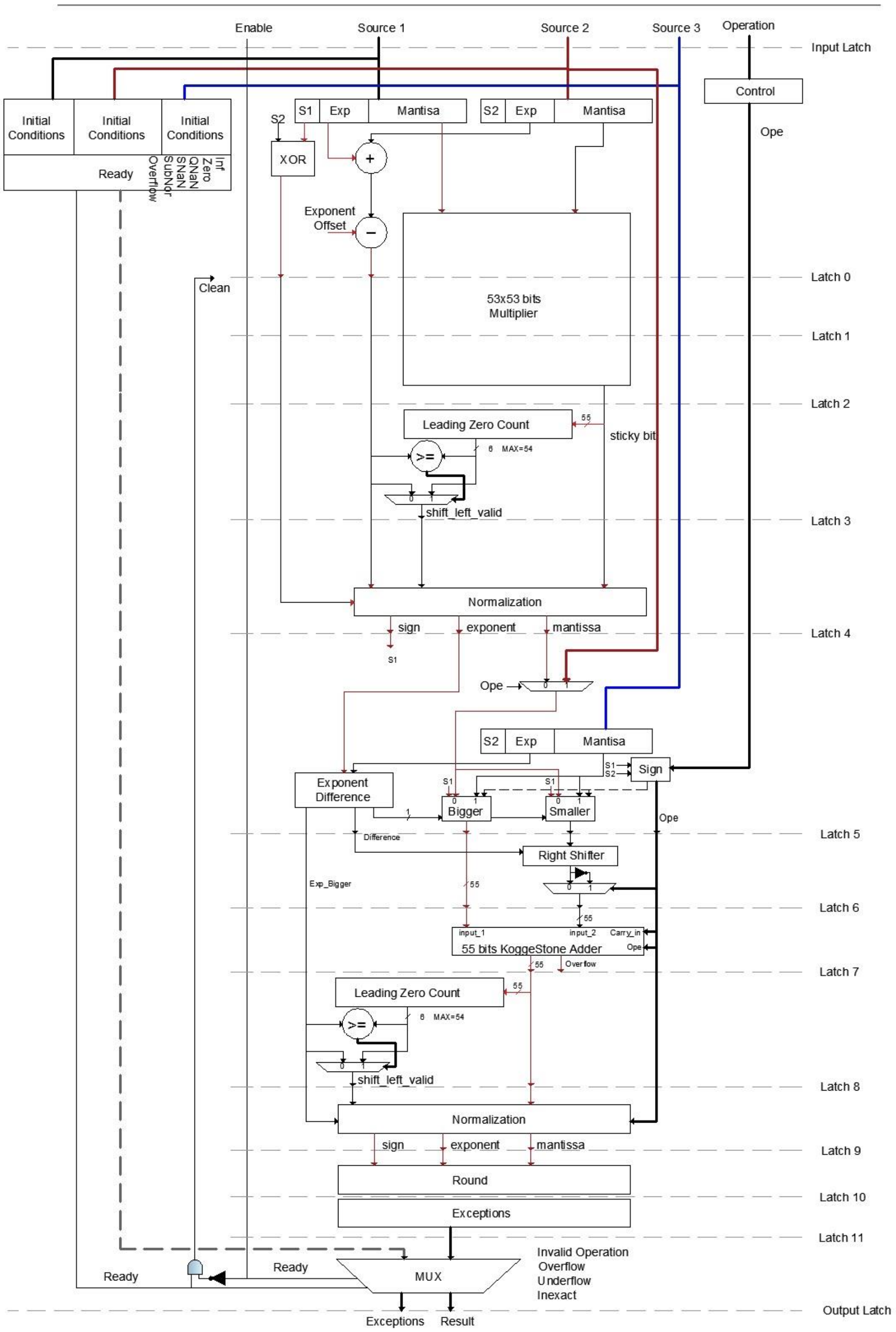

**Fig. 4.33** Design of Fused Multiply Add Unit

**FP Comparison unit**

In order to try to support all possible of the MIPS64 R6 instruction set for the execution of benchmarks for testing and verification, were implemented all FPU Comparison instructions, all FP Formatted Unconditional Operand Move Instructions and all FP Branch Instructions (40 instructions). Basically the implementation of these instructions is simpler than arithmetic instructions and all instruction are executed in one cycle.

**Table 4.6** FP Comparison Instructions

| Mnemonic | Instruction |
|---|---|
| MAX.fmt | Floating Point Maximum |
| MAXA.fmt | Floating Point Value with Maximum Absolute Value |
| MIN.fmt | Floating Point Minimum |
| MINA.fmt | Floating Point Value with Minimum Absolute Value |
| CLASS.fmt | Scalar Floating-Point Class Mask |
| CMP.cond.fmt | Floating Point Compare |

**Table 4.7** FP CMP.cond.fmt instructions

| | |
|---|---|
| AF | False Always False |
| UN | Unordered |
| EQ | Equal |
| UEQ | Unordered or Equal |
| LT | Ordered Less Than |
| ULT | Unordered or Less Than |
| LE | Ordered  Less than or Equal |
| ULE | Unordered or Less Than or Equal |
| SAF | Signalling always False |
| SUN | Signalling Unordered |
| SEQ | Ordered Signalling Equal |
| SUEQ | Signalling unordered orEqual |
| SLT | Ordered Signalling Less Than |
| SULT | Unordered or Less Than |
| SLE | Ordered Signalling Less Than or Equal |
| SULE | Signalling Unordered or Less Than or Equal |
| AF | False Always False |
| UN | Unordered |
| EQ | Equal |
| UEQ | Unordered or Equal |
| LT | Ordered Less Than |
| ULT | Unordered or Less Than |
| LE | Ordered  Less than or Equal |
| ULE | Unordered or Less Than or Equal |
| SAF | Signalling always False |
| SUN | Signalling Unordered |
| SEQ | Ordered Signalling Equal |
| SUEQ | Signalling unordered orEqual |
| SLT | Ordered Signalling Less Than |
| SULT | Unordered or Less Than |
| SLE | Ordered Signalling Less Than or Equal |
| SULE | Signalling Unordered or Less Than or Equal |

**Table 4.8 FPU Formatted Unconditional Operand Move Instructions**

| **Mnemonic** | **Instruction** |
|---|---|
| ABS.fmt | Floating-Point Absolute Value |
| NEG.fmt | Floating-Point Negate |
| MOV.fmt | Floating-Point Move |

**Table 4.9 FP Branch Instructions**

| **Mnemonic** | **Instruction** |
|---|---|
| BC1EQZ | Branch on FP condition Equal to Zero |
| BC1NEZ | Branch on FP condition Not Equal to Zero |

### 4.1.4. Bypass design

An important aspect to improve the IPC performance metric is to include the bypass network. Most processors today implement some form of bypass. A simple bypass network is used in our design to interconnect all output of our Floating Point Functional Units to all inputs in order to forward the values to improve the performance of the processor.

Figure 4.34 shown the Bypass Network. Three cycles before each functional unit obtains its result, the tag address is broadcast to the tag bus in order to notify to the wakeup mechanism that this value will be ready in the next 3 cycles. Then, in the best case, instruction perform wakeup in 1 cycle, is selected to be issue in other cycle, after that instruction read the operands from the register file (maybe read an erroneous value), and finally in the following cycle the correct value is present in the bypass network and is selected with the multiplexors that have each input of each functional unit. This previous steps increase significantly the performance in the current processors. Is possible anticipate 3 cycles for the FP execution engine using the bypass network.

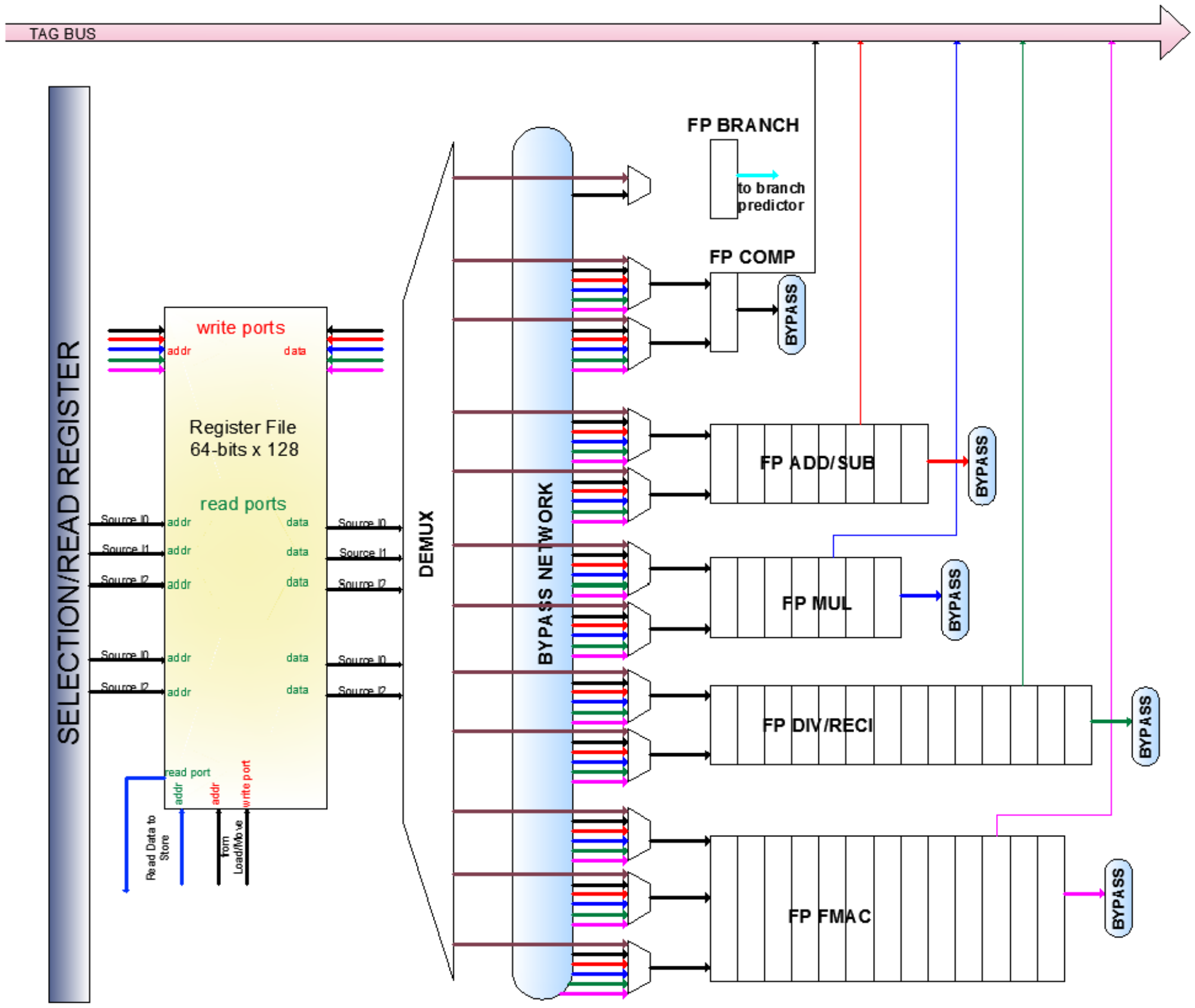

**Fig. 4.34** Bypass Network

### 4.1.5. Complete design

**Figure 4.35** shows the complete design of the "Out of Order FP Execution Engine" which was modeled in Verilog and the result are presented in Chapter 6.

The FP Issue queue can perform issue up to two instructions per clock cycle, after needs read the source operand to the register file or obtain the value from the bypass network in order to compute the operation. The latencies of each operation are list in Table 4.10.

**Table 4.10** Latency of FP functional units

| Functional Unit | Latency |
|---|---|
| FP Adder/Subtractor unit | 8 |
| FP Multiplier unit | 7 |
| FP Multiply Add unit | 13 |
| FP Divide/Reciprocal unit | 14 |
| FP Compare unit | 1 |
| FP Branch | 1 |

When an instruction is issue to Read register stage is sent to the Reorder Buffer the ROBEntry of this instruction in order to update the ROB structure. Same case when an instruction is executed, the same ROBEntry is sent to notify the state of the instruction in the window.

#### Recovery

As was mentioned in the background chapter, a superscalar processor with dynamic scheduling can execute instructions speculatively. When a misprediction occur the speculative state of the machine is incorrect because the processor has been fetching, renaming and executing instructions from wrong path. Therefore, when a branch misprediction is identified, the speculative processor state and the program counter should be restored to the point where the correct path starts.

In general, recovering the front-end (Fetch, Decode, Dispatch) implies flushing all intermediate buffers where instructions fetched from the wrong path are in-fly, restoring the history of the branch predictor and updating the program counter to resume fetching instructions from the correct path. By contrast, recovering the back-end implies removing all instructions belonging to the wrong path residing on any buffer like the Issue queue, Reorder buffer, etc. Moreover, RAT's should be restored as well in order to properly rename instructions from the correct path. Finally, back-end resources like physical registers or issue queue entries allocated by wrong-path instructions should also be reclaimed.

Lagarto II processor every cycle takes a snapshot of the current state, the number of snapshots that LagartoII can save is defined by the size of the Reorder Buffer which is 128 locations, means that can has 128 in fly instructions. Also Lagarto II can perform fetch decode and dispatch up to two instructions per clock cycle, that means that in 64 cycles the reorder buffer will be full. Then are needed 64 snapshots of the state of the processor to recovery if some recovery is necessary.

In the execution engine the snapshot is taken every cycle of the next three structures: The *Ready Bit Vector*, the *Fifo_Blocks* and the *valid bit* of the CAM blocks in the issue queue.

When new value is written in the *Register File*, the same location in the *Ready Bit Vector* is set, in order to notify to new instructions that needs this register as source operand that it is ready. The content of the *Register File* doesn't matter because can be written some values that were written during a speculative execution, but this values can be discarded only with the *Ready Bit*, and this location not be valid until a new correct value is written. Same case occurs with the *valid bit* of the CAM blocks, when a new instructions arrives to the Queue the *valid bit* of this locations is set, and when a misprediction occur, the recover mechanism give the address of the snapshot to recovery the correct state at the time of the branch predictor perform a miss prediction. Also are needed a snapshots of the *Fifo_Blocks* in order to recover the empty localities of each IQ block and the state of the pointers.

*Fifo_Blocks* and *valid bit* shares a first shadow memory of 34-bits x 64-entries. The content of each entry is shown in Figure 4.35.

| 33 31 | 30 28 | 27 25 | 24 22 | 21 19 | 18 16 | 15 13 | 12 10 | 9 7 | 6 4 | 3 1 | 0 |
|---|---|---|---|---|---|---|---|---|---|---|---|
| FIFO[0] | FIFO[1] | FIFO[2] | FIFO[3] | FIFO[4] | FIFO[5] | FIFO[6] | FIFO[7] | R_Pointer | W_Pointer | Counter | Valid_bit |
| 3 | 3 | 3 | 3 | 3 | 3 | 3 | 3 | 3 | 3 | 3 | 1 |

**Fig. 4.35** 1 entry of the first recovery shadow memory (*Fifo_Blocks and valid bit*).

A second memory only save the 128 *Ready-bit-Register,* then is needed a 128-bits x 64 entries Memory. Both are dual-port memory.

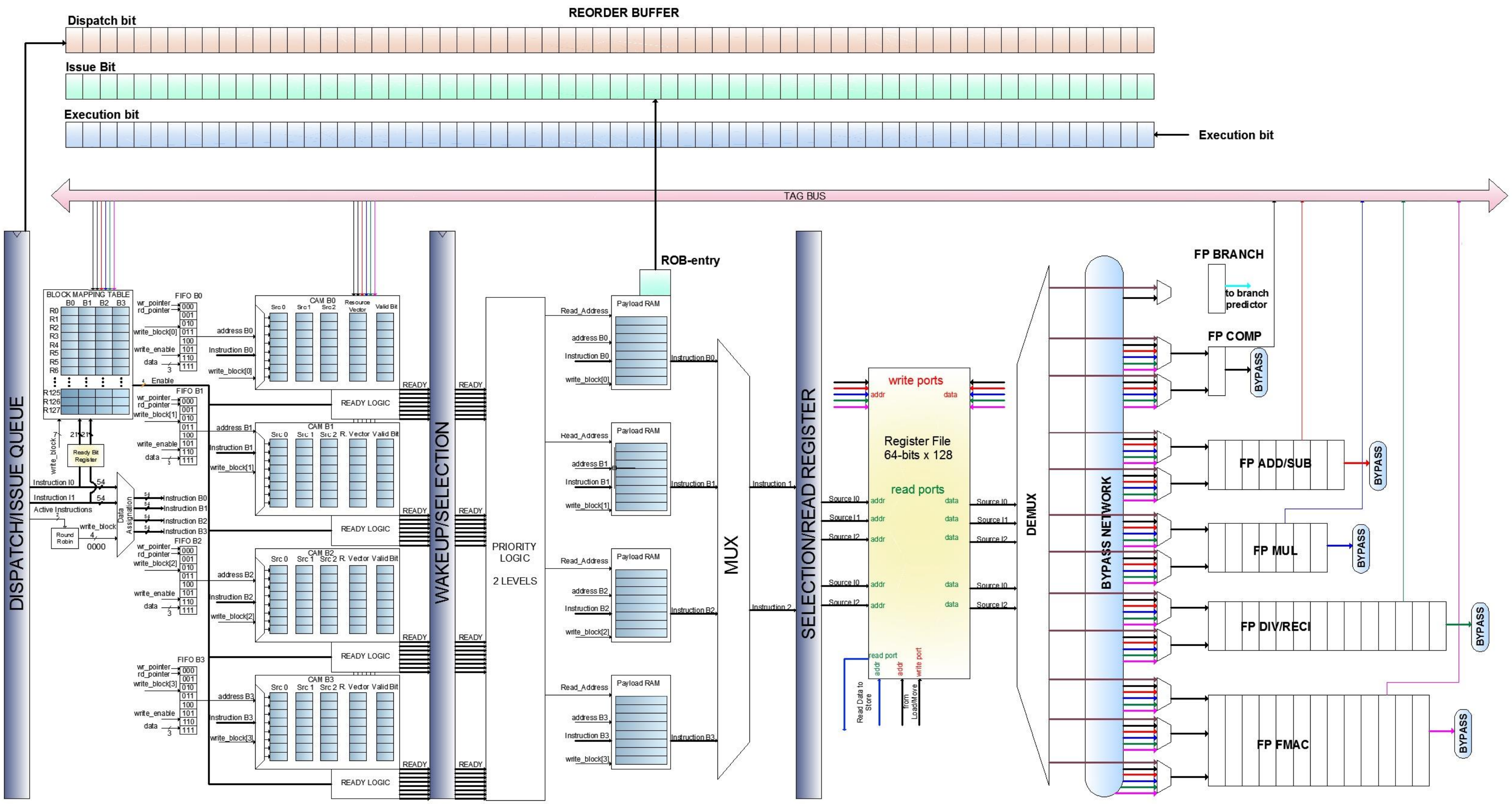

**Fig. 4.36** Out of Order Floating Point Execution Engine Version 1

## 4.2. Second Proposal

Many floating-point units can actually be thought of as collection of datapaths, one for each operation. In some current microprocessors the FMA unit entirely replaces all independent execution units. Good examples are presented in Itanium Processors [34] and more recent in AMD bulldozer microarchitecture [35], last is present until today in many processors as FX series. For some applications several units inside the CPU remain idle for a lot time, and these units could be combined, it can save area, register ports, save energy and cut cost according AMD.

Lagarto II processor only can issue two instructions, means that in the first design presented in Figure 4.35, five of the total seven units remain idle in the first stage, and in general in each same stage of all units.

A new redesign taken account the Bulldozer Microarchitecture [35] is proposed, where basically hardware of the Scalar Floating Point execution units with the SIMD execution units is shared. These architectures use a 128-bit FMAC, which under this block finally put two 64-bit MAC as shown in Figures 4.37 and 4.38. This 64-bit FMAC perform all arithmetic and logical operations.
FPU also contains two 128-bit integer units, which perform arithmetic and logical operations on AVX, MMX and SSE packed integer data.

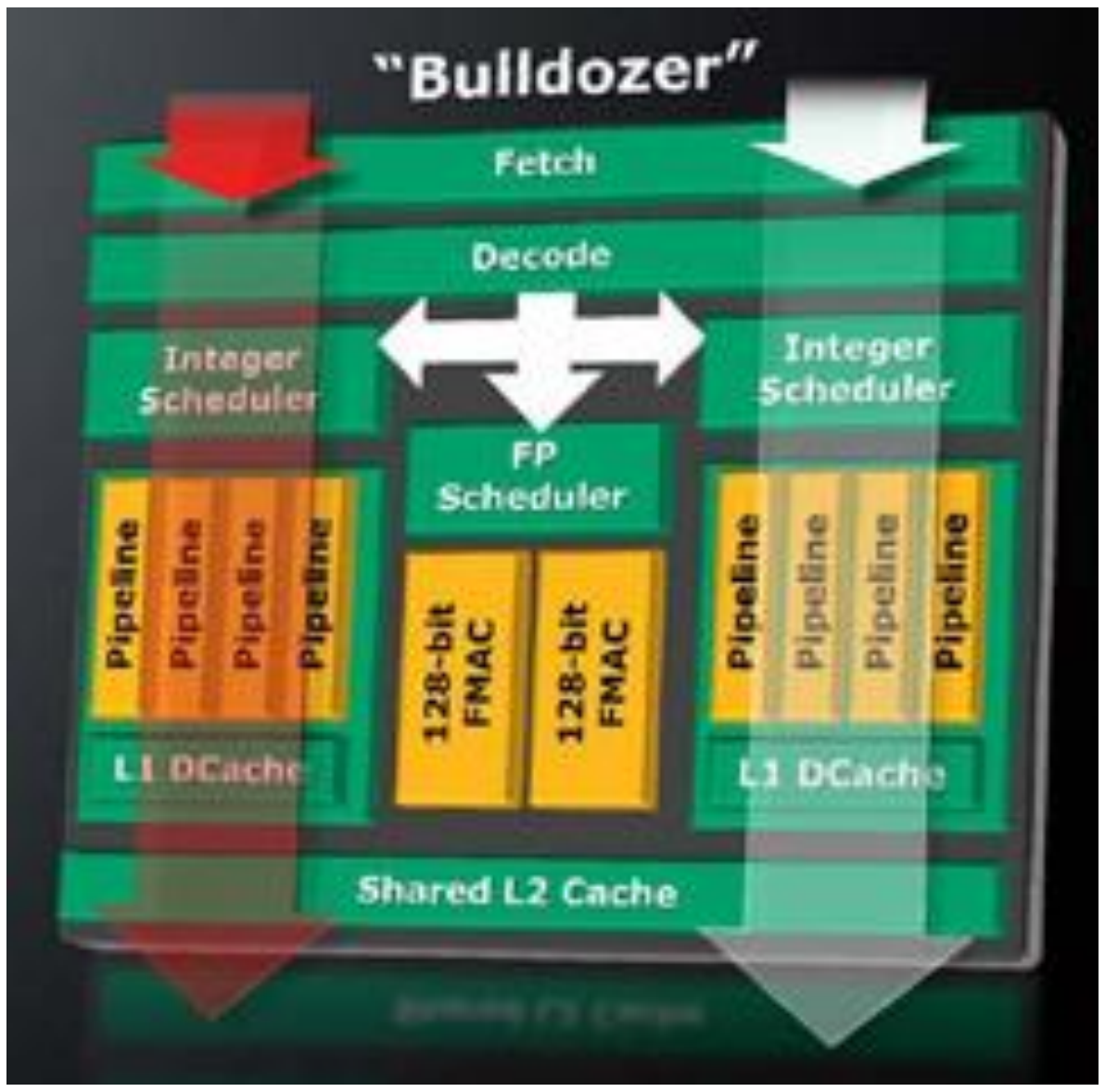

**Fig. 4.37** Bulldozer Microarchitecture

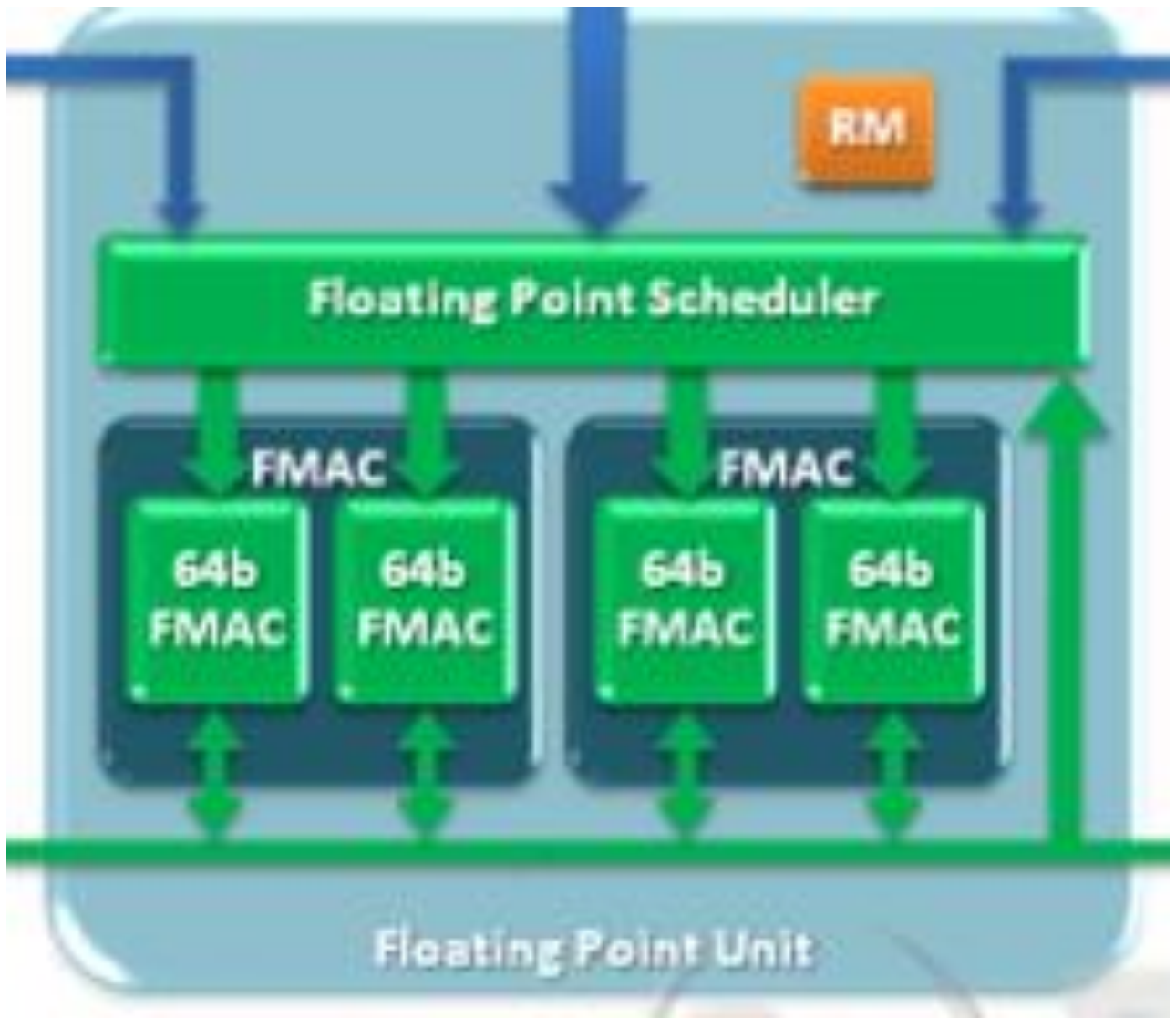

**Fig. 4.38** Floating-point Bulldozer Microarchitecture

A similar idea is proposed, but for our architecture which at the moment can only execute scalar FP instructions, one 128-FMAC is implemented, that means that only are needed 2x64-bit FMAC to execute 2 instructions per clock cycle using the same issue scheme. General design is presented in Figure 4.40.

With this design are obtained a various benefit in area, performance and energy efficiency which are described below:

- In the Issue Queue the number of tag to compare in the CAM blocks are reduced, the first design performs 6 comparisons per source, with new proposal only perform 3 comparisons when the Block Mapping Table designate (2 from FMAC and 1 from load/move). Also in the Block Mapping Table are deleted three read and three write ports. That implies reduction in area and improves in frequency the design.
- In the Register File 1 read port is added and three write ports are deleted. The design of the Register File will occupy less area and the operative frequency increases.
- The bypass logic becomes less complex.
- Execution Engine has less idle stages of executions units.
- The extra logic needed after the register file to send the data to corresponding functional unit is deleted.
- Complete design works a higher frequency.

Many improvements were obtained with these modifications, in fact the IPC is more or less the same, the maximum IPC for this design is two.

In order to perform all this changes, some modifications to the 64-bit FMAC are needed to execute all arithmetic and logical operations in the same unit. Basically is implemented an extra logic to compute comparisons and other instructions previously implemented in *Comparison* unit and put this in the penultimate stage. Design is presented in Figure 4.39. The new FMAC design can execute all instructions previously mentioned. Fused Multiply Accumulate instructions and Multiply have priority, addition/subtraction and compare instructions needs check if the two or eight stages are free in order to be issue to execute.

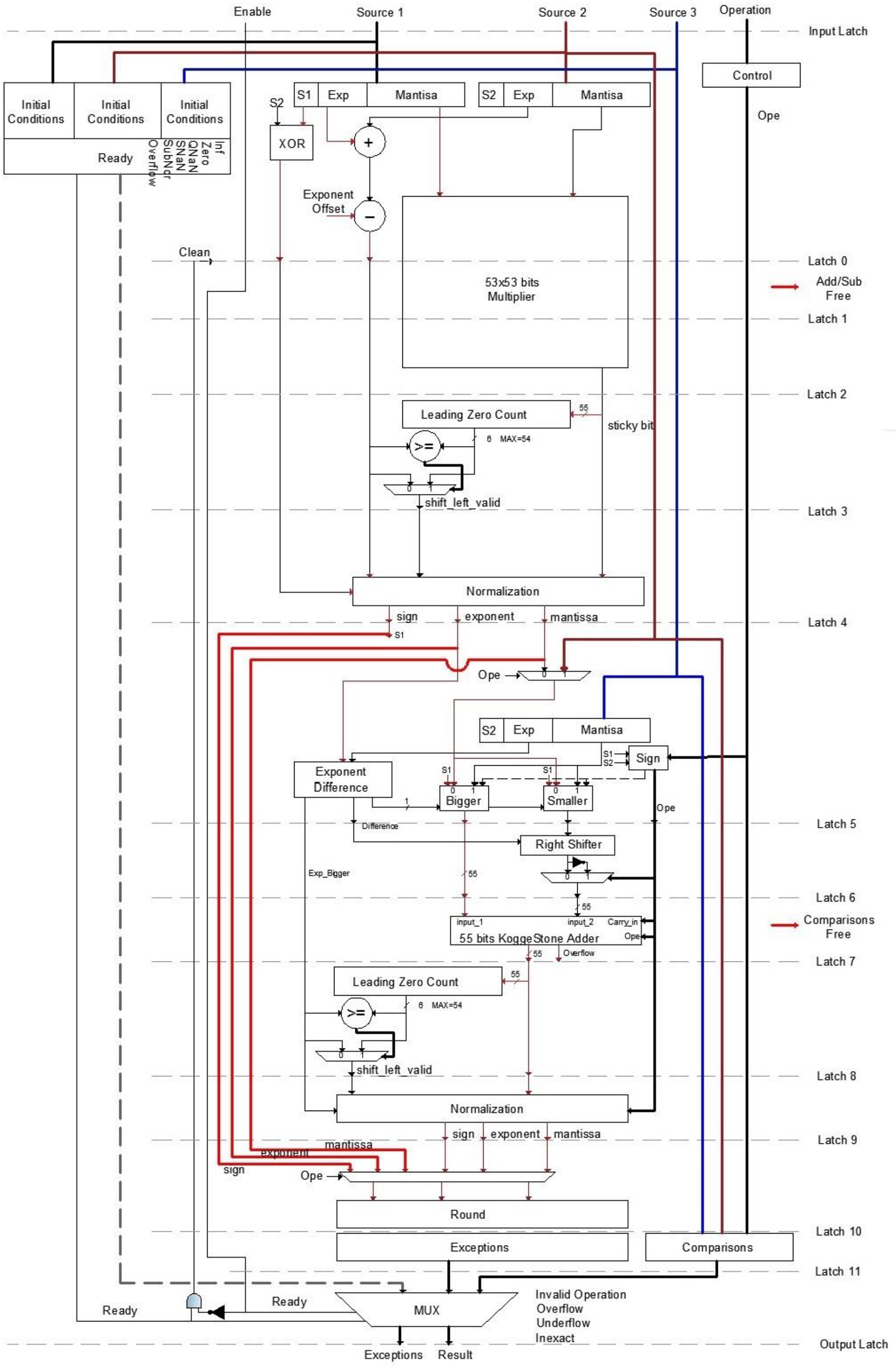

**Fig. 4.39 New FMAC design**

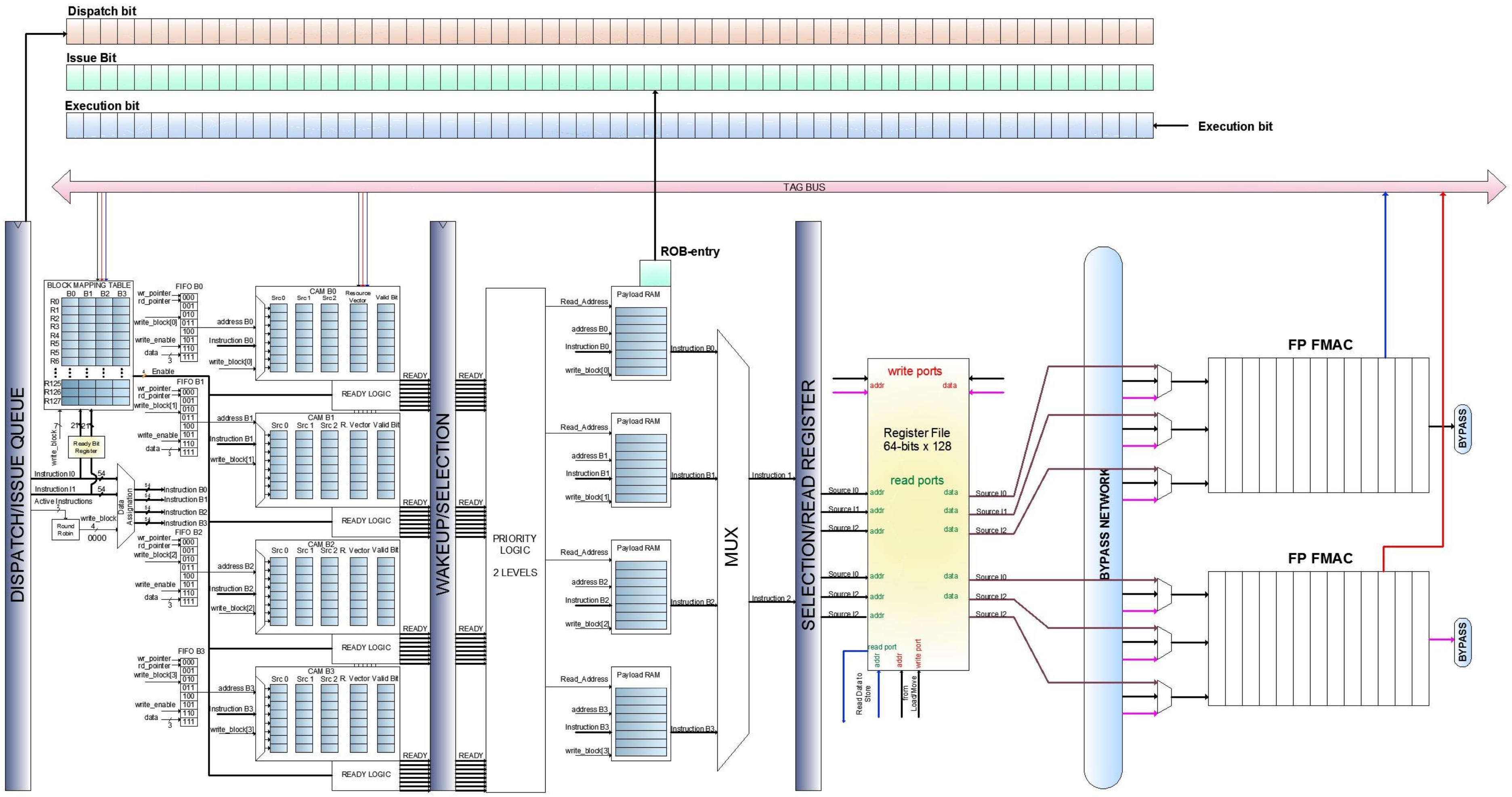

**Fig. 4.40** Out of Order Floating Point Execution Engine Version 2

# Chapter 5

# 5. Implementation

In this chapter is presented the implementation of proposed designs, each independent building blocks and the complete design. Both proposals were implemented in Hardware Description Languages (HDL-Verilog) and was used the Altera DE2-115 FPGA [39] as platform of proves. The main features of the Develop Platform Altera DE2-115 are listed below:

- Altera Cyclone IV FPGA
- 50 MHZ oscillator for clock sources.
- 114 480 Logic elements.
- 432 M9K memory blocks
- 3888 Kbits embedded memory
- 128 MB (32Mx32-bit) SDRAM
- 2 MB (1Mx16) SRAM
- 8MB (4Mx16) Flash

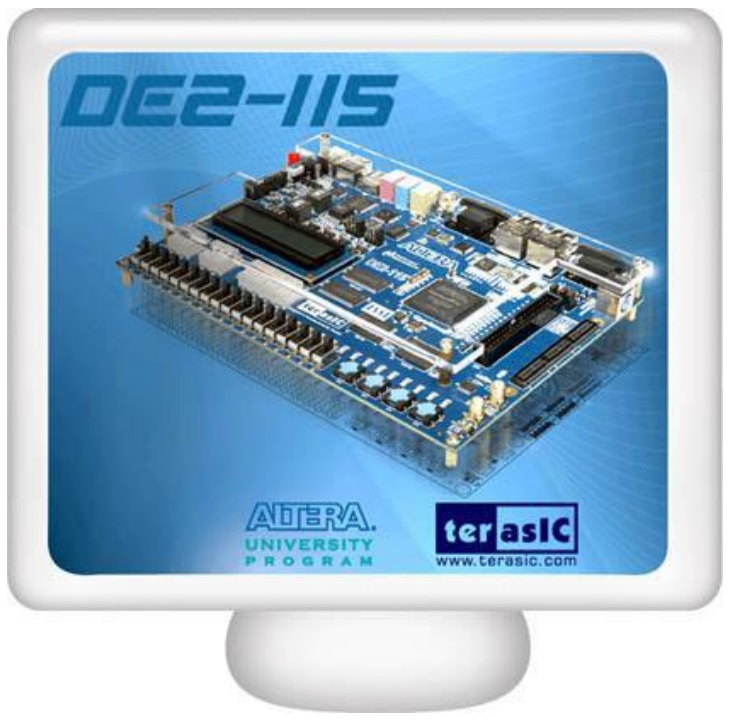

## 5.1. First Version

### 5.1.1. Issue Queue

In table 5.1 are shown the resource used in the implementation of the Out of Order Issue Queue. As was mentioned in chapter 3, the Issue Queue design is some of the elements of the processor which consumes more energy, even in new processor, for this reason the needed of choose a low power design is critical for Lagarto II.

First is presented the BMT implementation version which includes the low power mechanism. This design is using 10864 Logic elements, which is less than the 10 % of logic elements of Cyclone IV device. Also the Frequency achieve up to 92.04 MHZ in the worst case (Slow 1200mV 85C Model). Some of the objectives of this project is that the complete design (Lagarto II) reach a operating frequency between 80 and 90 MHz, but in the complete design the frequency become down because long wires from functional units to the Issue Queue , this results will be show latter.

**Table 5.1** Implementation results of FP Issue Queue

| Design | Area (Logic Elements) | Embedded Memory bits | FMAX (MHZ) Worst case ( Slow 1200mv 85C Model ) |
|---|---|---|---|
| **With BMT** | 10864 | 64512 | 92.04 |
| **Without BMT** | 8056 | 0 | 110.61 |

Furthermore, is presented use of resources of a design of the Issue Queue without the low power mechanism (BMT), the area is less than the first design because don't needs the extra logic to perform comparisons in the BMT and in the Wakeup Logic, also not use embedded memory and the frequency is better than the low power consumption design.
As shown in Chapter 3.1, comparisons per committed instructions for floating-point benchmarks for a 32- and 64-queue size, the averages are 12 and 17 comparisons per committed instruction, there are unnecessary comparisons that can be avoided, and using a proposed design (with BMT) only require 1.5 comparisons per committed instruction achieve a reduction near of 73 % for SPEC2000 benchmarks. The operative frequency is reduced by only 7%, but the energy saving is too much, which justifies the loss of performance, Lagarto II processor is a processor designed for mobile devices, so the power consumption is an important factor.
If a low power consumption in processors is required, the performance inevitably declines, basically the designer must decide which is more important. Therefore, today the big companies of processors like AMD or Intel have several processor families, families focused to mobile devices such as laptops or tablets, these processors use low power consumption techniques and inevitably they have less performance than processor families targeted to desktop market which do not care about power consumption as they are always connected to a power outlet.

### 5.1.2. Register Bank

As mentioned in chapter 3.2, the implementation of Multiport Memory in FPGAs has some inconvenient, the number of logic elements (LE) increases according to the number of read and write ports and the size of the memory, for this reason is necessary to use techniques in order to reduce needed the LE. The register file is a 64-bits x 128-entries multiport memory with 6-read and 6-write ports.

Table 5.2 show the comparison between the implementation of LVT and XOR designs versus the implementation using only Logic Elements. All designs have 6-read and 6-write ports. First one is the implementation using only Logic Elements, is clearly the abundant use of LE, near of the 33% of the total LE provided by the Altera DE2-115 FPGA, also de frequency is enough for the requirements. Second version was implemented using the LVT design, this proposal uses only 5,188 LE and 294, 912 embedded memory bits (7.4% of the total embedded memory bits), the reduction of needed LE is notable, also the operative frequency increase up to 221 MHZ.
Third version was implemented using the XOR design, this proposal uses the small amount of 1536 LE, but the embedded memory bits increase versus the LVT design up to 540, 672 em-bedded memory bits (14% of the total embedded memory bits, also the operative frequency obtained is 116.71 MHZ.

**Table 5.2** Implementation results of Register Bank

| Design | Area (Logic Elements) | Embedded Memory bits | FMAX (MHZ) Worst case ( Slow 1200mv 85C Model ) |
|---|---|---|---|
| **Logic Elements** | 37, 337 | 0 | 130.82 |
| **LVT design** | 5, 188 | 294, 912 | 221 |
| **XOR design** | 1, 536 | 540, 672 | 116.71 |

Finally, the chosen design was de XOR design, because the needed of LE's resources is small and the operative frequency full fit the requirements.

### 5.1.3. Execution Stage

**FP Adder/Subtractor**

Table 5.3 show the implementation results of the FP Adder/subtractor proposal versus the IP Core provided by Altera. Both designs contemplate 8 stages of pipeline.

**Table 5.3** Implementation results of FP Adder/Subtractor

| Design | Area (Logic Ele-ments) | Embedded Memory bits | Embedded Multiplier | FMAX (MHZ) Worst case ( Slow 1200mv 85C Model ) |
|---|---|---|---|---|
| **Altera IP Core** | 1804 | 0 | 0 | 116.36 |
| **Proposal IP Core** | 2537 | 0 | 0 | 124.83 |

Table 5.4 shows the comparisons between both designs (Proposal/Altera IP Core).

**Table 5.4** Comparisons between both FP adder/Subtractor

| Design | Advantages | Disadvantages |
|---|---|---|
| **Altera IP Core** | • Less area | • Not support subnormal numbers<br>• Only support one rounding mode<br>• Slower |
| **Proposal IP Core** | • Support subnormal numbers<br>• Early execution Logic<br>• 4 Rounding modes<br>• Faster | • More area |

Altera FP adder/subtractor IP Core consumes less area, but this is because is more simple, only implement one rounding mode and the most important, not support subnormal numbers which increase a lot the complexity of the design and of course the accuracy of the functional unit. The frequency obtained with the proposal is higher than the IP Core although the design is more complex due to be implemented some of the best high performance algorithms in the critical parts like Leading Zero Count and Kogge-Stone Adder.

**FP Multiplier**

Table 5.5 show the implementation results of the FP Multiplier proposal versus the IP Core provided by Altera. Proposal design contemplate 7 stages of pipeline, Altera IP Core contemplate 6 and 10 stages.

**Table 5.5** Implementation results of FP Multiplier

| Design | Stages | Area (Logic Elements) | Embedded Memory bits | Embedded Multiplier | FMAX (MHZ) Worst case ( Slow 1200mv 85C Model ) |
|---|---|---|---|---|---|
| **Altera IP Core** | 6 stages | 832 | 0 | 18 | 119 |
| **Altera IP Core** | 10 Stages | 1041 | 110 | 18 | 132.59 |
| **Proposal IP Core** | 7 Stages | 1932 | 0 | 30 | 118.89 |

Table 5.6 show the comparisons between both designs (Proposal/Altera IP Core).

**Table 5.6** Comparisons between both FP adder/Subtractor

| Design | Advantages | Disadvantages |
|---|---|---|
| **Altera IP Core** | • Less area<br>• Faster | • Not support subnormal numbers<br>• Only support one rounding mode |
| **Proposal** | • Support subnormal numbers<br>• Early execution Logic<br>• 4 Rounding modes | • More area<br>• Slower |

Altera FP Multiplier IP Core consumes less area, but this is because is more simple, only implement one rounding mode and the most important, not support subnormal numbers which increase a lot the complexity of the design and of course the accuracy of the functional unit. The frequency obtained with the proposal is slightly less than the IP Core, it is because the design is more complex due to the support of subnormal numbers and all rounding modes defined in the standard IEEE754.

**FP Divider**

Table 5.7 show the implementation results of the FP Divider proposal versus the IP Core provided by Altera. Proposal design contemplate 19 stages of pipeline, Altera IP Core contemplate 24 and 10 stages.

**Table 5.7** Implementation results of FP Divider

| Design | Stages | Area (Logic Elements) | Embedded Memory bits | Embedded Multiplier | FMAX (MHZ) Worst case ( Slow 1200mv 85C Model ) |
|---|---|---|---|---|---|
| **Altera IP Core** | 24 stages | 1, 344 | 6, 441 | 44 | 117.91 |
| **Altera IP Core** | 10 Stages | 1, 325 | 4, 709 | 44 | 88.94 |
| **Proposal IP Core** | 19 Stages | 3, 550 | 2, 048 | 62 | 118.55 |

Table 5.8 shows the comparisons between both designs (Proposal IP Core /Altera IP Core).

**Table 5.8** Comparisons between both FP adder/Subtractor

| Design | Advantages | Disadvantages |
|---|---|---|
| **Altera IP Core** | • Less area | • Not support subnormal numbers<br>• Only support one rounding mode |
| **Proposal IP Core** | • Support subnormal numbers<br>• Early execution Logic<br>• 4 Rounding modes<br>• Faster | • More area |

Altera FP divider IP core implement a similar algorithm using a memory for initial approximation. The memory used is bigger than the implemented in the proposal IP core. Also the proposal consumes more area than Altera IP core. Finally, the operative frequency is given by the lower stage which is the multiplier, the complete design of the proposal FP divider IP Core is faster than the Altera FP divider.

**FP Fused Multiply Accumulate**

Table 5.9 shows the implementation results of the FP Multiply Accumulate IP Core proposal. IP Core for FMAC is not provided by Altera. Proposal FMAC IP Core design contemplates 13 stages of pipeline.

**Table 5.9** Implementation results of FP Multiply Accumulate

| Design | Stages | Area (Logic Elements) | Embedded Memory bits | Embedded Multiplier | FMAX (MHZ) Worst case ( Slow 1200mv 85C Model ) |
|---|---|---|---|---|---|
| **Altera FMAC IP Core** | - | - | - | - | - |
| **Proposal FMAC IP Core** | 13 Stages | 3612 | 0 | 30 | 113.34 |

**FP Comparison Unit**

Table 5.10 shows the implementation results of the FP Comparison. This module compute 40 instructions like comparisons, mask, movements, absolute value and others.

**Table 5.10** Implementation results of FP ALU

| Design | Stages | Area (Logic Elements) | Embedded Memory bits | Embedded Multiplier | FMAX (MHZ) Worst case ( Slow 1200mv 85C Model ) |
|---|---|---|---|---|---|
| **Altera FP Comparison IP Core** | - | - | - | - | - |
| **Proposal FP Comparison IP Core** | 1 | 1068 | 0 | 0 | 166.06 |

## 5.1.4. Recovery

Recovery basically is a set of dual-port memories (1-read and 1-write). Table 5.11 shows the implementation results of the shadow memories needed for recovery from a speculative state.

**Table 5.11** Implementation results of recovery

| Design | Memory size | Embedded Memory bits | FMAX (MHZ) Worst case ( Slow 1200mv 85C Model ) |
|---|---|---|---|
| **Ready bit vector** | 128-bitsx64-entries Dual-port | 8192 | 311.33 |
| **FIFO blocks and Valid bit** | 4 x (34-bitsx64-entries) Dual-port | 8704 | 311.33 |

This memory blocks basically each cycle save a vector which contain information about the current state of the processor. Each memory has 64 entries because are needed 64 snapshots.

## 5.1.5. Complete design

Table 5.12 show the resource utilization of the complete FP engine, are used 26, 893 LE, near of the 23% of the total LE of the FPGA and 574, 976 embedded memory bits which is the 14.7 % of the total embedded memory.

**Table 5.12** Implementation results of the complete design

| Design | Area (Logic Elements) | Embedded Memory bits | FMAX (MHZ) Worst case ( Slow 1200mv 85C Model ) |
|---|---|---|---|
| **FP Engine** | 26 893 | 574 976 | 88.1 |

Also the operative frequency decreases up to 88.1 MHZ due to the interconnection (wires) between the outputs of the FP units and the Wakeup Logic in the issue queue and the interconnection of the Bypass Network. This frequency full fit the requirements.

Figure 6.1 shows the RTL viewer of the complete FP engine generated by Quartus II, which include the Issue Queue, the Read Register, the Ready Bit Vector, the functional units and the Bypass Network.

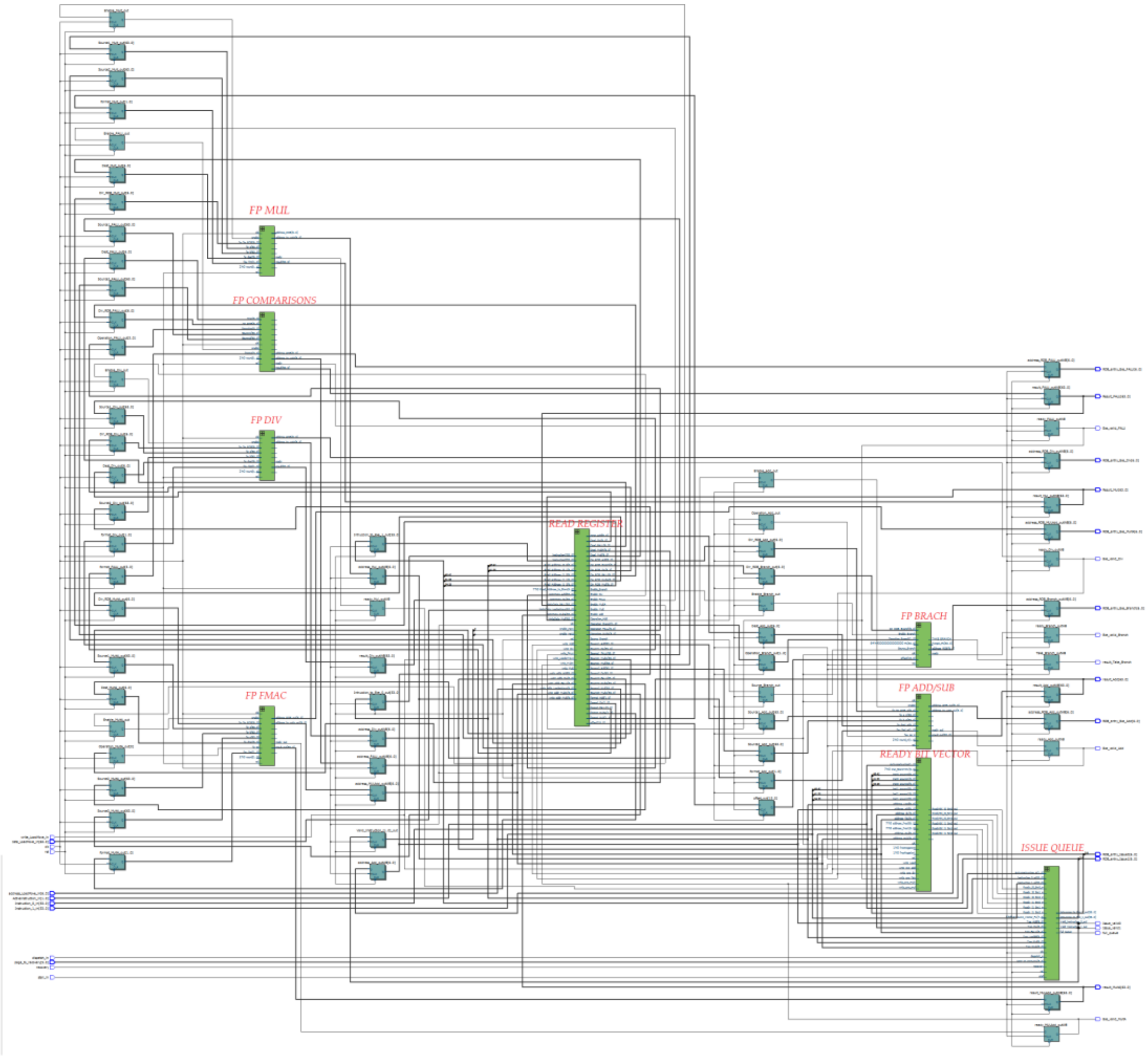

**Fig. 5.1** RTL viewer of the Complete FP Engine.

## 5.2. Second Version

An enhancement of the first version is presented. Improvements are shown in each sub-chapter. With this design are obtained a benefit in area, performance and energy efficiency.

### 5.2.1. Issue Queue

In Table 5.13 are presented the implementation results of the improvement of the Out of Order Issue Queue versus the first design.

**Table 5.13** Implementation results of FP Issue Queue

| Design | Area (Logic Elements) | Embedded Memory bits | FMAX (MHZ) Worst case ( Slow 1200mv 85C Model ) |
|---|---|---|---|
| **First design** | 10864 | 64 512 | 92.04 |
| **Improvement** | 6628 | 24 576 | 111.17 |

The area is reduced and the performance increased due to many causes as the number of tag to compare in the CAM blocks, with first design 6 comparisons per source are performed, with new proposal only 3 comparisons are performed. Also in the Block Mapping Table are deleted three read and three write ports. That implies reduction in area and improves the frequency of the design.

### 5.2.2. Register Bank

In Table 5.14 are shown the implementation results of the improvement of the Register File versus the first design.

**Table 5.14** Implementation results of Register Bank

| Design | Area (Logic Elements) | Embedded Memory bits | FMAX(MHZ) Worst case ( Slow 1200mv 85C Model ) |
|---|---|---|---|
| **XOR design** | 1, 536 | 540, 672 | 116.71 |
| **Improvement** | 576 | 196, 608 | 151.06 |

The area is reduced because are added 1 read port and deleted three write ports. The frequency increases due to the number of XOR operation in writes and read operation decrease almost by 2 times. Furthermore, the extra logic needed adjacent to the register file to send the data to corresponding functional unit is removed.

### 5.2.3. Fused Multiply Accumulate Unit (FMAC)

New proposal is based in the use of only FMAC units which can compute almost all FP arithmetic operations. Table 5.15 show the implementation results of the FP Multiply Accumulate proposal versus the first design proposed before.

**Table 5.15** Implementation results of FP Multiply Accumulate

| Design | Stages | Area (Logic Elements) | Embedded Memory bits | Embedded Multiplier | FMAX (MHZ) Worst case ( Slow 1200mv 85C Model ) |
|---|---|---|---|---|---|
| **First FMAC** | 13 Stages | 3612 | 0 | 30 | 113.34 |
| **New FMAC** | 13 Stages | 4680 | 0 | 30 | 110.34 |

In the new proposal are used 2 FMAC units, Lagarto II can perform issue up to 2 instructions per cycle, for these reason the new design can execute any combination of instructions unlike the first version which could only execute 2 different instructions in a given cycle. FP divider is executed by software with the new design.

### 5.2.4. Recovery

Recovery basically is the same that the presented in first version.

### 5.2.5. Complete design

Table 5.16 show the resource utilization of the new complete FP engine versus the first version, are used 14, 635 LE, near of the 13% of the total LE of the FPGA and 245 760 embedded memory bits which is the 6 % of the total embedded memory.

**Table 5.16** Implementation results of FP Multiplier

| Design | Area (Logic Elements) | Embedded Memory bits | FMAX (MHZ) Worst case ( Slow 1200mv 85C Model ) |
|---|---|---|---|
| **FP Engine First Version** | 26, 893 | 574, 976 | 88.1 |
| **FP Engine Second Version** | 14, 635 | 245, 760 | 100.07 |

Compared with the first version, the area reduction is huge, and the performance increases noticeably. This new proposal is obtained many benefits in area and energy consumption, which is one of the specific goals of this work due to Lagarto II processors is designed for mobile devices and the energy efficiency is important aspect to obtain more autonomy in the mobile devices.

Also the operative frequency increases up to 100.07 MHZ due to the bypass logic become less complex, and in general all blocks work at higher operative frequencies compared with the first version.

Figure 5.2 shown the RTL viewer of the complete new FP engine generated by Quartus II, which include the Issue Queue, the Read Register, the Ready Bit Vector, the FMAC's units and the Bypass Network.

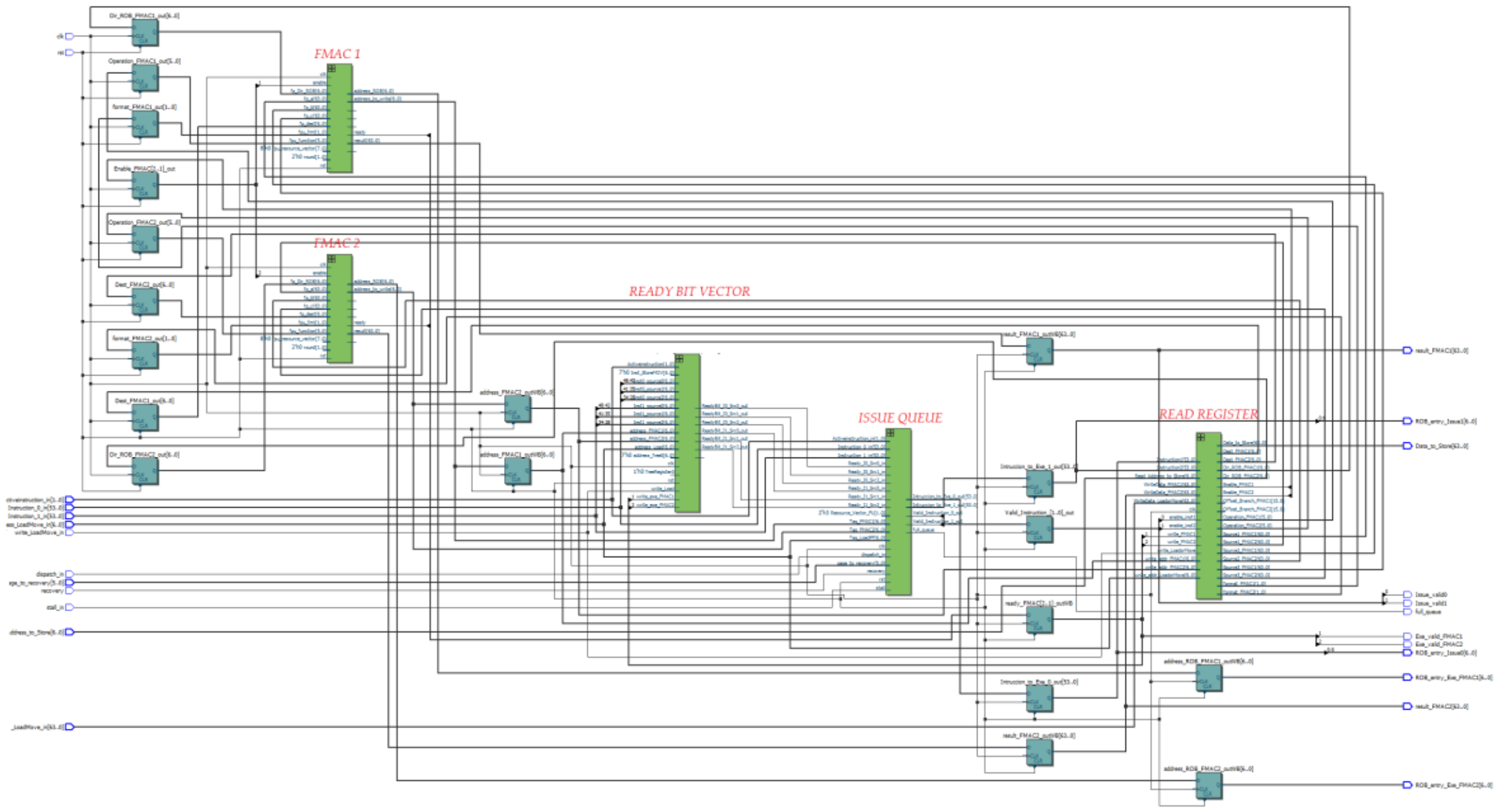

**Fig. 5.2** RTL viewer of the Complete FP Engine.

# Chapter 6

# 6. Testing

Each functional y unit was tested individually in order to check all exceptions, rounding modes and special cases, also complete design was testing with a set of programs in order to prove the correct functionality. Following is presented a little example in order to see easily the correct functionality of the complete design, checking the issue, read register, bypass logic and execution unit at detail. Furthermore, to prove the accuracy of the execution units is compared a little program in assembler language running in the current proposal versus the result provided by a program written in C language running in the Intel i5 processor.

The following code was written in MIPS assembler language, basically perform 3 arithmetic operations, the load/store instructions are emulated because this instruction are executed in the Load/store unit.

```
//Example
LDC1 $f10, 0x0080($0)          // Load Double Word
LDC1 $f11, 0x0088($0)          // Load Double Word
LDC1 $f12, 0x0090($0)          // Load Double Word
MADDF$f8, $f10, $f11, $f12     // Fused Multiply Add
LDC1 $f13, 0x0098($0)          // Load Double Word
LDC1 $f14, 0x00A0($0)          // Load Double Word
MSUBF$f9, $f12, $f13, $f14     // Fused Multiply Subtract
MADDF$f15, $f8, $f9, $f10      // Fused Multiply Add
SDC1 $f15 ,0x00A8($0)          // Store Double Word
```

The following code was written in C language, basically perform the same arithmetic operation that the last example in assembler language.

```
#include<stdio.h>
main()
{
double num1,num2,num3,num4.num5;
double result0, result1, result2;       // $f8,  $f9,  $t7
num1 = 899.5612547825644;               // $f10
num2 = 8979.56546454515;                // $f11
num3 = 7895.1212121289;                 // $f12
num4 = 124.2525465741;                  // $f13
num5 = 999.978569887878;                // $f14

result0= num1 * num2 + num3;
result1= num3 * num4 - num5;
result2= result0* result1+ num1;

printf("\n Final Result:  %lf", result2);
return 0;
}
```

With a bigger program follow the results is more complicated because the amount of data, for this reason only is show a little example, the main idea is check the back-end pipeline stages: Instruction Wakeup, Instruction Issue and the Read register. In this section of the instruction data path perhaps the values read from register file are erroneous but the bypass logic fix the correct value at input ports of the execution units.

Still is not possible execute complete benchmarks due to the Lagarto II processors is not complete, the back-end of the Lagarto II processor still is in development, but the current proves is possible observe the expected behavior.

Table 6.1 shows the values of the corresponding variables used in the proposed program (C and assembler), both uses the same data in order to compare the final result to prove the accuracy of the functional units of the proposal.

**Table 6.1** Values of the load operations in decimal and Floating-point representation.

| Variable | Decimal | Floating Point Representation (Double Precision) |
|---|---|---|
| num1,$f10 | 899.5612547825644 | 0 10000001000 1100000111000111110101110011001001011011111010110100 |
| num2,$f11 | 8979.56546454515 | 0 10000001100 0001100010011100100001100001001001000110100000111100 |
| num3,$f12 | 7895.1212121289 | 0 10000001011 1110110101110001111100000111110000100001000110000001 |
| num4,$f13 | 124.2525465741 | 0 10000000101 1111000100000001010011011100100011011000111001111100 |
| num5,$f14 | 999.978569887878 | 0 10000001000 1111001111111101010000011100011100110000101001001011 |

Table 6.2 shows the logic registers used in the assembler program.

**Table 6.2** Logic Register used in the assembler program.

| Name | Logic Register |
|---|---|
| $f8 | 8 |
| $f9 | 9 |
| $f10 | 10 |
| $f11 | 11 |
| $f12 | 12 |
| $f13 | 13 |
| $f14 | 14 |
| $f15 | 15 |

The following Figures show the simulation of the assembler program proposed before. This simulation was doing in Altera ModelSim Simulator.

Figure 6.1 show up to 15ps, basically at 3 ps enter to the queue the first instruction which is a Fused Multiply Add, in the following cycle enter to the queue 2 instructions, Fused Multiply Subtract and Fused Multiply Add. In next cycles arrive data from the Load/Store unit, the physical registers $f10, $f11 and $f13, also its corresponding tags (*Tag_LoadPF*) is used to read the Block Mapping Table in order to enable the comparisons with the source operands in the CAM Blocks enabled with the *EnableComparison* bits.

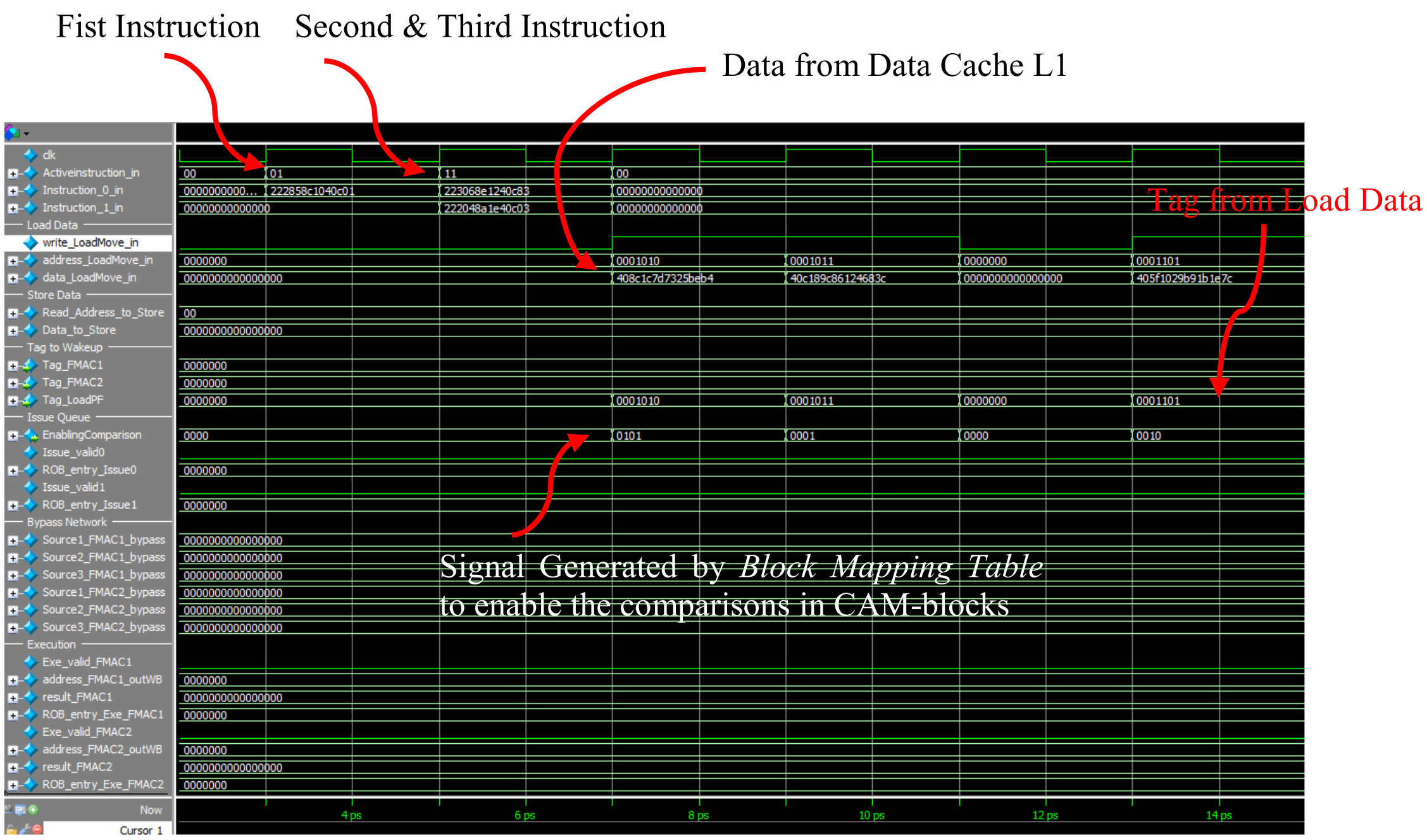

**Fig. 6.1** Simulation from 2ps to 15ps

At 7 ps is shown the first comparisons with the *EnableComparison* bits in CAM-blocks 0 and 2, means that the data from the first load instruction is needed by Instruction 1 and instruction 3 to be issue. The next tags from load data enable comparisons for CAM-blocks 0 and 1.

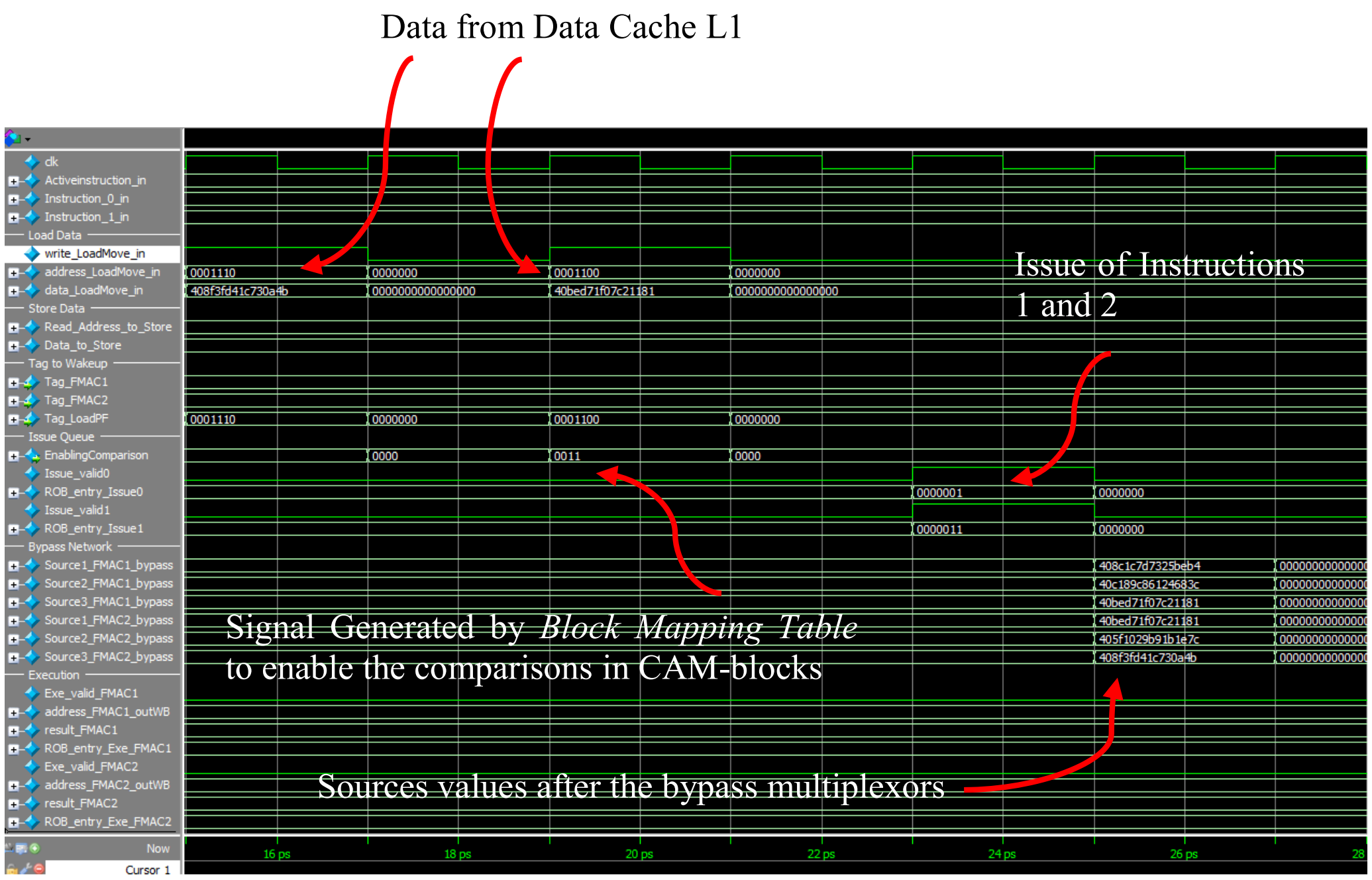

**Fig. 6.2** Simulation from 15ps to 28ps

Figure 6.2 shows the simulation from 15ps to 28ps; In the signals section named **Load Data** is shown the data arrive from the Load/Store unit, the physical registers \$f14 and \$f12, also its corresponding tags (*Tag_LoadPF*) is used to read the Block Mapping Table in order to enable the comparisons with the source operands in the CAM Blocks enabled with the *EnableComparison* bits, in this case, Blocks 0 and 1 are enable to comparisons.

As described previously, the requirements to issue an instruction to execute are that it should have all sources ready, the needed functional unit free and be selected by the *Priority Arbiter*. All conditions were complying by first and second arithmetic instructions, in time 23ps both instruction are Issue, the *Issue_valid* bit is enable and its corresponding *ROB_entry* is used to notify the *Reorder Buffer* that the instructions was Issue.

At 25ps are shows the final values that will arrive to the functional units ports, this values come from the *Register File* or the *Bypass Network*. In this case, this values come from the Register File due to all values were written in early cycles.

Figure 6.3 shows the simulation from 40ps to 53ps; At 43ps in the signal section named **Tag to Wakeup** are shown two values: Tag_FMAC1 and Tag_FMAC2 which were send by the functional units to notify the Wakeup Logic that these instructions will be complete in the next three cycles and the Wakeup process can start. Third instruction comply with all condition and is Issue at 47ps.

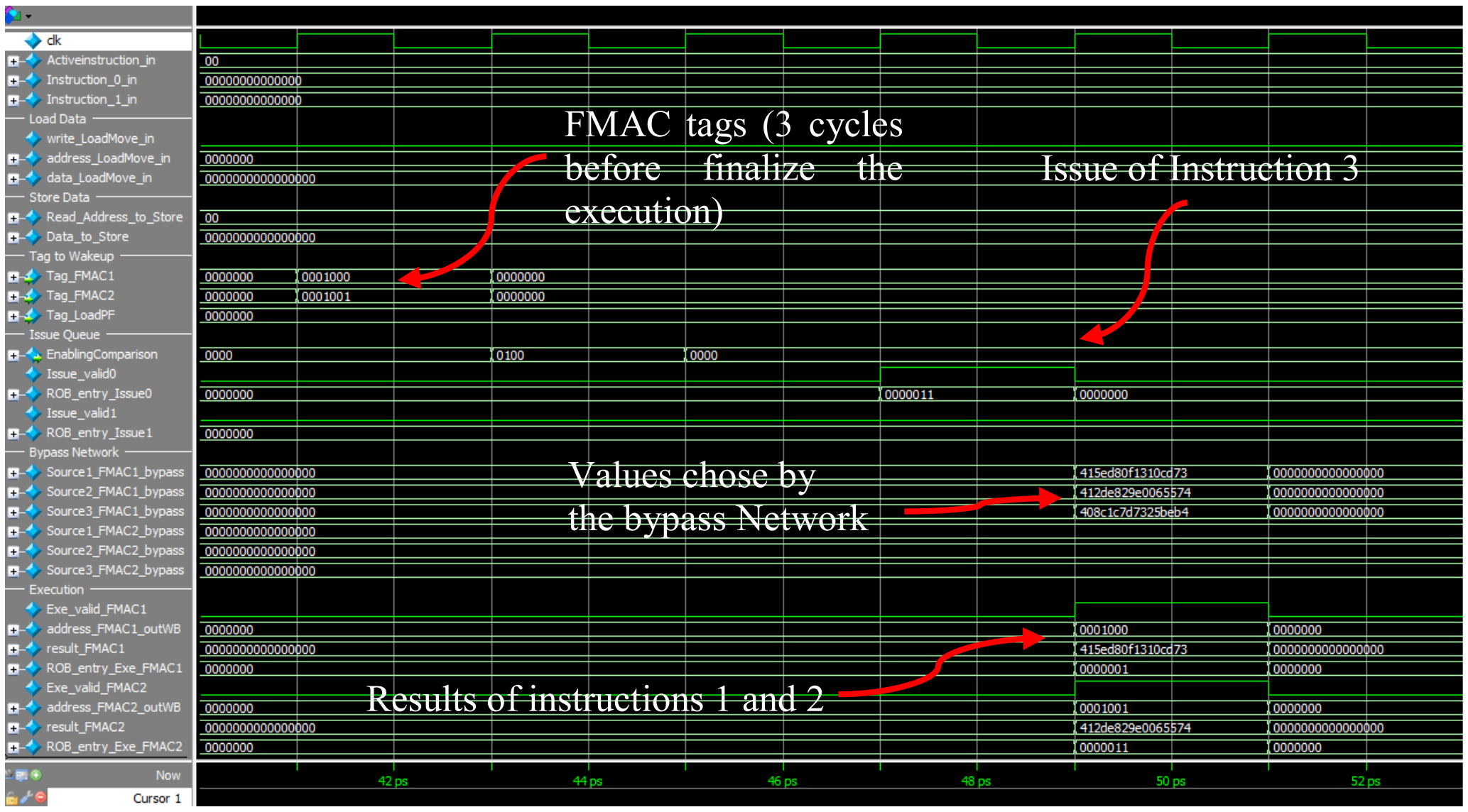

**Fig. 6.3** Simulation from 40ps to 53ps

Also at 49ps is shown the results of the first and second instruction, these values are needed to execute the instruction that was issue one cycle before, and then this values go through the *Bypass Network* to replace the old value read from the register file.

Finally, in Figure 6.4 is show the final result of the third instruction. This data is read by a Store instruction in order to save this data in memory.

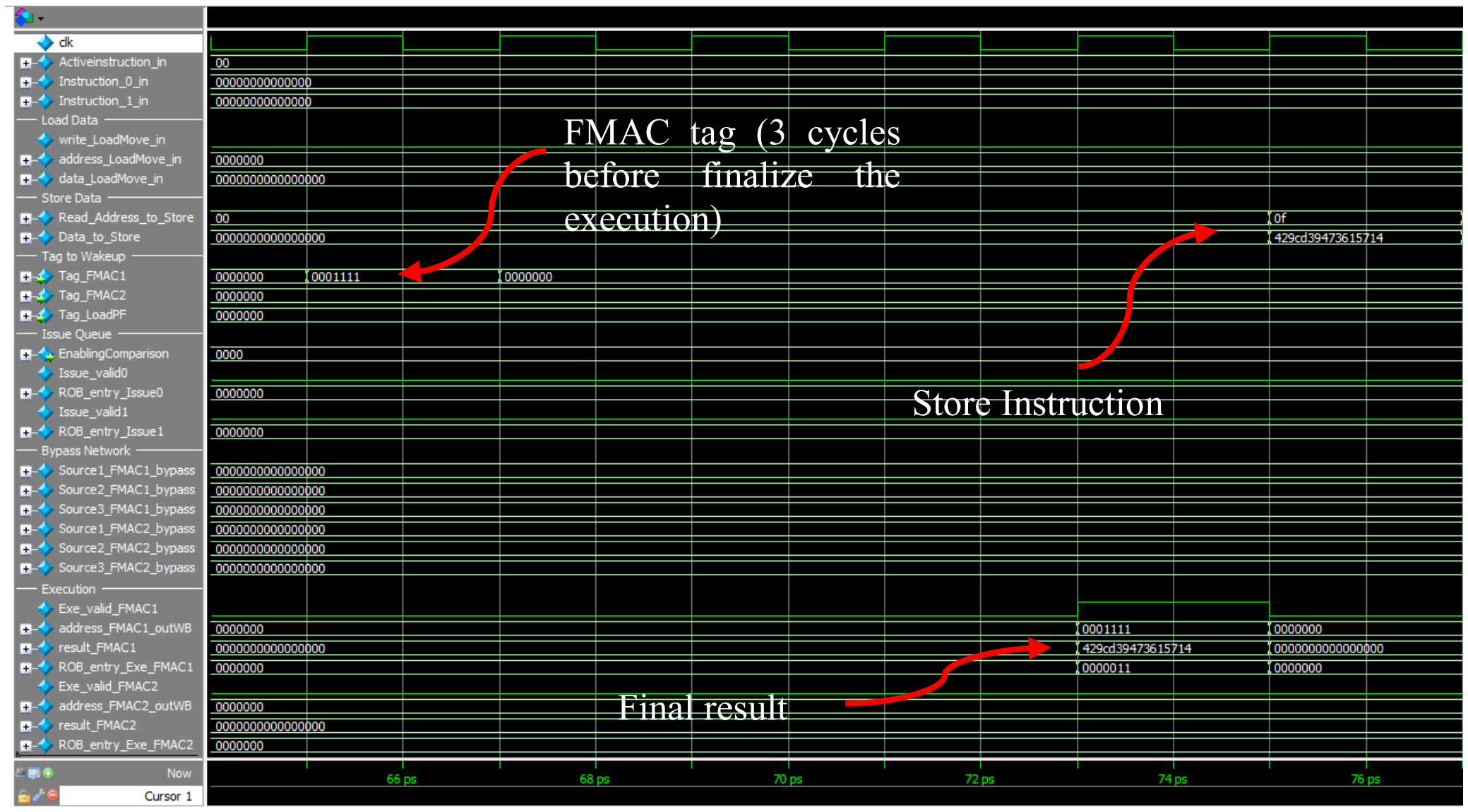

**Fig. 6.4** Simulation from 64ps to 78ps

The final result shown in Figure 6.4 corresponding to **0x429CD39473615714 (Hexadecimal)** which is the representation in the IEEE-754 double precision format. The conversion of this value to decimal representation correspond to **7923763566677.76953125**d.

Figure 6.5 show the final result of the previous example written in C language and executed in a Intel X86 architecture, this result is exactly the same obtained with the program written in Assembler language and running in the proposal FP Execution Engine design.

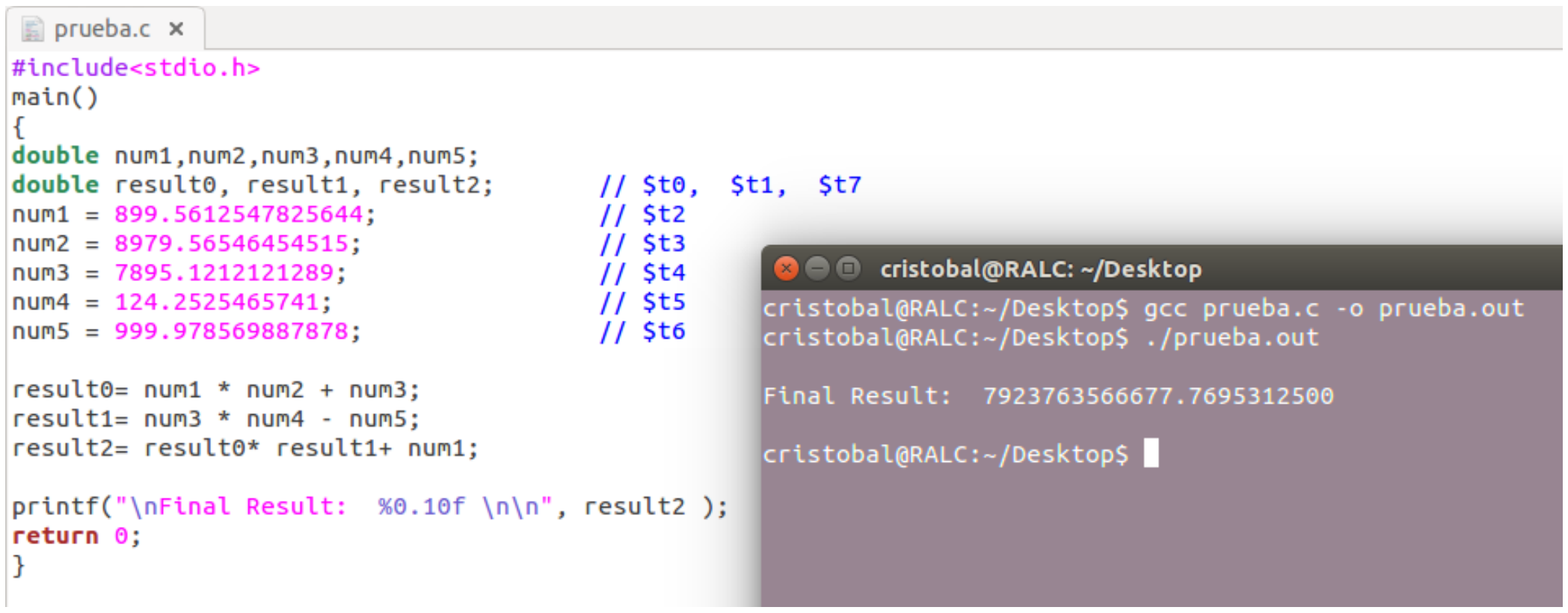

**Fig. 6.5** Result of the program written in C language.

# Chapter 7

# 7. Conclusions, Results, Future works and Research's Products

Embedded processors oriented to mobile devices needs a high performance in order to support the new applications, furthermore, these architectures need to use of low power consumption techniques in order to provide the greatest possible energy autonomy.
In superescalar processors with Out of Order execution the issue queue play an important role in the design because is one of the elements which consumes more power of the total power consumption in the processor. Lagarto II now has a high performance Issue Queue accompanied with low power consumption techniques. With the current design is expected save near of the 70% in energy only in the Issue Queue unlike use a traditional design using RAM-CAM Schemes.

In the complete design there are some other ways to save energy, in the register file design there are a large number of proposals in order to save energy, but this designs do not apply to implementation on FPGA, also implement multiport memory in the FPGA is not a trivial task, are needed a special techniques special for FPGA as the presented in this work.

High performance FP functional units were designed and implemented, also still we can perform improvements on energy consumption in these units. Usually the floating-point units have a large latency and techniques like Clock gating are used in order to reduce the dynamic power dissipation in stages that not doing any work in a given cycle.

The bypass network is an important element in the processor, in fact all processors today include a bypass network in order to increase the performance due to in combination with the wakeup logic, the back to back execution is supported.

The current thesis work accomplishes with the all goals, also were designed and implemented extra FP units in order to support all possible the FP instruction set.

## Results

High performance Floating-Point IP Cores for:

- Addition/Subtraction
- Multiplication
- Division

- Fused Multiply Accumulate

Multiport memories for FPGA.

Issue Queue with two stages:

- Low power consumption Wakeup Logic

Synchronization for back-to-back execution.

Testing of the complete design in order to prove the functionality and accuracy of the complete design.

# Future works

Second proposal was designed in order to in a near future the FP Execution Engine can execute SIMD instructions with the adding of a little extra logic. Sharing the FPU unit with SIMD unit Lagarto II could save a lot of area due to FP scalar hardware occupy a big area comparable with the FP SIMD unit. Furthermore, the plan is not only execute 128-bits SIMD instructions, with the adding of 2 extra FMAC units (similar to the Bulldozer microarchitecture) can execute 4 scalar floating-point instructions in the same cycle or 2x128-bit, 1x256-bit SIMD instruction or combine of these instructions, as is shown in Figure 7.1. Furthermore, integer SIMD instructions will be contained in the complete floating point unit. The Logic in the Issue queue will be simpler because the proposal was divided in four blocks, and performing Issue of 4 instructions the *Priority Arbiter* will be simpler due to can remove one level of priority. The register bank will be bigger due to it can read words of 128-bits for 128-bits SIMD instructions or 2-words of 128-bits for 256-bits SIMD instructions.

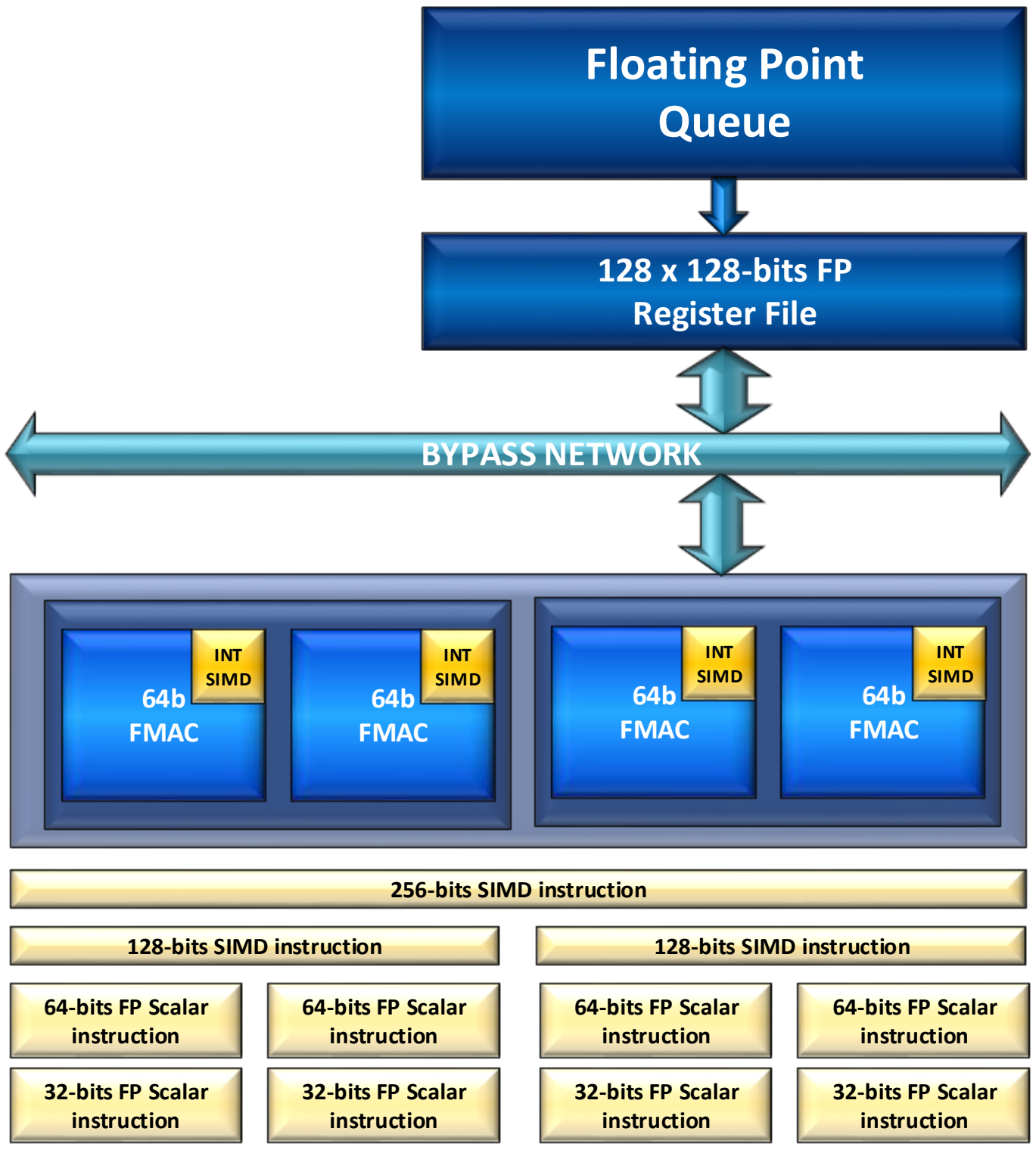

**Fig. 7.1** FP Scalar/SIMD units sharing hardware.

Finally, the current design was not proven in conjunction with Lagarto II, due to the processor still is not complete, for this reason complete benchmarks was not proven in order to obtain data of the real performance.

## Research products

- Prototype of the out-of-order floating-point execution engine.
- 4 High performance Floating-Point IP Cores
- Currently working on writing a paper for publish in Microprocessors and Microsystems: Embedded Hardware design (MICPRO) Journal

# APPENDICES

## APPENDIXE A MIPS 64 Revision 6 and the IEEE standard 754

The following information was extracted from *"MIPS Architecture For Programmers"* Volume I-A [40] and Volume II-A [41].

In the MIPS architecture, the FPU is implemented via Coprocessor 1, an optional processor implementing IEEE 754 floating point operations.
The FPU also provides a few additional operations not defined by the IEEE standard.

### FPU Data types

The FPU provides both floating-point and fixed-point data types

- The single and double precision floating-point data types are those specified by the IEEE standard.
- The fixed-point types are signed integers provided by the CPU architecture.

#### Floating Point Formats

The following floating point formants are provided by the FPU:

- 32-bit single-precision floating point.
- 64-bit double-precision floating point.

The floating point data types represent numeric values as well as other special entities, such as the following:

- Two infinities, +∞ and -∞.
- Signaling non-numbers(SNaNs).
- Quiet non-numbers(QNaNs).
- Numbers of the form: $(-1)^s 2^E \, b_0.b_1 b_2 \ldots b_{p-1}$ where
- *s* = 0 or 1
    - *E* = any integer between *E_min* and *E_max*, inclusive.
    - $b_i$= 0 or 1 (the high bit, $b_0$, is to the left of the binary point)
    - $p$ is the signed-magnitude precision

**Table A. 1** Parameters of Floating Point Data Types

| Parameter | Single | Double |
|---|---|---|
| **Bits of mantissa** | 24 | 53 |
| **Maximum exponent, E_max** | +127 | +1023 |
| **Minimum exponent, E_min** | -126 | -1022 |
| **Exponent bias** | +127 | +1023 |
| **Bits in exponent field, e** | 8 | 11 |
| **Representation of $b_0$ integer bit** | hidden | hidden |
| **Bits in fraction field, *f*** | 23 | 52 |
| **Total format width in bits** | 32 | 64 |

The single and double floating point data types are composed of three fields: sign, exponent and fraction as we can see in Figures A.1 and A.2.

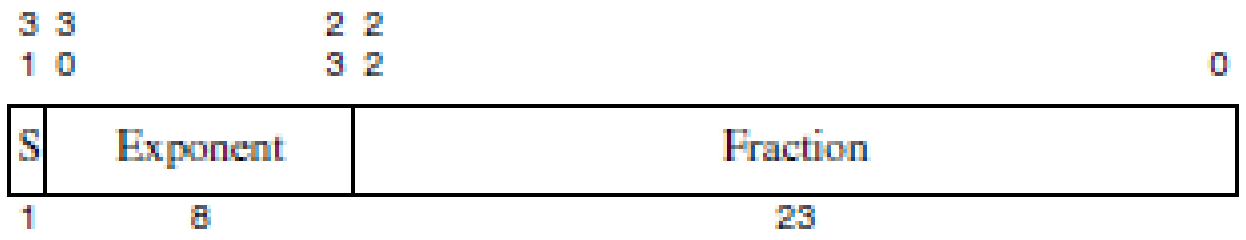

**Fig A. 1** Single-Precision Floating Point Format (S)

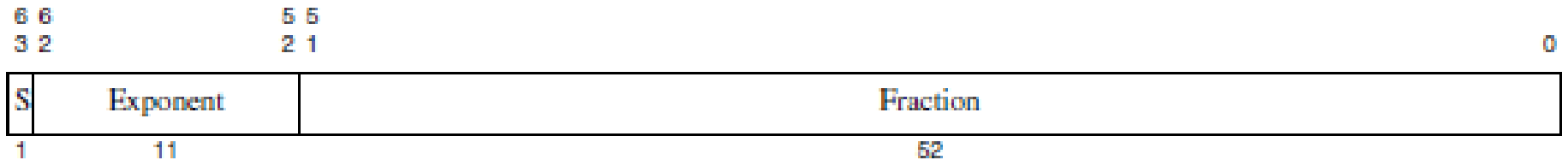

**Fig A. 2** Double-Precision Floating Point Format (D)

Values are encoded in the specified format by using unbiased exponent, fraction, and sign values listed in Table A.2

**Table A. 2** Value of Single or Double Floating Point Data Type Encoding

| Unbiased E | f | s | $b_1$ | Value V | Type of Value | Typical Single Bit Pattern[1] | Typical Double Bit Pattern[1] |
|---|---|---|---|---|---|---|---|
| *E_max* + 1 | ≠ 0 | | 1 | SNaN | Signaling NaN ($FIR_{Has2008}$=0 or $FCSR_{NAN2008}$=0) Not applicable to Release 6 | `0x7fffffff` | `0x7fffffff ffffffff` |
| | | | 0 | QNaN | Quiet NaN ($FIR_{Has2008}$=0 or $FCSR_{NAN2008}$=0) Not applicable to Release 6 | `0x7fbfffff` | `0x7ff7ffff ffffffff` |
| *E_max* + 1 | ≠ 0 | | 0 | SNaN | Signaling NaN ($FCSR_{NAN2008}$=1) | `0x7fbfffff` | `0x7ff7ffff ffffffff` |
| | | | 1 | QNaN | Quiet NaN ($FCSR_{NAN2008}$=1) | `0x7fffffff` | `0x7fffffff ffffffff` |
| *E_max* +1 | 0 | 1 | | - ∞ | minus infinity | `0xff800000` | `0xfff00000 00000000` |
| | | 0 | | + ∞ | plus infinity | `0x7f800000` | `0x7ff00000 00000000` |
| *E_max* to *E_min* | | 1 | | $-(2^E)(1.f)$ | negative normalized number | `0x80800000` through `0xff7fffff` | `0x80100000 00000000` through `0xffefffff ffffffff` |
| | | 0 | | $+(2^E)(1.f)$ | positive normalized number | `0x00800000` through `0x7f7fffff` | `0x00100000 00000000` through `0x7fefffff ffffffff` |
| *E_min* -1 | ≠ 0 | 1 | | $-(2^{E_min})(0.f)$ | negative denormalized number | `0x807fffff` | `0x800fffff ffffffff` |
| | | 0 | | $+(2^{E_min})(0.f)$ | positive denormalized number | `0x007fffff` | `0x000fffff ffffffff` |
| *E_min* -1 | 0 | 1 | | - 0 | negative zero | `0x80000000` | `0x80000000 00000000` |
| | | 0 | | + 0 | positive zero | `0x00000000` | `0x00000000 00000000` |

### Normalized and Denormalized Numbers

For single and double data types, each representable nonzero numerical value has just one encoding; numbers are kept in normalized form. The high-order bit of the p-bit mantissa, which lies to the left of the binary point, is "hidden," and not recorded in the Fraction field. The encoding rules permit the value of this bit to be determined by looking at the value of the exponent. When the unbiased exponent is in the range *E_min* to *E_max,* inclusive, the number is normalized and the hidden bit must be 1. If the numeric value cannot be normalized because the exponent would be less than *E_min*, then the representation is denormalized and the encoded number has an exponent of E_min-1 and the hidden bit has the value 0. Plus and minus zero are special cases that are not regarded as denormalized values.

### Reserved Operand Values – Infinity and NaN

A floating-point operation can signal IEEE exception conditions, such as those caused by uninitialized variables, violations of mathematical rules, or results that cannot be represented. If a program does not choose to trap IEEE exception conditions, a computation that encounters these conditions proceeds without trapping but generates a result indicating that an exceptional condition arose during the computation. To permit this, each floating-point format defines representations, listed in Table A.2, for plus infinity (+∞), minus infinity (-∞), quiet non-numbers (QNaN), and signaling non-numbers (SNaN).

### Infinity and Beyond

Infinity represents a number with magnitude too large to be represented in the format and exists to represent a magnitude overflow during a computation. A correctly signed ∞ is generated as the default result in division by zero and some cases of overflow.
Once created as a default result, ∞ can become an operand in a subsequent operation. The infinities are interpreted such that -∞ < (every finite number) < +∞. Arithmetic with ∞ is the limiting case of real arithmetic with operands of arbitrarily large magnitude, when such limits exist. In these cases, arithmetic on ∞ is regarded as exact and exception conditions do not arise. The out-of-range indication represented by ∞ is propagated through subsequent computations.

For some cases, there is no meaningful limiting case in real arithmetic for operands of ∞, and these cases raise the Invalid Operation exception condition.

### Signaling Non-Number (SNaN)

SNaN operands cause the Invalid Operation exception for arithmetic operations. SNaNs are useful values to put in uninitialized variables. An SNaN is never produced as a result value.

### Quiet Non-Number (QNaN)

QNaNs are intended to afford retrospective diagnostic information inherited from invalid or unavailable data and results. Propagation of the diagnostic information requires information contained in a QNaN to be preserved through arithmetic operations and floating-point format conversions.
QNaN operands do not cause arithmetic operations to signal an exception. When a floating-point result is to be delivered, a QNaN operand causes an arithmetic operation to supply a QNaN result. When possible, this QNaN result is one of the operand QNaN values. QNaNs do have effects similar to SNaNs on operations that do not deliver a floating-point result—specifically, comparisons.

**Table A. 3** Value supplied when a new Quiet NaN is created

| Format | New QNaN value ($FIR_{Has2008}$ = 0 or $FCSR_{NAN2008}$ = 0) | New QNaN value ($FCSR_{NAN2008}$ = 1) |
|---|---|---|
| Single floating-point | 0x7fbf ffff | 0x7fc0 0000 |
| Double floating-point | 0x7ff7 ffff ffff ffff | 0x7ff8 0000 0000 0000 |
| Word fixed point (result from converting any FP number too big to represent as a 32-bit positive integer) | 0x7fff ffff | 0x7fff ffff |
| Word fixed point (result from converting any FP NAN) | 0x7fff ffff | 0x0000 0000 |
| Word fixed point (result from converting any FP number too small to represent as a 32-bit negative integer) | 0x7fff ffff | 0x8000 0000 |
| Longword fixed point (result from converting any FP number too big to represent as a 64-bit positive integer) | 0x7fff ffff ffff ffff | 0x7fff ffff ffff ffff |
| Longword fixed point (result from converting any FP NAN) | 0x7fff ffff ffff ffff | 0x0000 0000 0000 0000 |
| Longword fixed point (result from converting any FP number too small to represent as a 64-bit negative integer) | 0x7fff ffff ffff ffff | 0x8000 0000 0000 0000 |

### Fixed Point Formats

The FPU provides two fixed-point data types:

- 32 bit Word fixed-point (type W)
- 64 bit Longword fixed-point (type L)

The fixed-point values are held in the 2's complement format used for signed integers in the CPU. Unsigned fixed-point data types are not provided by the architecture; application software may synthesize computations for unsigned integers from the existing instructions and data types.

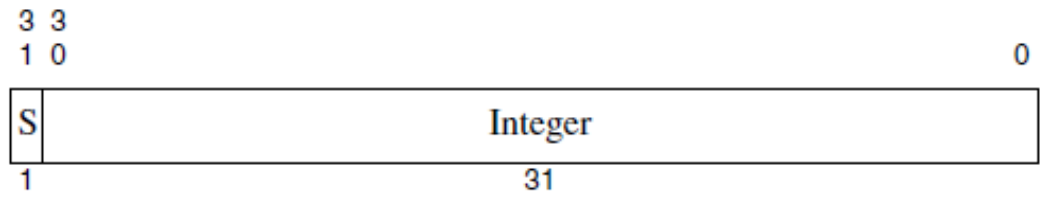

**Fig A. 3** Word Fixed Point Format (W)

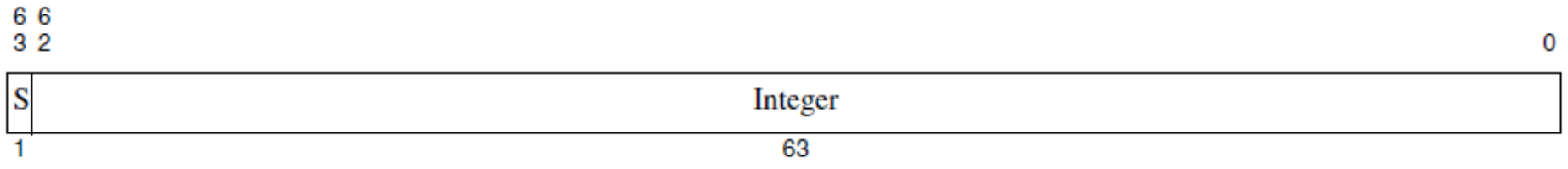

**Fig A. 4** LongWord Fixed Point Format (L)

## Floating Point Registers

This section describes the organization and use of the two types of FPU register sets:

• Floating Point General Purpose Registers (FPRs) are 32 or 64 bits wide. These registers transfer binary data between the FPU and the system, and are also used to hold formatted FPU operand values. A 32-bit FPU contains 32, 32-bit FPRs, each of which is capable of storing a 32-bit data type. A 64-bit floating point unit contains 32, 64-bit FPRs, each of which is capable of storing any data type.

• Floating Point Control Registers (FCRs) are 32 bits wide and are used to control and provide status for all floating-point operations.

In Release 6 the 32-bit register model does not support 64-bit data types (stored in even-odd pairs of registers), and 64-bit operations are required to signal the Reserved Instruction exception.

### FPU Register Models

The MIPS architecture supports two FPU register models:

- 32-bit FPU register model: 32 , 32-bit registers
  - 32-bit data types stored in any register
  - Pre-release 6 : 64-bit data types stored in even-odd pairs of registers

  In release 6 the 32-bit register model does not support 64-bit data types (Stored in even-odd pairs of registers), and 64-bit operation are required to signal the Reserved Instruction exception.

- 64-bit FPU register model: 32 , 64-bit registers, with all formats supported in a register.

Release 6 supports both FPU register models. However, with a 64-bit FPU ($FIR_{F64}$=1), Release 6 requires the 64-bit FPU register model and does not support the 32-bit FPU register model, i.e., $Status_{FR}$=1 is required. With a 32-bit FPU ($FIR_{F64}$=0, 32-bit FPRs), Release 6 does not support 64-bit data types and requires instructions manipulating such data types to signal a Reserved Instruction exception. In particular, Release 6 does not support even-odd register pairs.

In Table A.4 we show the availability and compliance requirements of FPU register widths, register models, and data types.

**Table A. 4** FPU Register Models Availability and Compliance

| ISA | MIPS32 | | | MIPS64 | | |
|---|---|---|---|---|---|---|
| FPU Type | 32-bit FPU $FIR_{F64}$=0 | 64-bit FPU $FIR_{F64}$=1 | | 32-bit FPU $FIR_{F64}$=0 | 64-bit FPU $FIR_{F64}$=1 | |
| FPU Register Width | 32 | 64 | | 32 | 64 | |
| 32-bit Data Formats S/W | 32-bit data formats S and W required[1] whenever FPU is present: $FIR_S$=1 and $FIR_W$=1 | | | | | |
| Support for 64-bit Data Types D/L | See below | $FIR_D$=1 and $FIR_L$=1 | | See below | $FIR_D$=1 and $FIR_L$=1 | |
| FPU Register Model | 32-bit | 32-bit $Status_{FR}$=0 | 64-bit $Status_{FR}$=1 | 32-bit | 32-bit $Status_{FR}$=0 | 64-bit $Status_{FR}$=1 |
| 64-bit Data Storage[2] | See below | **[even/odd register pairs]** | **[true 64-bit FPRs]** | See below | **[even/odd register pairs]** | **[true 64-bit FPRs]** |
| Release 1[3] | **[even-odd register pairs]** 64-bit data formats D/L use even/odd pairs: $FIR_D$=1 and $FIR_L$=1 required | Not Available | | Not Available | Required | Required |
| Release 2 | | Required | Optional | | | |
| Release 3 | | | | | | |
| Release 5 | | | Required | | | |
| Release 6 | **[strictly 32-bit]**[4] 64-bit data formats D/L not available: $FIR_D$=0 $FIR_L$=0 | Not Available | | Available **[strictly 32-bit]**[4] | Not Available | |

Where "Required" means "required if an FPU of specified type is present". "Available" means that the feature is available to implement, i.e., is optional. "Not available" means that the feature cannot be implemented.

## Floating point control registers (FCRs)

The MIPS64 Architecture supports the following Floating Point Control Registers (FCRs):

- *FIR:* FP Implementation and Revision Register
- *FCSR:* FP Control/Status register
- *FEXR:* FP Exceptions Register
- *FENR:* FP Enables register

Access to the FCRs is not privileged; they can be accessed by any program that can execute floating point instructions.

**Floating Point Implementation Register (FIR, CP1 Control Register 0)**

The Floating Point Implementation Register (FIR) is a 32-bit read-only register that contains information identifying the capabilities of the floating point unit, the floating point processor identification, and the revision level of the floating point unit.

Figure A.5 shows the format of the FIR register and Table A.5 describes the FIR register fields.

| 31 30 | 29 | 28 | 27 24 | 23 | 22 | 21 | 20 | 19 | 18 | 17 | 16 | 15 8 | 7 0 |
|---|---|---|---|---|---|---|---|---|---|---|---|---|---|
| 0 | FREP | UFRP | Impl | Has 2008 | F64 | L | W | 3D | PS | D | S | ProcessorID | Revision |

**Fig A. 5** FIR Register Format

**Table A. 5** FIR Register Field Descriptions

| Field | Description |
|---|---|
| Revision | Specifies the revision number of the floating point unit. This field allows software to distinguish between one revision and another of the same floating point processor type. |
| Processor ID | Identifies the floating point processor |
| S | Indicates that the single precision (S) floating point data type and instructions are implemented.<br>0 – not implemented<br>1 - implemented |
| D | Indicates that the double precision (D) floating point data type is implemented.<br>0 – not implemented<br>1 - implemented |
| PS | Indicates that the paired-single (PS) floating point data type and instructions are implemented:<br>0 – not implemented<br>1 – implemented<br>Note: In release 6 PS data type is removed. |
| 3D | Indicates that MIPS 3D is implemented:<br>0 – not implemented<br>1 – implemented |
| W | Indicates that the word fixed-point (W) data type and instructions are implemented:<br>0 – not implemented<br>1 – implemented |
| L | Indicates that the longword fixed-point (L) data type and instructions are implemented:<br>0 – not implemented<br>1 – implemented |
| F64 | Indicates that the floating point unit has registers and data paths that are 64-bits wide.<br>0 - FPU is 32 bits<br>1 – FPU is 64 bits |
| Has2008 | Indicates that one or more IEEE-754-2008 features are implemented. If this bit is set, the ABS2008 and NAN2008 field within the FCSR register also exist. |
| Impl | These bits are implementation-dependent and are not defined by the architecture. |
| UFRP | Indicates user-mode FR switching is supported. |
| FREP | User mode access of FRE is supported.<br>0 – Support for emulation of $status_{FR}$=0 handling on a 64-bit FPU with $status_{FR}$=0 only is not available.<br>1 – Support for emulation of $status_{FR}$=0 handling on a 64-bit FPU with $status_{FR}$=1 only is available. |
| 0 | Reserved |

**Floating Point Control and Status Register (FCSR, CP1 Control Register 31)**

The Floating Point Control and Status Register (FCSR) is a 32-bit register that controls the operation of the floating point unit, and shows the following information:

- Selects the default rounding mode for FPU arithmetic operations
- Selectively enables traps of FPU exceptions conditions
- Controls some denormalized number handling options
- Reports any IEEE exceptions that arose, cumulatively, in complete instructions.
- Release 6 removes the FP condition codes.

The access to FCSR is no privileged; it can be read or written by any program that has access to the floating point unit.

Figure A.6 shows the format of the FCSR register and Table A.6 describes the FCSR register fields.

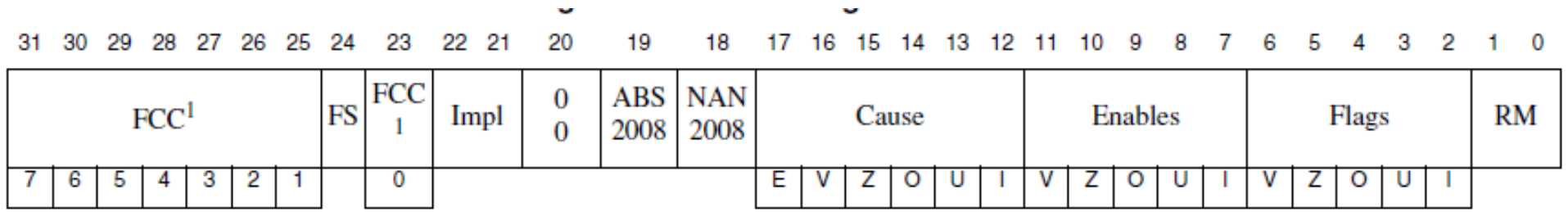

**Fig A. 6** FCSR Register Format

**Table A. 6** FCSR Register Field Descriptions

| Fields | Description |
|---|---|
| RM | Rounding Mode. This field indicates the rounding mode used for most floating point operation (some operations use a specific rounding mode). See the Table A.7 for the meaning of the encodings of this field. |
| Flags | Flag bits. This field shoes any exception conditions that have occurred for completed instructions since the flag was last reset by software.<br>When a FPU arithmetic operation raises an IEEE exception condition that does not result in a Floating point exception, the corresponding bits in the Flags field are set, while the others remain unchanged. Arithmetic operation that result in a Floating point exception do not update the *Flag* bits.<br>This field is never reset by hardware and must be explicitly reset by software.<br>Refer to Table A.8 for the meaning of each bit. |
| Enables | Enable bits. These bits control whether or not a exception is taken when an IEEE exception condition occurs for any of the five conditions. The exception occurs when both an *Enables* bit and the corresponding *Cause* bit are set either during an FPU arithmetic operation or by moving a value to FCSR or one of its alternative representations. Note that *Cause* bit E has not corresponding *Enables* bit; the non-IEEE Unimplemented Operation exception is defined by MIPS as always enabled.<br>Refer to Table A.8 for the meaning of each bit. |

| Fields | Description |
|---|---|
| Cause | Cause bits. These bits indicate the exception conditions that arise during execution of an instruction and is set to 0 otherwise. By reading the registers, the exception condition caused by the proceeding FPU arithmetic instruction can be determined. Refer to Table A.8 for the meaning of each bit. |
| NAN2008 | Quiet and Signaling NaN encodings recommended by the IEEE standard 754-2008, i.e.,a quiet NaN is encoded with the first bit of the fraction field being 0. MIPS legacy FPU encodes NaN values with the opposite polarity, i.e.,a quiet NaN is encoded with the first bit of the fraction being0 and signaling NaN is encoded with the first bit of the fraction field being 1.<br><br>Refer to Table A.3 for the quiet NaN encoding values.<br>This fields exist if $FIR_{HAS2008}$ is set.<br><br>0 – MIPS legacy NaN econding<br>1 – IEEE 754-2008 NaN encoding |
| ABS2008 | ABS.fmt and NEG.fmt instructions compliant with IEEE Standard 754-2008.<br>The IEEE 754-2008 standard requires that the ABS and NEG functions are non-arithmetic and accept NAN inputs without trapping.<br>This fields exist if $FIR_{HAS2008}$ is set.<br><br>0 – ABS and NEG intructions are arithmetic and trap for NAN inputs. MIPS legacy behavior<br>1 – ABS and NEG intructions are non-arithmetic and accept NAN inputs without trapping. IEEE-754-2008 behavior. |
| 0 | Reserved for future use; reads as zero. |
| Impl | Available to control implementation-dependent features of the floating point unit. If these bits are not implemented, they must be ignored on write and read as zero. |
| FS | Flush to Zero (Flush subnormals).<br>0 – Input subnormal values and tiny non-zero result are not altered. Un-implemented Operation exception may be signaled as needed.<br>1 – When FS is one, subnormal results are flushed to zero. The unimple-mented Operation exception is not signaled for this reason. |
| FCC | Floating point Condition Codes, removed in realease 6. |

**Table A. 7** Rounding Mode Definitions

| RM Field Encoding | Meaning |
|---|---|
| 0 | RN – Round to Nearest<br>Rounds the result to the nearest representable value. When two representable values are equally near, the result is rounded to the value whose least significant bit is zero (that is, even) |
| 1 | RZ - Round Toward Zero<br>Rounds the result to the value closest to but not greater than in magnitude than the result. |
| 2 | RP - Round Towards Plus Infinity<br>Rounds the result to the value closest to but not less than the result. |
| 3 | RM - Round Towards Minus Infinity<br>Rounds the result to the value closest to but not greater than the result. |

Most arithmetic operations do not result in a number that can be represented exactly. In such cases the result need to be rounded to a number that can be represented in a given format. IEEE-754 standard define four rounding modes listed in Table A.7.

The most popular mode is round toward nearest, ties to even. This rounding mode generally introduces the smallest error as the result of round toward nearest is the number closest to the exact value. However, certain applications such as interval arithmetic perform better on simpler rounding mode like round toward zero. For this reason, IEEE-754 includes directed rounding modes as well.

**Table A. 8** Cause, Enable, and Flag Bit definitions

| Bit Name | Bit Meaning |
|---|---|
| E | Unimplemented Operation (this bit exist only in the cause field) |
| V | Invalid operation |
| Z | Divide by Zero |
| O | Overflow |
| U | Underflow |
| I | Inexact |

**Floating Point Exception Register (FEXR, CP1 Control Register 26)**

The floating Point Exception Register (FEXR) is an alternative way to read and write the Cause and Flags fields that also appear in FCSR. Figure A.7 shows the format of the FEXR register; Table A.9 describes the FEXR register fields.

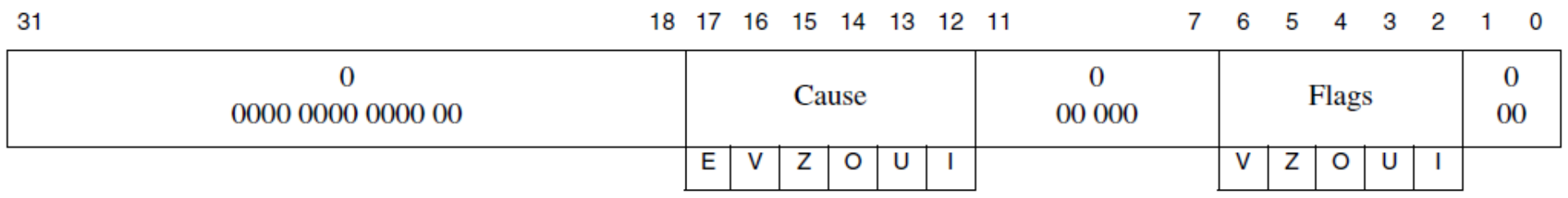

**Fig A. 7** FEXR Register Format

**Table A. 9** FENR Register Format

| Fields | Description |
|---|---|
| 0 | Must be written as zero; return zero on read |
| Cause | Cause bits. Refer to the description of this field in the FCSR Register. |
| Flags | Flags bits. Refer to the description of this field in the FCSR register. |

**Floating Point Enables Register (FENR, CP1 Control Register 28)**

The floating Point Enables Register (FENR) is an alternative way to read and write the Enables, FS and RM fields that also appear in FCSR. Figure A.8 shows the format of the FENR register; Table A.10 describes the FENR register fields.

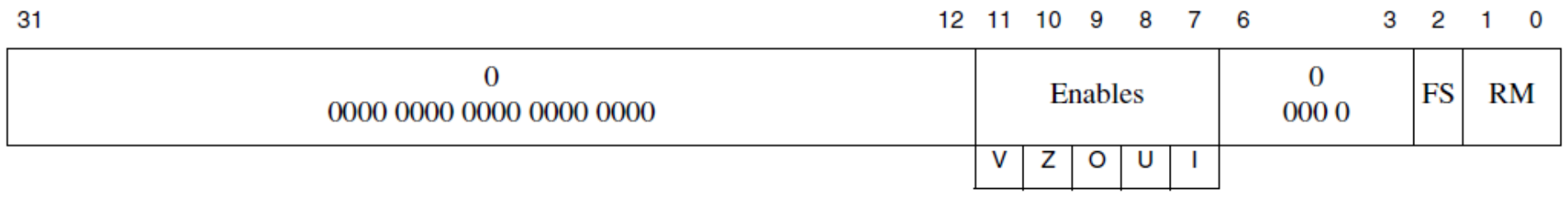

**Fig A. 8** FENR Register

**Table A. 10** FENR Register Field Description

| Fields | Description |
|---|---|
| 0 | Must be written as zero; returns zero on read |
| Enables | Enable bits. Refer to the description of this field in the FCSR register. |
| FS | Flush to zero bit. Refer to the description of this field in the FCSR register. |
| RM | Round mode. Refer to the description of this field in the FCSR register. |

## FPU Exceptions

FPU exceptions are implemented in the MIPS FPU architecture with the *Cause*, *Enable* and *Flag* fields of the *Control/Status* register. The *Flag* bits implement IEEE exception status flags, and the *Cause* and *Enable* bits control exception trapping. Each field has a bit for each of the five IEEE exception conditions and the Cause field has an additional exception bit, Unimplemented Operation, used to trap for software emulation assistance.

A trap occurs before the instruction that causes the trap, or any following instruction, can complete and write its results. If desired, the software trap handler can resume execution of the interrupted instruction stream after handling the exception.
A floating point trap is generated any time both a *Cause* bit and its corresponding *Enable* bit are set.

### Exceptions Conditions

The following five exception conditions defined by the IEEE standard are described below:

- Invalid Operation Exception
- Division by Zero Exception
- Underflow Exception
- Overflow Exception
- Inexact Exception

Also MIPS include a specific exception condition called Unimplemented Operation that is used to signal a need for software emulation of an instruction.

At the program's direction, an IEEE exception condition can either cause a trap or not cause a trap. The IEEE standard specifies the result to be delivered in case the exception is not enabled and no trap is taken. The MIPS architecture supplies these results whenever the exception condition does no result in precise trap. The default action taken depends on the type of exception condition, and in the case of the Overflow, the current rounding mode. The default results are summarized in Table A.11

**Table A. 11** Exceptions

| Bit | Description | Default Action |
|---|---|---|
| V | Invalid Operation | Supplies a Quiet NaN |
| Z | Divide by zero | Supplies a signed infinity |
| U | Underflow | Supplies a rounded result. |
| I | Inexact | Supplies a rounded result. If caused by an overflow without the overflow trap enabled, supplies the overflowed result. |
| O | Overflow | Depends on the rounding mode, as shown below. |
| | 0 (RN) | Supplies an infinity with the sign of the intermediate result. |
| | 1(RZ) | Supplies an format's largest finite number with the sign of the intermediate result. |
| | 2(RP) | For positive overflow values, supplies positive infinity. For negative overflow values, supplies the format's most negative finite number. |
| | 3(RM) | For positive overflow values, supplies the format's largest finite number. For negative overflow values, supplies minus infinity. |

**Invalid Operation Exception**

The Invalid Operation exception is signaled if one or both of the operands are invalid for the operation to be performed. The result, when the exception condition occurs without a precise trap, is a quiet NaN.

These are invalid operations:

- One or both operands are a signaling NaN (except for non-arithmetic FPU instructions such as MOV.fmt).
- Addition or subtraction: magnitude subtraction of infinities, such as (+∞) + (-∞) or (-∞) - (-∞).
- Multiplication: 0 × ∞, with any signs.
- Division: 0/0 or ∞/∞, with any signs.
- Square root: An operand of less than 0 (-0 is a valid operand value).
- Conversion of a floating point number to a fixed-point format when either an overflow or an operand value of infinity or NaN precludes a faithful representation in that format.
- Some comparison operations in which one or both of the operands is a QNaN value. (The detailed definition of the compare instruction, C.cond.fmt, in Volume II has tables showing the comparisons that do and do not signal the exception.)

**Division By Zero Exception**

An implemented divide operation signals a Division By Zero exception if the divisor is zero and the dividend is a finite nonzero number.
The result, when no precise trap occurs, is a correctly signed infinity.
Divisions (0/0) and (∞/0) do not cause the Division By Zero exception. The result of (0/0) is an Invalid Operation exception.
The result of (∞/0) is a correctly signed infinity.

**Underflow Exception**

Basically two events contribute to underflow:

- Tininess: the creation of a tiny nonzero result between $\mp 2^{E_min}$ which, because it is tiny, may cause some other exception later such as overflow on division.
- Loss of accuracy: the extraordinary loss of accuracy during the approximation of such tiny numbers by denormalized numbers.

The MIPS architecture specifies that tininess be detected after rounding.
The MIPS architecture specifies that loss of accuracy is detected as inexact result.

Alternative Flush to Zero Underflow
When register $FCSR_{FS}$=1 every tiny non-zero result is replaced with zero of the same sign.

**Overflow Exception**

An Overflow exception is signaled when the magnitude of a rounded floating point result, were the exponent range unbounded, is larger than the destination format's largest finite number.
When no precise trap occurs, the result is determined by the rounding mode and the sign of the intermediate result.

**Inexact Exception**

An Inexact exception is signaled if one of the following occurs:
- The rounded result of an operation is not exact
- The rounded result of an operation overflows without an overflow trap

**Unimplemented Operation Exception**

The Unimplemented Operation exception is a MIPS-defined exception that provides support for software emulation.
This exception is not IEEE-compliant.

Operations that are not fully supported in hardware cause an Unimplemented Operation exception so that software may perform the operation.

# APPENDIXE B FPU Instruction Set (Release 6)

The FPU instructions comprise the following functional groups:

- Data Transfer Instructions
- Arithmetic Instructions
- Conversion Instructions
- Formatted Operand-Value Move Instructions
- FPU Conditional Branch Instructions

## Data Transfer Instructions

Data is transferred between registers and the rest of the system with dedicated load, store, and move instructions.

Data Transfer instructions are listed in Table B.1 and B.2.

**Table B. 1** FPU Loads and Stores

| Mnemonic | Instruction | Defined in MIPS ISA |
|---|---|---|
| LDC1 | Load Doubleword to Floating Point | MIPS32 |
| LWC1 | Load Word to Floating Point | MIPS32 |
| SDC1 | Store Doubleword to Floating Point | MIPS32 |
| SWC1 | Store Word to Floating Point | MIPS32 |

Load and Store Instructions are executed in the Load/Store Queue, therefore only we mention about that.

**Table B. 2** FPU Move To and From Instructions

| Mnemonic | Instruction | Defined in MIPS ISA |
|---|---|---|
| CFC1 | Move control Word From Floating Point | MIPS32 |
| CTC1 | Move control Word to Floating Point | MIPS32 |
| DMFC1 | Doubleword Move From Floating Point | MIPS64 |
| DMTC1 | Doubleword Move to Floating Point | MIPS64 |
| MFC1 | Move Word From Floating Point | MIPS32 |
| MFHC1 | Move Word from High Half of Floating Point Register | MIPS32 R2 |
| MTC1 | Move Word To floating Point | MIPS32 |
| MTHC1 | Move Word to High Half of Floating Point Register | MIPS32 R2 |

Move To and From Instructions are not implement.

## Arithmetic Instructions

**FPU IEEE-Approximate arithmetic operations**

FPU IEEE-Approximate arithmetic operations are listed in Table B.3.

**Table B. 3** FPU IEEE Arithmetic Operations

| Mnemonic | Instruction | Defined in MIPS ISA |
|---|---|---|
| ADD | Floating Point Add | MIPS32 |
| CMP.cond.fmt | Floating Point Compare (setting FPR) | Release 6 |
| DIV.fmt | Floating Point Divide | MIPS32 |
| MUL.fmt | Floating Point Multiply | MIPS32 |
| SQRT.fmt | Floating Point Square Root | MIPS32 |
| SUB.fmt | Floating Point Subtrac | MIPS32 |

Instructions in green color was implemented.

**FPU Approximate arithmetic operations**

Two operations, Reciprocal Approximation (RECIP) and Reciprocal Square Root Approximation (RSQRT), may be less accurate than the IEEE specification.
FPU Approximate arithmetic operations are listed in Table B.4.

**Table B. 4** FPU-Approximate Arithmetic Operations

| Mnemonic | Instruction | Defined in MIPS ISA |
|---|---|---|
| RECIP.fmt | Floating Point Reciprocal Approximation | MIPS64 |
| RSQRT.fmt | Floating Point Reciprocal Square Root Approximation | MIPS64 |

**FPU Fused Multiply-Accumulate Instructions (Release 6)**

Release 6 provides IEEE 2008 compliant fused multiply-accumulate add and subtract instructions. These instructions are listed in Table B.5.

**Table B. 5** FPU Fused Multiply-Accumulate Instructions

| Mnemonic | Instruction | Defined in MIPS ISA |
|---|---|---|
| MADDF.fmt | Fused Floating Point Multiply Add | MIPS32 Release 6 |
| MSUBF.fmr | Fused Floating Point Multiply Subtract | MIPS32 Release 6 |

**Floating Point Comparison Instructions**

Floating point comparison instructions are listed in Table B.6.

**Table B. 6** Floating Point Comparison Instructions

| Mnemonic | Instruction | Defined in MIPS ISA |
|---|---|---|
| CLASS.fmt | Scalar Floating Point Class Mask | MIPS32 Release 6 |
| CMP.cond.fmt | Floating Point Compare | MIPS32 Release 6 |
| MAX.fmt | Floating Point Maximum | MIPS32 Release 6 |
| MAXA.fmt | Floating Point Value with Maximum Absolute Value | MIPS32 Release 6 |
| MIN.fmt | Floating Point Minimum | MIPS32 Release 6 |
| MINA.fmt | Floating Point Value with Minimum Absolute Value | MIPS32 Release 6 |

## Conversion Instructions

These instructions perform conversions between floating point and fixed point data types. Some conversion instructions use the rounding mode specified in the Floating Control/Status register (FCSR), while others specify the rounding mode directly.

**Table B. 7** FPU Conversion Operations Using the FCSR Rounding Mode

| Mnemonic | Instruction | Defined in MIPS ISA |
|---|---|---|
| CVT.D.fmt | Floating Point convert to Double Floating Point | MIPS32 |
| CVT.L.fmt | Floating Point convert to Long Fixed Point | MIPS64 |
| CVT.S.fmt | Floating Point convert to Single Floating Point | MIPS32 |
| CVT.”.fmt | Floating Point convert to Word Fixed Point | MIPS64 |
| RINT.fmt | Scalar Floating Point convert round to integer | MIPS32 release 6 |

**Table B. 8** FPU Conversion Operations Using a Directed Rounding Mode

| Mnemonic | Instruction | Defined in MIPS ISA |
|---|---|---|
| CEIL.L.fmt | Floating Point Ceiling to Long Fixed Point | MIPS64 |
| CEIL.W.fmt | Floating Point Ceiling to Word Fixed Point | MIPS32 |
| FLOOR.L.fmt | Floating Point Floor to Long Fixed Point | MIPS64 |
| FLOOR.W.fmt | Floating Point Floor to Word Fixed Point | MIPS32 |
| ROUND.L.fmt | Floating Point Round to Long Fixed Point | MIPS64 |
| ROUND.W.fmt | Floating Point Round to Word Fixed Point | MIPS32 |
| TRUNC.L.fmt | Floating Point Truncate to Long Fixed Point | MIPS64 |
| TRUNC.W.fmt | Floating Point Truncate to Word Fixed Point | MIPS32 |

## Formatted Operand-Value Move Instructions

These instructions move formatted operand values among FPU general registers.

There are four kinds of move instructions:

- Unconditional move
- Instructions which modify the sign bit (ABS.fmt and NEG.fmt when FCSRABS2008=1)
- FPU conditional select instructions, based on testing bit 0 of an FPT

**Table B. 9** FPU Formatted Unconditional Operand Move Instructions

| Mnemonic | Instruction | Defined in MIPS ISA |
|---|---|---|
| ABS.fmt | Floating Point Absolute Value | MIPS32 |
| MOV.fmt | Floating Point Move | MIPS32 |
| NEG.fmt | Floating Point Negate | MIPS32 |

**Table B. 10** FPU Conditional Select Instructions

| Mnemonic | Instruction | Defined in MIPS ISA |
|---|---|---|
| SEL.fmt | Floating Point Select | MIPS32 Release 6 |
| SELEQZ.fmt | Floating Point Select if condition Equal to Zero | MIPS32 Release 6 |
| SELNEZ.fmt | Floating Point Select if condition is Not Equal to Zero | MIPS32 Release 6 |

The Floating-point instruction format can be check in *"MIPS Architecture For Programmers"* Volume II-A [41].